# Exploring the Quantum Universe

# Pathways to Innovation and Discovery in Particle Physics

Report of the 2023 Particle Physics Project Prioritization Panel

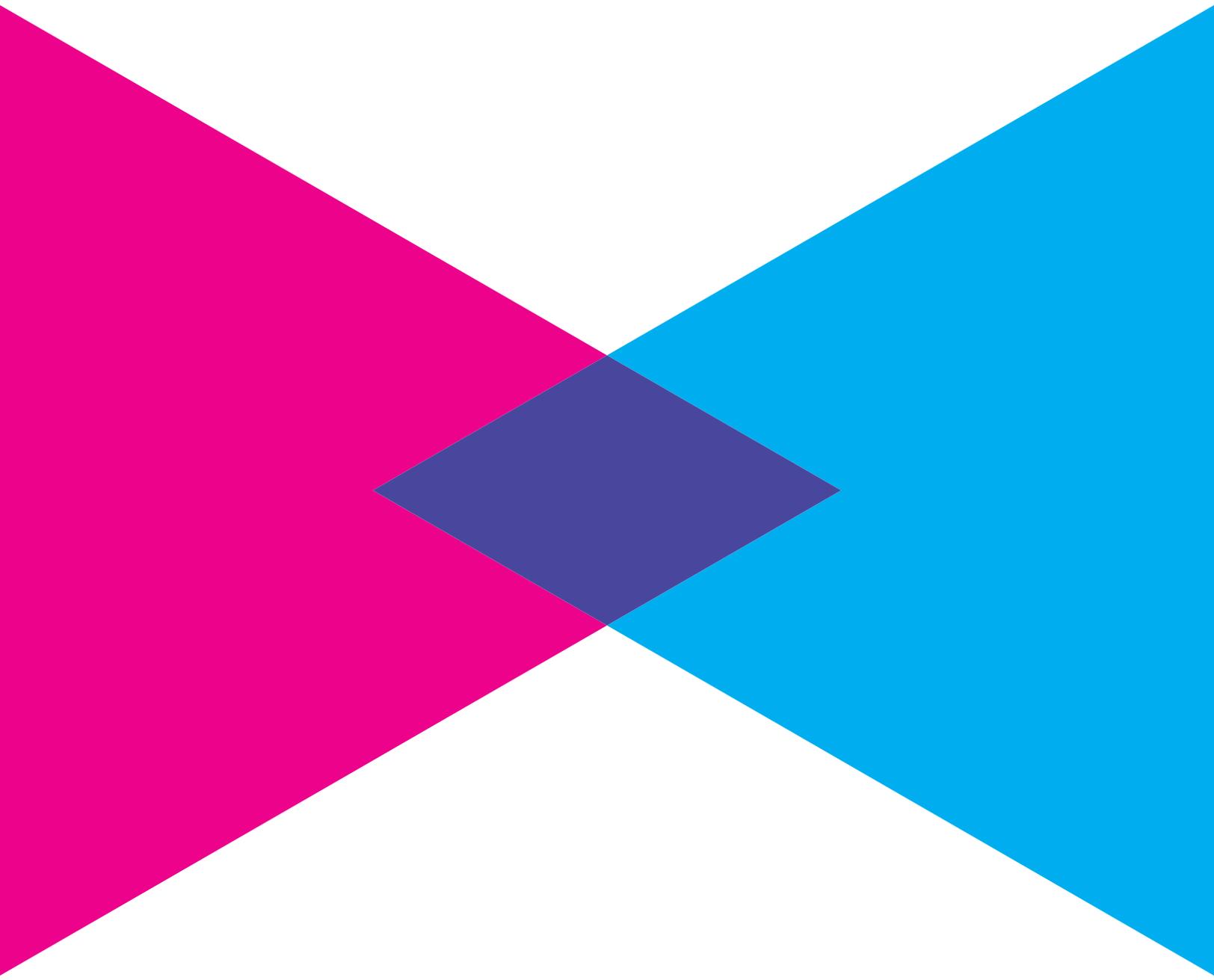

A strategic plan for the High Energy Physics Advisory Panel

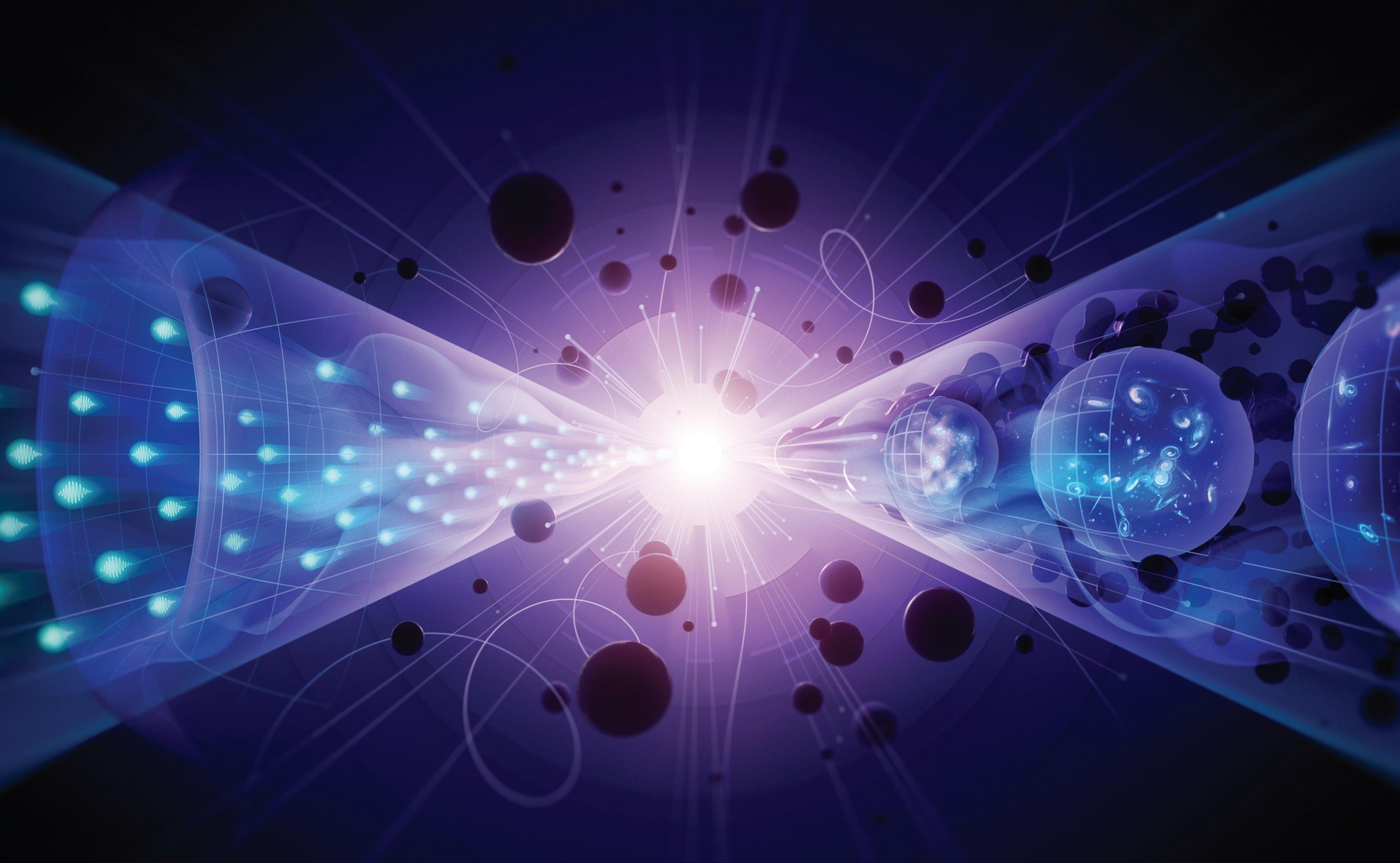

Now more than ever, particle physics is an international, even global, endeavor. The experiments needed to address the most profound questions of our field often require resources and cooperation at a global scale and can take more than a decade to design and build. We found the scope of our charge and the responsibility it represents humbling. Throughout our deliberations, we were aware that the impact of our recommendations would be felt past the end of the next decade and beyond the borders of the US particle physics program. The recommended program reflects the consensus of the panel.

The 2014 P5 report laid the foundations for the current particle physics program. Embracing its recommendations positioned the US as a leader and strong international partner in efforts that encompass neutrino and flavor physics, the study of dark matter and cosmic evolution, and collider experiments. The community stands on the threshold of realizing the enormous scientific potential of the 2014 P5 program.

At the same time, the community-driven planning process organized by the Division of Particles and Fields of the American Physical Society produced a spectrum of exciting new ideas for the future. We thank the particle physics community members—in the US and abroad—for their dedication and thoughtful input, not only through the community planning process, but in numerous town halls, talks, and private communications.

The enthusiasm and engagement of early career participants, both in the planning process and in recent town halls, has been truly inspiring. They are the future leaders who will bring to life the goals and aspirations outlined in this report.

We strove to craft a balanced program in terms of scientific focus, project timescales, and the interplay between ongoing initiatives and the innovation essential for the future. Adhering to fiscal constraints means that not every ambitious endeavor can be immediately realized. Agile, adaptable, and forward-looking projects are essential to the balance. Sustained progress over the next decade requires enhanced investment in research, theoretical frameworks, critical infrastructure, and emerging technologies. Also crucial is the commitment to build a respectful and inclusive community. We hope the resulting program enables early career researchers to spearhead progress and shape the future.

We are excited to present our vision for US particle physics, one that builds on recent investments and successes, while opening pathways to innovation and discovery in the quest to explore the quantum universe.

Respectfully submitted,
2023 Particle Physics Project Prioritization Panel

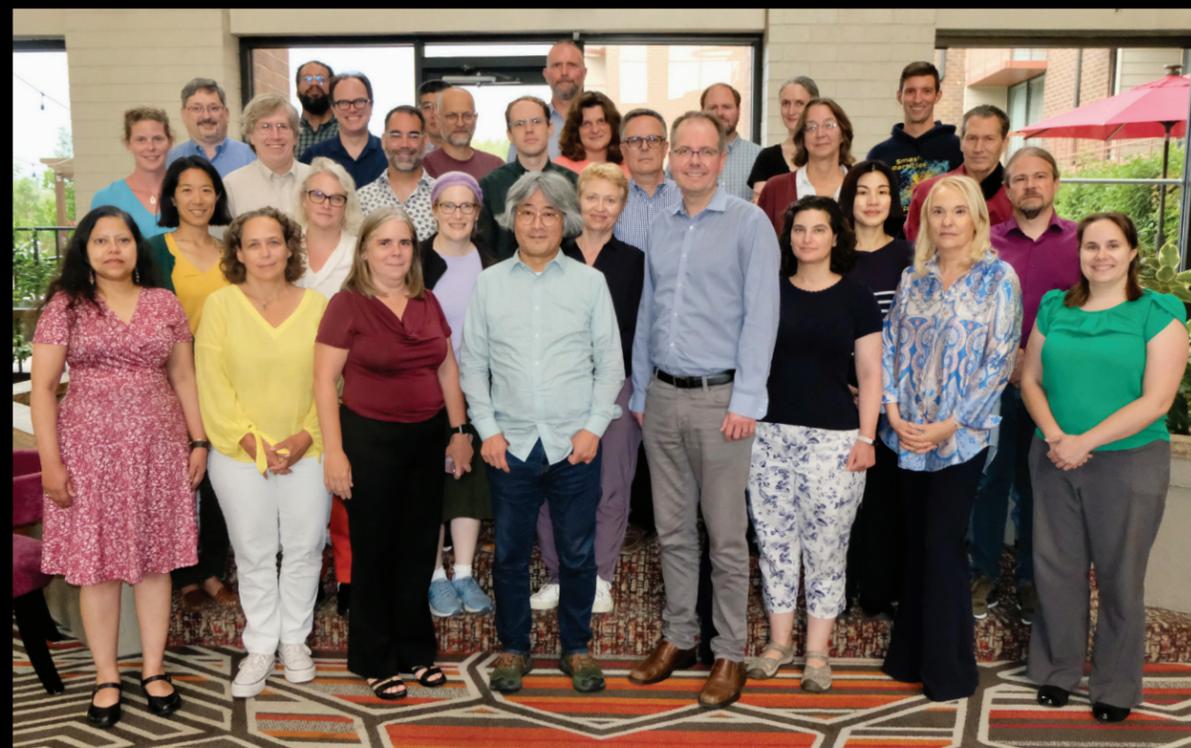

2023 Particle Physics Project Prioritization Panel in Denver, August 2023. *Photo: Rowena Smith*

Art Direction, Design and Development
Sandbox Studio, Chicago

Editing
James Dawson

Copy Editing
Marty Hanna

Mural Illustration
Olena Shmahalo

Spot Illustrations
Abigail Malate

Exploring the Quantum Universe: Pathways to Innovation and Discovery in Particle Physics

Particle physics studies the smallest constituents of our vast and complex universe. At such small scales, the fundamental principles of quantum physics prevail. Remarkably, the entire observable universe, now billions of light years across, was once small enough to be quantum in nature. Its quantum history is imprinted on its large-scale structure.

Past successes in particle physics have revolutionized our understanding of the universe and prompted a new set of questions. Collectively, these questions have spurred the construction of state-of-the-art facilities, from particle accelerators to telescopes, that will illuminate the profound connections between the very small and the very large. We stand on the threshold of a new era of insight and discovery.

The 2023 Particle Physics Project Prioritization Panel (P5) was charged with developing a 10-year strategic plan for US particle physics, in the context of a 20-year global strategy and two constrained budget scenarios. An essential source of input was the 2021 Snowmass Community Planning Exercise organized by the Division of Particles and Fields of the American Physical Society. The panel received additional input from several other sources, including town hall meetings, laboratory visits, and individual communications. We found this input aligned with three overarching science themes. Within each theme we identified two focus areas, or science drivers, that represent the most promising avenues of investigation for the next 10 to 20 years.

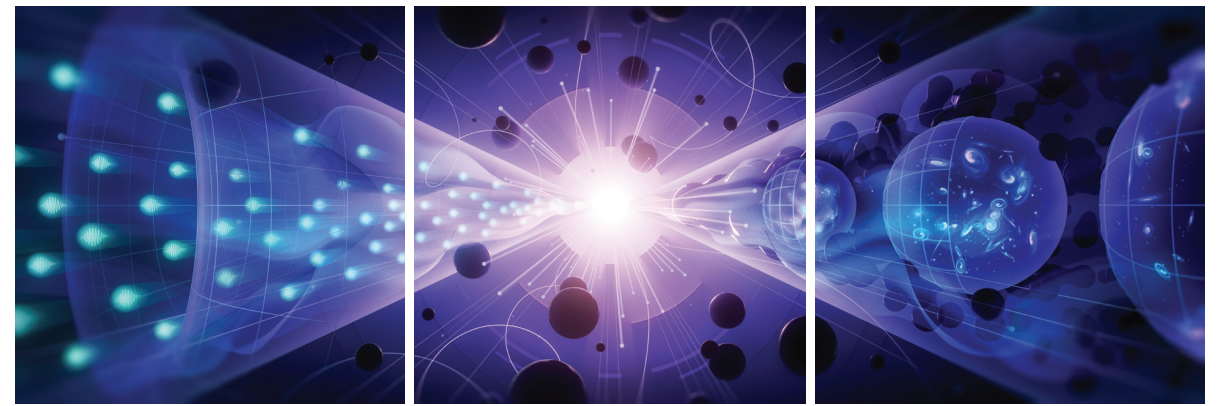

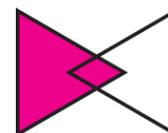 Decipher the Quantum Realm

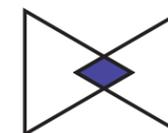 Explore New Paradigms in Physics

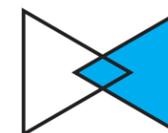 Illuminate the Hidden Universe

Elucidate the Mysteries of Neutrinos

Search for Direct Evidence of New Particles

Determine the Nature of Dark Matter

Reveal the Secrets of the Higgs Boson

Pursue Quantum Imprints of New Phenomena

Understand What Drives Cosmic Evolution

Exploring the Quantum Universe: Pathways to Innovation and Discovery in Particle Physics

The community presented P5 with more inspiring and ambitious projects than either budget scenario could accommodate. To guide the necessary choices, the panel categorized projects as small, medium, and large, based on their construction costs to the particle physics program. In the large and medium categories, initiatives were first prioritized based on individual scientific merit, then design maturity. An expert subcommittee independently reviewed the costs, technical risks, and schedule of large projects. The final prioritization holistically considered the cost of construction, commissioning, operations, and related research support, distributed over a 10- to 20-year period.

This report outlines an ambitious program balanced across science drivers. A mix of large-, medium-, and small-scale experiments ensures a continuous stream of ground-breaking research with exceptional discovery potential. The time-sequenced plan is summarized in Figure 1. We recommend continuing specific projects, strategically advancing some to the construction phase, and delaying others. Where necessary, individual phases or elements of a large-scale project are prioritized separately.

A significant consideration in the prioritization process was the execution of projects begun in the past decade. In addition to operating facilities producing excellent science, three major facilities are currently under construction: the **High-Luminosity Large Hadron Collider (HL-LHC)**, the **Deep Underground Neutrino Experiment (DUNE)**, and the **Vera C. Rubin Observatory (Rubin)**. Each plays a crucial role in a different facet of the particle physics program. Our partnership in the HL-LHC at CERN will begin to unlock the secrets of the Higgs boson. DUNE is the centerpiece of a decades-long program to reveal the mysteries of elusive neutrinos. This US-hosted international project will exploit a unique underground laboratory (now nearing completion) and neutrino beams produced at Fermi National Accelerator Laboratory (Fermilab). The Rubin Observatory Legacy Survey of Space and Time (LSST) anchors an ongoing program of cosmic surveys. Realizing the full scientific potential of these and other ongoing projects is our highest priority.

The panel identified several crucial areas in cosmic evolution, neutrinos, and dark matter, where next-generation facilities with dramatically enhanced capabilities could have revolutionary impact.

The **Cosmic Microwave Background Stage IV experiment (CMB-S4)** will use telescopes sited in Chile and Antarctica to study the oldest light from the beginning of the universe. Achieving its goals requires unique US infrastructure at the South Pole.

Early implementation of a planned accelerator upgrade (Main Injector Ramp and Target; ACE-MIRT) at Fermilab advances the timeline of the DUNE program. The **re-envisioned second phase of DUNE**, which adds a third underground detector module and an upgraded near detector complex, further expands DUNE's power and scope as a neutrino laboratory. We recommend research and development (R&D) toward an advanced fourth detector that could ultimately expand DUNE's physics program.

A comprehensive program that includes a **Generation 3 (G3) Dark Matter experiment** will probe the enigmatic nature of dark matter, which makes up a significant portion of the universe's mass and energy and has been one of the most enduring mysteries in modern physics. The recommended program also invests in multi-messenger observatories with dark matter sensitivity, including **IceCube Gen-2**, and **small-scale dark matter experiments** using innovative technologies.

In the area of colliders, the panel endorses an **offshore Higgs factory**, located in either Europe, including CERN, or Japan, to advance studies of the Higgs boson following the HL-LHC while maintaining a healthy onshore particle physics program. The US should actively engage in design studies to establish the technical feasibility and cost envelope of Higgs factory designs. We recommend that a targeted collider panel review the options after feasibility studies converge. At that point, it is recommended that the US commit funds commensurate with its involvement in the LHC and HL-LHC.

In addition to these major initiatives, the panel recommends support for a series of **current and future mid-scale projects** related to cosmic evolution, neutrinos, dark matter, and quantum imprints of new phenomena.

Small- and mid-scale projects play an essential role, complementing major facilities and **ensuring the scientific balance of the proposed US program**. Notably, small projects can rapidly seize on new opportunities as they arise. To preserve this agility, the panel recommends that the Department of Energy (DOE) create a new, competitive program named **Advancing Science and Technology through Agile Experiments (ASTAE)** to support a **portfolio of small-scale and agile experiments**. This program complements the successful **Mid-Scale Research Infrastructure (MSRI)** and **Major Research Instrumentation (MRI)** programs at the National Science Foundation (NSF). We recommend continued support for these vital components of the NSF research portfolio.

The panel recommends dedicated R&D to explore a suite of promising future projects. One of the most ambitious is a future collider concept: a **10 TeV parton center-of-momentum (pCM) collider** to search for direct evidence and quantum imprints of new physics at unprecedented energies. Turning this concept into a cost-effective, realistic collider design demands that we aggressively develop multiple innovative accelerator and detector technologies. This process will establish whether a proton, electron, or muon accelerator is the optimal path to our goal.

As part of this initiative, we recommend **targeted collider R&D** to establish the feasibility of a **10 TeV pCM muon collider**. A key milestone on this path is to design a muon collider demonstrator facility. If favorably reviewed by the collider panel, such a facility would open the door to building facilities at Fermilab that test muon collider design elements while producing exceptionally bright muon and neutrino beams. By taking up this challenge, the US blazes a trail toward a new future by advancing critical R&D that can benefit multiple science drivers and ultimately bring an unparalleled global facility to US soil.

**Investing in the scientific workforce and enhancing computational and technological infrastructure is crucial.** To achieve this goal, funding agencies should support programs that foster a supportive, collaborative work environment; help recruit and retain diverse talent; and reinforce professional standards. Targeted increases in support for **theory**, **general accelerator R&D (GARD)**, **instrumentation**, and **computing** will bolster areas where US leadership has begun to erode. These areas align with national initiatives in **artificial intelligence and machine learning (AI/ML)**, **quantum information science (QIS)**, and **microelectronics**, creating valuable synergies. Such increased support maximizes the return on scientific investments, fosters innovation, and benefits society in domains from medicine to national security.

The impact of the more constrained budget scenario is severe. It forces the US to cede leadership in many initiatives to other regions of the world and reduces investments in workforce and future technologies to critical levels. Specifically, the G3 dark matter exploration would be relocated outside the US, and elements of DUNE would be descoped or delayed. This limiting of DUNE's physics reach would negatively impact the reputation of the US as an international host, and more limited contributions to an offshore Higgs factory would tarnish our standing as a partner for future global facilities. Conversely, a modest increase in budget over the less constrained scenario allows for additional pathways to discovery by accelerating science, leveraging existing investments, and solidifying US scientific leadership in the international context.

We have crafted a well-balanced program that is technically and financially feasible for the coming 10 years and builds toward a longer-term vision of US leadership and scientific discovery in particle physics. As this program is realized, we look forward to sharing new insights into the quantum universe.

# 1

## Introduction

# 1.1

# Overview and Vision

Curiosity-driven research is at the core of particle physics, a field of science in which we study the building blocks of the subatomic world. In examining these point-like particles and their interactions, we *decipher the quantum realm*. We also look out into the universe, beyond the visible stars, by building instruments that can *illuminate the hidden universe*. By studying the very small and the very large, realms that are beyond the limits of normal human perception, we expand our understanding of the world around us and begin to grasp our place in the cosmos. Going beyond phenomena that we can probe using current experiments, we can use theoretical principles to test our current physics understanding and predict new particles and new phenomena; in this way we *explore new paradigms in physics*.

Within each of these broad themes, we identify compelling questions that define our priorities and drive what instruments we build and what experiments we design. These science drivers change over time, as new discoveries are made and our understanding deepens.

Informed by the community-driven Snowmass planning process, we have identified a new set of three **science themes** and six **science drivers**. The drivers evolved from those in the 2014 P5 report.

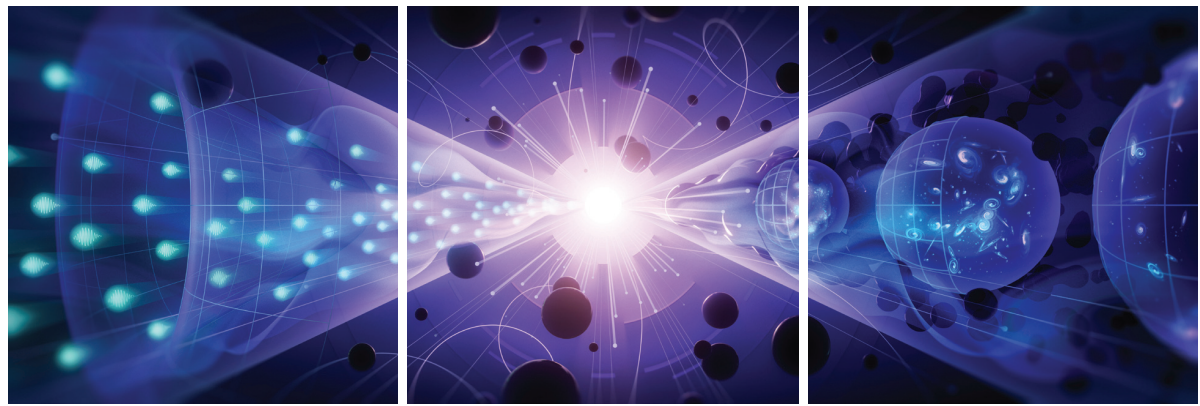

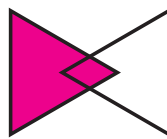 **Decipher the Quantum Realm**

Elucidate the Mysteries of Neutrinos

Reveal the Secrets of the Higgs Boson

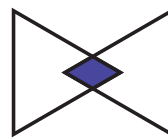 **Explore New Paradigms in Physics**

Search for Direct Evidence of New Particles

Pursue Quantum Imprints of New Phenomena

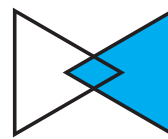 **Illuminate the Hidden Universe**

Determine the Nature of Dark Matter

Understand What Drives Cosmic Evolution

The US particle physics program would not exist without the sustained support of the Department of Energy (DOE) and the National Science Foundation (NSF). These agencies have nurtured a world-class scientific enterprise. From the discovery of the Higgs boson and the revelation of the neutrino masses to the stunning realization that the expansion of the universe is now accelerating, this effort has revolutionized our understanding of how the universe works. With each new insight into the universe, scientific exploration encounters escalating challenges. The tools and technologies required to meet those challenges become increasingly sophisticated.

Particle physics now operates on a global scale. In isolation, no nation has the financial resources, workforce, or technical capacity required to tackle all of the most pressing questions of our field. Yet that same scientific program is within the grasp of a collaborative global effort. The US is a major player on that stage, currently hosting an international program in the study of neutrinos and charged leptons and in dark matter physics and cosmology. We are a major partner in offshore high-energy collider facilities in Europe and Japan. Each of these initiatives represents a cornerstone in our collective effort to push the boundaries of understanding in particle physics.

Since the 2014 P5 report, the US has made significant investments in expanding its capabilities for groundbreaking discoveries in accelerators and deep underground laboratories. The High-Luminosity Large Hadron Collider (HL-LHC) upgrade is proceeding successfully with critical US contributions. This project addresses key questions about the Higgs boson while searching for new particles and phenomena. Concurrently, the construction of the Deep Underground Neutrino Experiment (DUNE) and the Long Baseline Neutrino Facility (LBNF) is establishing a world-leading experiment for precision neutrino studies. This international mega-project on US soil positions the US as a potential host for future projects. The commissioning of the world's largest digital camera for the Legacy Survey of Space and Time (LSST) is underway, set to be deployed at the nearly completed Vera C. Rubin Observatory. These projects hold immense potential for producing groundbreaking scientific discoveries in the coming decade.

Mid-scale projects recommended for construction in the previous report, such as the Generation 2 dark matter searches, are either nearing completion or producing exciting new results. Highlights from the mid-scale program include an early data release from the Dark Energy Spectroscopic Instrument (DESI) and the short baseline neutrino (SBN) experiments setting stringent limits on the existence of sterile neutrinos. The Muon g-2 experiment measuring the anomalous magnetic moment of the muon has observed a discrepancy between the measured value and the value predicted by the Standard Model of particle physics, a result that spurs further theoretical developments.

We envision a new era of scientific leadership, centered on decoding the quantum realm, unveiling the hidden universe, and exploring novel paradigms. Balancing current and future large- and mid-scale projects with the agility of small projects is crucial to our vision. We emphasize the importance of investing in a highly skilled scientific workforce and enhancing computational and technological infrastructure. Acknowledging the global nature of particle physics, we recognize the importance of international cooperation and sustainability in project planning. We seek to open pathways to innovation and discovery that offer new insights into the mysteries of the quantum universe.

# 1.2

# The Particle Physics Landscape

Particle physics has been increasingly successful in recent decades in describing matter, its interactions, and the 14-billion-year evolution of the universe. Our understanding is captured by the Standard Model and the $\Lambda$CDM model, a minimal paradigm of cosmology. The development of these paradigms based on detailed experimental investigations and deep theoretical principles is a triumph and hallmark for the field.

We strive to answer the questions these paradigms cannot yet address. Here we discuss which existing projects remain central to answering key questions in the current scientific landscape as embodied in the 2023 science drivers. We highlight where experimental results demand new initiatives and how theoretical developments of the decade influence our path forward. Finally, we note key investments in accelerator technology, detector instrumentation, computing, and theory crucial to the long-term future of the field.

## 1.2.1 — Decipher the Quantum Realm

The Standard Model is a remarkable achievement. It provides a comprehensive description of all known fundamental particles and their interactions. In that framework, the Higgs boson is the key to understanding the origin of particle mass, because particles interacting with the Higgs field acquire mass through the Higgs mechanism. However, not all particles behave that way. Neutrinos, the tiniest and most elusive matter particles, once assumed to be massless, appear to defy the predictions of the Standard Model. Their ability to oscillate between flavors is linked to their mass. Yet the connection between the Higgs mechanism and neutrino masses remains a mystery.

Elucidating the mysteries of neutrinos and revealing the secrets of the Higgs boson are essential for understanding the complete picture of particle physics. Each of these science drivers, with its potential to challenge the Standard Model, can act as a key to unlock the quantum realm.

### Elucidate the Mysteries of Neutrinos

Neutrino interactions are rare, and their behavior is unique. With technology available today we can detect neutrinos from various sources and produce them in large numbers by accelerating intense proton beams on a target. We can aim the produced neutrinos at detectors from a few kilometers (short-baseline) to more than a thousand kilometers (long-baseline) away to observe how their nature oscillates as they travel.

Experiments in SBN oscillation are more compact in scope, allowing us to pursue specific questions and test new technologies. The SBN effort as recommended by the 2014 P5 is well underway. The design of long-baseline experiments, with detectors placed both close to and far from the beam, allows them to act as many experiments in one. For example, the large volume of a far detector allows it to detect neutrinos from supernovae and the sun in parallel with long-baseline measurements.

We stand on the verge of precision neutrino physics. Recent results from the NOvA and T2K neutrino experiments have confirmed the benchmarks future neutrino experiments must meet. Following the recommendations of the previous P5, the DUNE experiment is under construction, with LBNF leveraging the Fermi National Accelerator Laboratory (Fermilab) proton complex to deliver an intense neutrino beam to the nearly complete Sanford Underground Research Facility (SURF) in South Dakota. DUNE will use massive, cryogenic liquid-argon (LAr) time-projection chambers (LArTPCs) and the intense beam to comprehensively determine the structure of neutrino mixings and the pattern of their masses. The realization and operation of the ProtoDUNE detector prototypes at the CERN Neutrino Platform successfully demonstrated the LArTPC technology for DUNE. This program is poised to acquire large neutrino interaction datasets to challenge the validity of the neutrino oscillation framework.

Both short- and long-baseline programs offer opportunities to search for signatures of unexpected neutrino interactions. They are complemented by small experiments like COHERENT at the Oak Ridge National Laboratory, which recently announced the discovery of a formerly undetected neutrino interaction mode, coherent elastic neutrino-nucleus scattering. A program of neutrino experiments at different scales can ensure that we have the tools and understanding of neutrino interactions and production to enable the potential for discovery of the large facilities being constructed.

If a departure from the current neutrino oscillation framework were to be found, there would be interest in muon storage rings as a neutrino source offering precise and well-characterized neutrino beams. A so-called neutrino factory, whose technology requires research and development (R&D), could propel the neutrino program to new heights of precision oscillation studies.

### Reveal the Secrets of the Higgs Boson

The Higgs boson, a unique elementary particle devoid of spin, can interact with all known matter particles. This particle was discovered at the Large Hadron Collider (LHC) with crucial contributions from the US community, just prior to the last P5 report. The existence of the Higgs raises questions about why it is "frozen" in the universe, imbued in a field that gives mass to all elementary particles. To understand the pervasive influence of the Higgs boson, the interactions of the Higgs field with itself, which determine its potential energy, must be thoroughly studied. To date, measurements of the Higgs at the LHC agree with the predictions of the Standard Model. The ATLAS (A Toroidal LHC Apparatus) and CMS (Compact Muon Solenoid) experiments have achieved precise measurements of the Higgs boson mass, confirmed its spin as zero, and measured its lifetime. However, many questions remain about its nature.

Machines like the LHC collide high-energy beams of protons to investigate their fundamental structure and produce particles of higher mass. The HL-LHC upgrade, starting in 2029, will continue to push the boundaries of our understanding of the Higgs boson by increasing the rate of particle collisions to obtain on the order of 400 million Higgs bosons. It is possible that the Higgs could decay to entirely novel particle families connected to physics beyond the Standard Model. Upgraded detectors and advances in software and computing, including artificial intelligence/machine learning (AI/ML), will enable the experiments to detect rare events with higher efficiency and greater purity.

These studies will eventually be limited by the challenges of using proton beams, which are composite objects made of quarks and gluons. The next step is to use electron and positron beams to construct a Higgs factory, which would allow precision measurements of the Higgs boson properties and searches for exotic decays, possibly into dark matter.

Precision studies of the Higgs self-interaction and searches for possible new spinless particles related to the Higgs require much larger energies per fundamental particle (parton) interaction than previously considered: on the order of 10 TeV or more. For lepton colliders, this is the nominal collision energy in the center-of-momentum (CM) frame. For proton colliders, the parton-parton interaction energy is roughly a tenth of the CM energy. We refer henceforth to a 10 TeV parton-center-of-momentum (pCM) collider, since this term applies equally to colliders of all types. Realizing such a collider has impacts beyond the Higgs science driver.

## 1.2.2 — Illuminate the Hidden Universe

When we look backward in cosmic time, we see a universe very different from the one we know today. The universe has evolved from early moments of rapid expansion (cosmic inflation), which left behind the seeds of its future structure, to intermediate periods dominated by radiation (potentially including unknown light particle species) and dark matter, to our current epoch of accelerated expansion, driven by an unknown component we call dark energy.

The ΛCDM paradigm captures the physics that governs this evolution, which is outside the Standard Model. The two science drivers discussed in this section investigate, and potentially challenge, different aspects of this paradigm.

### Determine the Nature of Dark Matter

Our observations of the universe tell us that dark matter exists, but we have yet to determine its nature. Cosmic surveys, including LSST and DESI, probe the distribution of dark matter on a variety of length scales and yield essential data about its properties. The remaining efforts, which use a blend of underground facilities, telescopes, quantum sensors, and accelerator-based probes, focus on detecting particle dark matter candidates.

Until now, the study of dark matter has focused on a class of theoretically well-motivated candidates called weakly interacting massive particles (WIMPs). Generation-2 WIMP dark matter direct-detection experiments recommended in 2014 are in progress. Several are already refining our understanding of dark matter. The next generation of direct WIMP searches will need to be so sensitive that even the elusive neutrinos will be background noise. Reaching this so-called neutrino fog is a milestone, at which we cover a large fraction of WIMP theories. Navigating through the neutrino fog will require innovative thinking and R&D of novel technologies. Testing the full range of WIMP theories could require a collider with at least 10 TeV pCM.

These endeavors complement efforts to study non-WIMP dark matter candidates such as the quantum chromodynamics (QCD) axion and hidden sector particles. Since the 2014 P5 report, both theoretical and experimental efforts to study these theoretically well-motivated candidates have matured. These candidates can be searched for even with small-scale experiments.

A class of heavy WIMP dark matter candidates would produce astrophysical signals that reflect their nature. Searches for these signals are part of a broader multi-messenger astrophysics program that maps our universe with light, neutrinos, and gravitational waves. Gamma-ray observatories and the neutrino observatory IceCube have begun to place interesting constraints on these candidates, with substantial advances promised by the far more sensitive next-generation observatories.

### Understand What Drives Cosmic Evolution

Light from the early universe, known as the cosmic microwave background (CMB), carries the imprint of quantum fluctuations left behind by cosmic inflation. Precision measurements of the polarization of the CMB have already shaped our understanding of inflation and constrained certain neutrino properties. Future surveys, built on well-established techniques, aim to study the imprint of gravitational waves from the inflation era on the CMB. These observations will also constrain new light particle species that may have influenced the universe's evolution at an early stage.

Galaxy surveys using both imaging and spectroscopy have generated major advances in robust, precise constraints on the accelerated expansion rate of the universe. Current data reveal challenges explaining both early- and late-universe observations within the ΛCDM model. More and higher precision data are needed to understand the implications. The complementary spectroscopic and imaging surveys DESI and LSST will take that next step in exploring the limits of the current cosmological paradigm and determining whether it needs to be modified. DESI is already yielding results, while the LSST is expected to begin in 2025. These experiments will be at their most powerful when combined with each other and with current and future CMB measurements.

Early results from both DESI and LSST will shape future priorities. Together with a potential DESI upgrade, they will inform the design of a next-generation spectroscopic survey by telling us which potential science goals—inflation, late-time cosmic acceleration, light relics, neutrino masses, and dark matter—should be emphasized. LSST science results will drive future survey concepts for the Rubin Observatory. A new technique based on line intensity mapping (LIM) could provide a more complete view of the intermediate period between the period of inflation and our current era of accelerated cosmic expansion. Further study is needed to establish its feasibility.

### 1.2.3 — Explore New Paradigms in Physics

To gain a deeper understanding of the quantum universe, we must map out unexplored territory beyond the Standard Model and ΛCDM and understand how these two paradigms fit together. Particle accelerators allow us to search for new particles directly and to seek quantum imprints of new phenomena currently beyond our direct reach. This theme is the evolution of the "Explore the Unknown" science driver from 2014.

#### Search for Direct Evidence of New Particles

Direct searches for new heavy particles using high-energy accelerators have historically been a strong driver of progress in particle physics. A decade of direct searches for new particles at the LHC has produced a treasure trove of data and a wealth of innovative analyses that leverage AI/ML techniques. The discovery of the Higgs boson at the electroweak scale was a triumph, but its small mass remains a mystery. Theoretical and experimental studies indicate that a comprehensive study of the electroweak scale requires colliders with energy of at least 10 TeV pCM, larger than previously assumed.

There is new interest in searching for relatively light but weakly coupled new particles, such as dark matter particles from hidden sector models. Weak coupling implies their production is rare. In this search, our strategy is not to push for the highest possible energy, but for higher-intensity accelerators. Large datasets from a Higgs factory will enable direct searches for feebly coupled light states. In the case of hidden sector dark matter, accelerator-based searches using existing beam dumps are sensitive to benchmark models in the MeV−GeV mass ranges. Intensity upgrades at the proton beamline at Fermilab and focused theory efforts may uncover exciting new lines of inquiry.

#### Pursue Quantum Imprints of New Phenomena

New phenomena at high energies may be probed through their low-energy quantum imprints. Quantum imprints can manifest in a diverse range of systems, from the relatively light muons and bottom, charm, and strange quarks, to the heaviest known fundamental particles, the top quark and the Higgs boson. The pursuit of these subtle effects requires large data samples. This in turn requires accelerators producing high-intensity beams and precision detectors that can handle the intensity.

The LHCb and Belle II experiments use, respectively, a proton accelerator in Europe and an electron accelerator in Japan for their precision studies of bottom quark systems. Modest upgrades to these experiments will expand their already unprecedented datasets.

The successful completion of the Muon g-2 experiment has left us with a tantalizing hint of new physics. New techniques it developed for producing and transporting muons are applicable to other current and future experiments. Another initiative, Mu2e, searches for a muon changing to an electron without producing any neutrinos. Detection of this charged lepton flavor violation effect would be a revolutionary indication of new physics. Mu2e is nearing completion and poised to begin its first run in 2026, with data-taking con-tinuing intermittently until the end of the decade. Future experiments of this type depend on upgrades to Fermilab's accelerator complex. R&D would be required to establish the feasibility of these upgrades.

In addition to studying the nature of the Higgs boson, a Higgs factory would be a highly sensitive probe of quantum imprints of new phenomena. Precision measurements of Higgs couplings could yield information about extended Higgs sectors. The large samples of W and Z bosons produced at a Higgs factory would support exceptionally precise studies of electroweak interactions, studies that indirectly probe an energy scale well beyond the HL-LHC.

### 1.2.4 — Interconnected Opportunities

The science themes and drivers above provide a useful organizing principle, and they are deeply interconnected. For example, studies of cosmic evolution provide critical information not only about dark energy and inflation, but also about light relic particles in the early universe, dark matter, and the scale of the neutrino masses. Complementary measurements obtained across science drivers provide a powerful tool for testing paradigms.

It is perhaps not surprising that a common thread across many drivers is the need for more powerful accelerators. Future studies of neutrinos and charged lepton flavor violation would likely need higher-intensity neutrino, muon, and proton beams. Revealing the secrets of the Higgs boson, characterizing WIMP dark matter, and searching for direct evidence of new particles ultimately requires access to the electroweak scale provided by a collider with pCM energy of 10 TeV.

We do not yet have a technology capable of building a 10 TeV pCM energy machine, but the case for one is clear. Extensive R&D is required to develop cost-effective options. Possibilities include proton beams with high-field magnets, muon beams that require rapid capture and acceleration of muons within their short lifetime, and conceivably electron and positron beams with wakefield acceleration. All three approaches have the potential to revolutionize the field.

A demonstrator facility along the path to a 10 TeV pCM muon collider could fit into the evolution of the accelerator complex at Fermilab. Such a demonstrator might produce intense muon and neutrino beams in addition to performing critical R&D; it could leverage expertise in muon and neutrino beam facilities developed over the past decade. The improved accelerator complex could also support beam-dump and fixed-target experiments for direct searches and quantum imprints of new physics. This R&D path therefore aligns with five of the six science drivers.

# 1.3
## Enabling Capabilities and National Initiatives

The US has world-renowned capabilities in particle physics. The national laboratories operate some of the most powerful and sophisticated particle accelerators and detectors in the world and attract some of the brightest minds in science and engineering. Fermilab, the only single-purpose laboratory for particle physics in the US, and Argonne (ANL), Brookhaven (BNL), Lawrence Berkeley (LBNL) and SLAC National Laboratories, with their complementary strengths, provide unique infrastructure and technical capabilities for innovation and discovery in particle physics. US universities play an important role in experimental and theoretical research and in educating the next generation of scientists.

US national initiatives in quantum science, microelectronics, and AI/ML as well as general accelerator and detector R&D have an outsized impact on the field. They lead to new approaches in experiment and theory, and they inspire new experiments and detector upgrades. The US program leads in the application of AI/ML to particle physics, and recent advances in computing are beginning to revolutionize detector development, data taking, analysis, simulation, and accelerator design. The field continues to make substantial contributions to the initiatives on quantum information science and microelectronics. The US is also leading theoretical developments in particle physics, which has been crucial in providing guidance to experiments, interpreting data, and uncovering fundamental theories.

New particle detectors with enhanced sensitivity and accelerators that provide beams at higher energies and intensities have been key to the advancement of the field. Particle physics is the steward of accelerator R&D, together with our partners in nuclear physics, basic energy sciences (BES), accelerator R&D and production, and applied science.

In addition, the development of infrastructure supporting scientific research at the South Pole and the creation of a deep underground laboratory in the US have opened new facilities for discovery science. The South Pole station and surrounding science laboratories are a one-of-a-kind research facility maintained by the US. The unique location provides an unparalleled view of the universe and allows for science that is not accessible elsewhere. The infrastructure and support of the South Pole station and its science program are critical for the study of the CMB and astrophysical neutrinos.

The construction of SURF in the US enables the precision studies of neutrinos and the search for dark matter in an environment shielded by Earth. With SURF, the US has created a premier underground laboratory that is built on a decades-old distinguished history. The realization of this facility adds unparalleled infrastructure capability to the suite of national laboratories in the US. SURF enables the US to be an international host for the neutrino and dark matter experiments recommended in this report.

# 1.4
## Impact on Society

Pushing the frontiers of human knowledge in particle physics requires global scientific collaborations, state-of-the-art research facilities, and infrastructure to support the ambitious projects that advance science. In collaborating on and coordinating large international efforts, scientists connect irrespective of their backgrounds, making particle physics an enterprise that transcends borders, boundaries, cultures, and societies.

Discoveries of new laws of nature that deepen our understanding of the inner workings of the universe not only excite scientists but continue to amaze the public and inspire many people to pursue careers in science, technology, engineering, and mathematics—the STEM fields. Nurturing and developing the next generation of scientists and training a highly-skilled workforce are important benefits to society.

Particle physics has a long-proven record of creating new technologies that have revolutionized our daily experience, such as the world wide web and life-saving medical applications—from cancer treatment to medical imaging—derived from particle accelerator and detector technologies. The tools developed for particle physics research also impact society through their adoption by other scientific fields. Applications in chemistry and materials science, for instance, have led to innovations in drug discovery and the design and realization of novel materials.

Particle physics provides a training ground for a skilled workforce that drives not only fundamental science, but quantum information science, AI/ML, computational modeling, finance, national security, and microelectronics. The US has led the way in many groundbreaking discoveries in particle physics and is poised to continue its leadership role with sustained investment.

# 1.5
## Process and Criteria

The 2023 P5 is charged with developing a fiscally viable 10-year strategic plan for US particle physics. The charge further specified that this 10-year plan should fit in the context of a 20-year vision (see Appendix 1).

The recommended program must reflect the scientific interests of the particle physics community. The 2021 Snowmass Community Planning Exercise, organized by the Division of Particles and Fields of the American Physical Society, provided initial input to the deliberations of the P5 panel. To fully capture the views of the community, the panel solicited additional input through town hall meetings, laboratory visits, and individual communications. The panel was especially encouraged by the active participation of early career members in the community-driven planning process. They represent the future of our field and are essential to the realization of the goals and aspirations detailed in this report.

During the panel's deliberations, there was consensus that the overall program should enable US leadership in core areas of particle physics. The program should leverage unique US facilities and capabilities, engage with core national initiatives to develop key technologies, and develop a skilled workforce for the future that draws on US talent. Effective engagement and leadership in international endeavors were also considerations.

The community presented P5 with more inspiring and ambitious projects than fiscal reality could accommodate. In selecting projects, the panel considered a project's individual scientific merit and potential for transformational discovery as well as how well the project met criteria for the overall program.

The DOE provided the panel with two budget scenarios for High Energy Physics (HEP) derived from realistic near-term budget projections. The *baseline* scenario assumes budget levels for HEP for fiscal years 2023 through 2027 that are specified in the Creating Helpful Incentives to Produce Semiconductors (CHIPS) and Science Act of 2022. The baseline budget scenario then increases by 3% per year from fiscal year 2028 through 2033. The less favorable scenario assumes increases of 2% per year from fiscal year 2024 to 2033.

The panel was asked to develop DOE programs consistent with these scenarios. Prioritization of projects that would receive funding from the DOE therefore had to consider both project cost and the uncertainties in that cost related to the project's technical readiness and design maturity. For projects where the US contribution is expected to come entirely from NSF, the panel was only asked to consider scientific relevance to the US particle physics program. We note that in a number of cases, the NSF science case for a jointly funded project extends beyond particle physics into astrophysics.

The panel categorized projects as small (<$50M), medium ($50M–250M), and large (>$250M) based on the US contribution to their construction cost. In the large and medium categories, initiatives were first prioritized based on individual scientific merit, then assessed on project maturity and technical risk. The balance of project timescales was also considered, in order to ensure delivery of scientific results throughout the decade and opportunities for scientists at all career levels. The final prioritization holistically considered the cost of construction, commissioning, operations, and related research support, distributed over a 10- to 20-year period. As per the charge, the panel generally did not consider individual small projects, but did note areas where such projects could be particularly effective. As part of this process, the panel established budget profiles in FY23 dollars with assumptions on inflation as described in section 8.

To help the P5 panel better understand costs, schedules, and risks of the large projects, a subcommittee was convened with project management and technical experts from the community. The subcommittee provided an independent assessment of the cost range, schedule, and risks of the major projects. That input was used to assess the most likely cost scenarios and reduce the chance of unexpected budget overruns from current and new projects into the next decade.

The panel also considered broader questions of how this program should be carried out. We discuss ethical conduct in the field and present recommendations to build a more inclusive and respectful community that draws on all talent in the nation and beyond.

P5 worked completely independently from the "Elementary Particle Physics: Progress and Promise" (EPP2024)" panel that will deliver its report later in 2024, although we shared the same inputs from the community. EPP is convened by the National Academy of Sciences and charged to deliver a view of elementary particle physics over even longer time scales and unconstrained by financial considerations.

# 1.6
# Roadmap to the Report

Section 1 introduces the science drivers and our vision for particle physics as well as the process and criteria for the P5 deliberations.

The **recommendations** are presented in section 2. Brief discussions expand on the infrastructure and expertise required to carry them out as well as the importance of international and inter-agency partnerships. The subsequent sections discuss how the program can be adapted to alternative budget scenarios, both for less and more favorable funding. Sections 3, 4, and 5 describe the impact of these recommendations on the three science themes but are not intended to be comprehensive reviews of the science opportunities.

Additional **area recommendations** are made in section 6, which highlight theoretical, computational, and technological areas where sustained investments can advance the future of science and technology. In those recommendations, the panel explicitly indicates the increase in annual funding needed to achieve the field's 20-year goals—increases that should be achieved through a ramp, lasting no more than five years, between current and new funding levels. Section 7 expands on the recommendation supporting a technologically advanced workforce. Additional budgetary considerations are described in section 8.

> **"** By studying the very small and the very large, realms that are beyond the limits of normal human perception, we expand our understanding of the world around us and begin to grasp our place in the cosmos. **"**

# 2

## The Recommended Particle Physics Program

# 2.1

# Overview

A particle physics program that tackles the most important questions in each of the science drivers maximizes its potential for groundbreaking scientific discovery. Executing such a program requires a balanced portfolio of large, medium, and small projects, coupled with substantial investments in forward-looking R&D and the development of a skilled workforce for the nation.

Building upon the foundations laid by the previous P5, our recommended program completes ongoing projects and capitalizes on their momentum. A suite of new initiatives at a range of scales includes major projects that will shape the scientific landscape over the next two decades. The prioritized time sequencing of recommended projects and R&D, summarized in Figure 1, reflects our current understanding of the scientific landscape and its associated uncertainties.

The overall program is carefully constructed to be compatible with the baseline budget scenario provided by DOE. To achieve that, we recommend continuing specific projects, strategically advancing some to the construction phase, and delaying others. As shown in Figure 1, in some cases individual phases or elements of large-scale projects had to be prioritized separately. The process and criteria by which the recommended initiatives were selected are laid out in section 1.5.

We note our commitment to the ongoing projects described in section 1. The scientific direction and balance of the recommended new initiatives are summarized below.

DUNE will comprehensively explore the quantum realm of neutrinos, potentially unearthing new physics beyond current theoretical frameworks. Early implementation of the accelerator upgrade ACE-MIRT advances the DUNE program significantly, hastening the definite discovery of the neutrino mass ordering. This upgrade in conjunction with the deployment of the third far detector and a more capable near detector are indispensable components of the re-envisioned next phase of DUNE. R&D for an advanced fourth detector enables the expansion of the physics program of LBNF. These substantial initiatives find synergy with smaller-scale experiments to elucidate the mysteries of neutrinos.

A Higgs factory is the next step toward fully revealing the secrets of the Higgs boson within the quantum realm. We advocate substantial US participation in the design and construction of accelerators and detectors for an offshore facility, and we advocate investment of effort to support development of the Future Circular Collider-electron ($e^-$) positron ($e^+$) (FCC-ee) and the International Linear Collider (ILC), along with a parallel and increasingly intensive program of R&D pursuing revolutionary accelerator and detector technologies. These are crucial for a leadership role in the design and construction of the Higgs factory and for our aspiration to lead and potentially host a next high-energy collider facility beyond the Higgs factory.

There is a compelling physics case for constructing a 10 TeV or more pCM collider. Such a collider would search for direct evidence and quantum imprints of new particles and forces at unprecedented energies. There are several approaches: a 10 TeV muon collider, a 100 TeV proton-proton collider such as FCC-hh at CERN, or possibly a 10 TeV high-energy $e^+e^-$ or γ-γ collider based on the wakefield acceleration technology. Any of them would enable a comprehensive physics portfolio that includes ultimate measurements in the Higgs sector, a broad search program providing access to new hidden sectors by producing a substantially higher mediator mass or probing even smaller coupling, and opportunities to produce new particles directly. All options for a 10 TeV pCM collider are new technologies under development and R&D is required before we can embark on building a new collider.

Further insights will be gained over this decade through collaboration and planning with international partners and dedicated R&D efforts aimed at addressing technical challenges. A panel in the latter part of the decade will be able to harness this information to make further decisions on the path toward future colliders.

A transformative next-generation CMB experiment, CMB-S4, will plumb the secrets of the primordial universe during and immediately after a period of rapid expansion; it should reveal signatures of new physics at energies far beyond the reach of colliders. The ongoing galaxy survey program, enhanced by the Dark Energy Spectroscopic Instrument initiative DESI-II, investigates the cause for a more recent era of accelerated expansion. Together, they promise revolutionary insight into the drivers of cosmic evolution.

In parallel, a strong R&D effort builds toward the ultimate next-generation wide-field spectroscopic survey Spec-S5, which will study the possible time evolution of dark energy and provide a test of inflation complementary to CMB-S4. Development of the emerging technology of line-intensity mapping could create a 3D map of the universe and enable theoretically clean and powerful tests of cosmology.

Another suite of experiments pursues the undetermined nature of the dark matter that gravitationally influences our universe. Select dark matter experiments searching for WIMPs will reach critical discovery potential for a broad range of WIMP masses. Smaller-scale experiments will survey the wider set of dark matter theories and their parameter space. In total, these efforts promise an unprecedented view of the hidden universe.

A new DOE portfolio of agile projects across all science drivers complements existing opportunities within NSF. This portfolio plays a pivotal role in achieving a balanced and forward-looking program.

Realizing the full potential of the experimental landscape requires not just targeted R&D, but substantial strategic investments in theory and infrastructure. Just as hints of new physics revealed by experiment drive new theoretical developments, theory guides experimental inquiries and enriches our understanding of fundamental principles. A coordinated effort that develops shared cyberinfrastructure, harnesses emerging technologies, and leverages national initiatives such as AI/ML, microelectronics, and quantum information science (QIS) benefits all aspects of our scientific program. Equally crucial, key facilities must be maintained and developed in alignment with the long-term vision outlined in this report.

An ambitious, effective scientific program thrives when pursued by a vibrant scientific community. We therefore endorse strategic initiatives to collectively amplify and strengthen the workforce while fostering a healthy working environment. These initiatives are designed to uphold ethical conduct of research, dismantle barriers to entry and retention, recruit broadly, and pave new pathways of opportunity. This commitment nurtures an advanced technological workforce not only adept in particle physics but also equipped to contribute to the technological advancements essential for the nation.

The vision outlined in this report provides opportunities for paradigm-shifting discoveries. By deciphering the quantum realm, illuminating the hidden universe, and exploring new paradigms in physics, we step further into our quantum universe. Some of the priorities are designed to adapt naturally as this landscape evolves over the next decade, while others are designed to drive that evolution.

# 2.2

# Recommendations

To drive US particle physics forward and maintain strong global leadership, we advocate for a comprehensive and balanced program that strategically addresses the three science themes and their six interwoven drivers. The numerical order of the recommendations listed below is not meant to reflect their relative priority; instead it is used to group them thematically. The lists under the recommendations are not prioritized, except for the list of major projects under Recommendation 2. Each recommendation is stated in boldface, followed by concise, lettered explanations of how the recommendation can be realized. The impact of alternative budget scenarios on the different elements of the program is discussed in section 2.6.

A full list of Recommendations is provided at the end of the report. That list includes Area Recommendations (section 6) in addition to those here.

**Recommendation 1: As the highest priority independent of the budget scenarios, complete construction projects and support operations of ongoing experiments and research to enable maximum science.**

We reaffirm the previous P5 recommendations on major initiatives:

a. HL-LHC (including the ATLAS and CMS detectors, as well as the Accelerator Upgrade Project) to start addressing why the Higgs boson condensed in the universe (*reveal the secrets of the Higgs boson*, section 3.2), to *search for direct evidence for new particles* (section 5.1), to *pursue quantum imprints of new phenomena* (section 5.2), and to *determine the nature of dark matter* (section 4.1).

b. The first phase of DUNE and PIP-II to open an era of precision neutrino measurements that include the determination of the mass ordering among neutrinos. Knowledge of this fundamental property is a crucial input to cosmology and nuclear science (*elucidate the mysteries of neutrinos*, section 3.1).

c. The Vera C. Rubin Observatory to carry out the LSST, and the LSST Dark Energy Science Collaboration, to *understand what drives cosmic evolution* (section 4.2).

In addition, we recommend continued support for the following ongoing experiments at the medium scale (project costs > $50M for DOE and > $4M for NSF), including completion of construction, operations, and research:

d. NOvA, SBN, T2K, and IceCube (*elucidate the mysteries of neutrinos*, section 3.1).

e. DarkSide-20k, LZ, SuperCDMS, and XENONnT (*determine the nature of dark matter*, section 4.1).

f. DESI (*understand what drives cosmic evolution*, section 4.2).

g. Belle II, LHCb, and Mu2e (*pursue quantum imprints of new phenomena*, section 5.2).

The agencies should work closely with each major project to carefully manage the costs and schedule to ensure that the US program has a broad and balanced portfolio.

**Recommendation 2: Construct a portfolio of major projects that collectively study nearly all fundamental constituents of our universe and their interactions, as well as how those interactions determine both the cosmic past and future.**

These projects have the potential to transcend and transform our current paradigms. They inspire collaboration and international cooperation in advancing the frontiers of human knowledge. Plan and start the following major initiatives in order of priority from highest to lowest:

a. CMB-S4, which looks back at the earliest moments of the universe to probe physics at the highest energy scales. It is critical to install telescopes at and observe from both the South Pole and Chile sites to achieve the science goals (section 4.2).

b. A re-envisioned second phase of DUNE with an early implementation of an enhanced 2.1 MW beam—ACE-MIRT—a third far detector, and an upgraded near-detector complex as the definitive long-baseline neutrino oscillation experiment of its kind (section 3.1).

c. An offshore Higgs factory, realized in collaboration with international partners, in order to reveal the secrets of the Higgs boson. The current designs of FCC-ee and ILC meet our scientific requirements. The US should actively engage in feasibility and design studies. Once a specific project is deemed feasible and well-defined (see also Recommendation 6), the US should aim for a contribution at funding levels commensurate to that of the US involvement in the LHC and HL-LHC, while maintaining a healthy US onshore program in particle physics (section 3.2).

d. An ultimate Generation 3 (G3) dark matter direct detection experiment reaching the neutrino fog, in coordination with international partners and preferably sited in the US (section 4.1).

e. IceCube Gen-2, for study of neutrino properties complementary to DUNE and for indirect detection of dark matter covering higher mass ranges, using non-beam neutrinos as a tool. (section 4.1).

The prioritization principles behind these recommendations can be found in sections 1.6 and 8.1.

IceCube-Gen2 also has a strong science case in multi-messenger astrophysics together with gravitational wave observatories. We recommend that NSF expand its efforts in multi-messenger astrophysics, a unique program in the NSF Division of Physics, with US involvement in the Cherenkov Telescope Array (CTA; recommendation 3c), a next-generation gravitational wave observatory, and IceCube-Gen2.

**Recommendation 3: Create an improved balance between small-, medium-, and large-scale projects to open new scientific opportunities and maximize their results, enhance workforce development, promote creativity, and compete on the world stage.**

To achieve this balance across all project sizes we recommend the following:

a. Implement a new small-project portfolio at DOE, Advancing Science and Technology through Agile Experiments (ASTAE), across science themes in particle physics with a competitive program and recurring funding opportunity announcements. This program should start with the construction of experiments from the Dark Matter New Initiatives (DMNI) by DOE-HEP (section 6.2).

b. Continue Mid-Scale Research Infrastructure (MSRI) and Major Research Instrumentation (MRI) programs as a critical component of the NSF research and project portfolio.

c. Support DESI-II for cosmic evolution, LHCb upgrade II and Belle II upgrade for quantum imprints, and US contributions to the global CTA Observatory for dark matter (sections 4.2, 5.2, and 4.1).

The Belle II recommendation includes contributions towards the SuperKEKB accelerator.

**Recommendation 4: Invest in a comprehensive initiative to develop the resources—theoretical, computational, and technological—essential to realizing our 20-year strategic vision. This includes an aggressive R&D program that, while technologically challenging, could yield revolutionary accelerator designs that chart a realistic path to a 10 TeV pCM collider.**

Fulfilling this vision requires a substantial investment vital to the effective implementation and success of the particle physics program. The initiative requires the following enhancements to current investments:

a. Support vigorous R&D toward a cost-effective 10 TeV pCM collider based on proton, muon, or possible wakefield technologies, including an evaluation of options for US siting of such a machine, with a goal of being ready to build major test facilities and demonstrator facilities within the next 10 years (sections 3.2, 5.1, 6.5, and Recommendation 6).

b. Enhance research in theory to propel innovation, maximize scientific impact of investments in experiments, and expand our understanding of the universe (section 6.1).

c. Expand the General Accelerator R&D (GARD) program within HEP, including stewardship (section 6.4).

d. Invest in R&D in instrumentation to develop innovative scientific tools (section 6.3).

e. Conduct R&D efforts to define and enable new projects in the next decade, including detectors for an $e^+e^-$ Higgs factory and 10 TeV pCM collider, Spec-S5, DUNE FD4, Mu2e-II, Advanced Muon Facility, and line intensity mapping (sections 3.1, 3.2, 4.2, 5.1, 5.2, and 6.3).

f. Support key cyberinfrastructure components such as shared software tools and a sustained R&D effort in computing, to fully exploit emerging technologies for projects. Prioritize computing and novel data analysis techniques for maximizing science across the entire field (section 6.7).

g. Develop plans for improving the Fermilab accelerator complex that are consistent with the long-term vision of this report, including neutrinos, flavor, and a 10 TeV pCM collider (section 6.6).

We recommend specific budget levels for enhanced support of these efforts and their justifications as Area Recommendations in section 6.

**Recommendation 5: Invest in initiatives aimed at developing the workforce, broadening engagement, and supporting ethical conduct in the field. This commitment nurtures an advanced technological workforce not only for particle physics, but for the nation as a whole.**

The following workforce initiatives are detailed in section 7:

a. All projects, workshops, conferences, and collaborations must incorporate ethics agreements that detail expectations for professional conduct and establish mechanisms for transparent reporting, response, and training. These mechanisms should be supported by laboratory and funding agency infrastructure. The efficacy and coverage of this infrastructure should be reviewed by a HEPAP subpanel.

b. Funding agencies should continue to support programs that broaden engagement in particle physics, including strategic academic partnership programs, traineeship programs, and programs in support of dependent care and accessibility. A systematic review of these programs should be used to identify and remove barriers.

c. Comprehensive work-climate studies should be conducted with the support of funding agencies. Large collaborations and national laboratories should consistently undertake such studies so that issues can be identified, addressed, and monitored. Professional associations should spearhead field-wide work-climate investigations to ensure that the unique experiences of individuals engaged in smaller collaborations and university settings are effectively captured.

d. Funding agencies should strategically increase support for research scientists, research hardware and software engineers, technicians, and other professionals at universities.

e.   A plan for dissemination of scientific results to the public should be included in the proposed operations and research budgets of experiments.

The funding agencies should include funding for the dissemination of results to the public in operation and research budgets.

**Recommendation 6: Convene a targeted panel with broad membership across particle physics later this decade that makes decisions on the US accelerator-based program at the time when major decisions concerning an offshore Higgs factory are expected, and/or significant adjustments within the accelerator-based R&D portfolio are likely to be needed. A plan for the Fermilab accelerator complex consistent with the long-term vision in this report should also be reviewed.**

The panel would consider the following:

a.   The level and nature of US contribution in a specific Higgs factory including an evaluation of the associated schedule, budget, and risks once crucial information becomes available.

b.   Mid- and large-scale test and demonstrator facilities in the accelerator and collider R&D portfolios.

c.   A plan for the evolution of the Fermilab accelerator complex consistent with the long-term vision in this report, which may commence construction in the event of a more favorable budget situation.

# 2.3

# The Path to 10 TeV pCM

Realization of a future collider will require resources at a global scale and will be built through a world-wide collaborative effort where decisions will be taken collectively from the outset by the partners. This differs from current and past international projects in particle physics, where individual laboratories started projects that were later joined by other laboratories. The proposed program aligns with the long-term ambition of hosting a major international collider facility in the US, leading the global effort to understand the fundamental nature of the universe.

There are multiple complementary technologies that could potentially reach the 10 TeV pCM scale, and the work to determine how to economically reach that goal must go forward. This is why we recommend pursuing revolutionary R&D in areas such as high-field magnets, a multi-megawatt proton driver, wakefield accelerator technology, and muon cooling (Recommendation 4a).

In particular, a muon collider presents an attractive option both for technological innovation and for bringing energy frontier colliders back to the US. The footprint of a 10 TeV pCM muon collider is almost exactly the size of the Fermilab campus. A muon collider would rely on a powerful multi-megawatt proton driver delivering very intense and short beam pulses to a target, resulting in the production of pions, which in turn decay into muons. This cloud of muons needs to be captured and cooled before the bulk of the muons have decayed. Once cooled into a beam, fast acceleration is required to further suppress decay losses.

Each of these steps presents considerable technical challenges, many of which have never been confronted before. This P5 plan outlines an aggressive R&D program to determine the parameters for a muon collider test facility by the end of the decade. This facility would test the feasibility of developing a muon collider in the following decade.

With a 10 TeV pCM muon collider at Fermilab as the long-term vision, a clear path for the evolution of the current proton accelerator complex at Fermilab emerges naturally: a booster replacement with a suitable accumulator/buncher ring would pave the way to a muon collider demonstration facility (Recommendation 4g, 6). The upgraded facility would also generate bright, well-characterized neutrino beams bringing natural synergies with studies of neutrinos beyond DUNE. It would also support beam dump and fixed target experiments for direct searches of new physics. Another synergy is in charged lepton flavor violation. The current round of searches at Mu2e can reveal quantum imprints of new physics at the 100 TeV energy scale, beyond the reach of direct searches at collider facilities in the foreseeable future. An intense muon facility may push this search even further.

Although we do not know if a muon collider is ultimately feasible, the road toward it leads from current Fermilab strengths and capabilities to a series of proton beam improvements and neutrino beam facilities, each producing world-class science while performing critical R&D toward a muon collider. At the end of the path is an unparalleled global facility on US soil. This is our Muon Shot.

# 2.4

# Stewardship of Key Infrastructure and Expertise

Successful completion of the recommended major projects depends on critical US infrastructure (section 6.6), including particular research sites and facilities. DOE National Laboratories are critical research infrastructure that must be maintained and enhanced based on the needs of the particle physics community. This is particularly true for Fermilab as the only dedicated US laboratory for particle physics. The South Pole, a unique site that enables the world-leading science of CMB-S4 and IceCube-Gen2, must be maintained as a premier site of science to allow continued US leadership in these areas. SURF, a deep underground research laboratory supported by the South Dakota Science and Technology Authority, private foundation funds, and DOE, is a critical addition to the suite of US research infrastructure, providing new space and essential infrastructure for DUNE and potentially a G3 dark matter experiment.

In other cases, the infrastructure is technological and intellectual. The GARD program is critical in supporting a broad range of accelerator science and technology (AS&T) for DOE's Office of Science, separate from the targeted R&D toward future colliders. Along with NSF-funded fundamental accelerator science, GARD supports a broad workforce of essential accelerator expertise. The program also provides stewardship of AS&T for DOE's Office of Science. This program and the balance across the different research thrusts should be reviewed regularly to ensure alignment with the goals in particle physics. Reviews should be conducted by broad teams, not only specialists.

## 2.5

# International and Inter-Agency Partnerships

Major facilities like Fermilab in the US, CERN in Europe, and KEK in Japan have led the worldwide effort to advance accelerator-based studies of particle physics. These facilities have enabled many groundbreaking experiments, and their continued leadership roles as host laboratories for future accelerators, cutting-edge experiments, and hubs for international collaborations are important for progress in the field.

Successful completion of the recommended major projects depends on significant coordination and collaboration among US agencies and international partners. Large international projects such as a Higgs factory and DUNE require not only DOE and NSF, but also the US Department of State and other entities in the federal government to work with global partners to establish the complex frameworks involved.

In the case of the Higgs factory, crucial decisions must be made in consultation with potential international partners. The FCC-ee feasibility study is expected to be completed by 2025 and will be followed by a European Strategy Group update and a CERN council decision on the 2028 timescale. The ILC design is technically ready and awaiting a formulation as a global project. A dedicated panel should review the plan for a specific Higgs factory once it is deemed feasible and well-defined; evaluate the schedule, budget and risks of US participation; and give recommendations to the US funding agencies later this decade (Recommendation 6). When a clear choice for a specific Higgs factory emerges, US efforts will focus on that project, and R&D related to other Higgs factory projects would ramp down.

Parallel to the R&D for a Higgs factory, the US R&D effort should develop a 10 TeV pCM collider, such as a muon collider, a proton collider, or possibly an electron-positron collider based on wakefield technology. The US should participate in the International Muon Collider Collaboration (IMCC) and take a leading role in defining a reference design. We note that there are many synergies between muon and proton colliders, especially in the area of development of high-field magnets. R&D efforts in the next 5-year timescale will define the scope of test facilities for later in the decade, paving the way for initiating demonstrator facilities within a 10-year timescale (Recommendation 6).

For studies of cosmic evolution and astrophysical studies of dark matter, inter-agency coordination and cooperation between DOE, NSF, and NASA using complementary observational approaches has been very productive in building a world-leading scientific program. Such coordination and cooperation should continue.

The field of particle physics is not an isolated endeavor, and it benefits from and contributes to neighboring areas in nuclear physics, astrophysics and astronomy, condensed matter physics, precision physics, computing, instrumentation, material science, fusion, and others. At the same time, it provides important theoretical and technological input to these areas, as well as medical, security, and many other fields, some as seemingly unrelated as archaeology. Funding agencies are urged to reach across the traditional boundaries to enhance collaboration, maximize science, and develop a strong workforce for the nation overall.

## 2.6

# Adapting to Alternative Budget Scenarios

The program recommendations are built considering the baseline budget scenario for DOE. This scenario assumes budget levels for HEP for fiscal years 2023 through 2027 that are specified in the CHIPS and Science Act of 2022. The baseline budget scenario then increases by 3% per year from fiscal year 2028 through 2033. We assume 3% inflation throughout our exercise, so it provides an initial increase over five years followed by an essentially flat budget in later years. In this scenario, hard choices were required as described in section 8.2. The recommended program is well-balanced and forward-looking, enabling scientific breakthroughs and maintaining scientific and technological leadership.

Two other scenarios were considered by the panel. Figure 2 summarizes the projects recommended under all three scenarios.

### 2.6.1 — Less Favorable Budget Scenario

We are charged to discuss a less favorable budget scenario that forces us to make more drastic and challenging choices. This scenario assumes budget increases of 2% per year during fiscal years 2024 to 2033 for DOE HEP, which is an erosion against an assumed 3% annual inflation rate. Under this scenario, some interesting scientific opportunities are still achievable, but scientific progress is significantly slowed.

In this scenario, we would aim for a program that covers most areas of particle physics for the next 10 years, maintaining continuity and exploiting the ongoing projects in Recommendation 1 as our highest priority. The agencies should launch the same major initiatives as outlined in Recommendation 2, some of them with significantly reduced scope:

a.   CMB-S4 without reduction in scope.

b.   DUNE Third Far Detector (FD3), but defer ACE-MIRT and the More Capable Near Detector (MCND). Infrastructure required to accommodate international contributions remains a priority.

c.   Contribution to an offshore Higgs factory delayed and at a reduced level.

d.   Reduced participation in a G3 dark matter experiment sited outside the US, and no SURF expansion.

e.   IceCube-Gen2 without reduction in scope.

The rationale for this prioritization is given in section 8.3. Recommendations 3 and 4 are crucial for maintaining the health and balance of the field. While these recommendations still apply, they receive reduced support in scenarios between the baseline and less favorable conditions. Reductions to all items in these two recommendations should be proportionate. Research must be supported at least at the current level. Recommendation 5 is deemed a high priority and is supported in all scenarios. Recommendation 6 applies in all scenarios.

This less favorable scenario will lead to a loss of US leadership in many areas, especially the science of the G3 dark matter experiment, and will damage our reputation as a reliable international host for DUNE and as a partner for a Higgs factory. We still make enhanced investments in the future, but at a significantly reduced level for small-scale experiments, including ASTAE, theory, computing, instrumentation, and collider R&D. In this scenario, it would be increasingly difficult to maintain US competitiveness as an international partner in accelerator technology. See section 8.3 for more details.

## 2.6.2  –  More Favorable Budget Scenario

In a budget outlook more favorable than the baseline budget scenario, we urge the funding agencies to support additional scientific opportunities. Even a small increase in the overall budget enables a large return on the investment, serving as a catalyst to accelerate scientific discovery and to unlock new pathways of inquiry. The opportunities include R&D, small projects, and the construction of advanced detectors for flagship projects in the US. They are listed below in four categories from small to large in budget size:

a.   R&D
   i.   Increase investment in detector R&D targeted toward future collider concepts for a Higgs factory and 10 TeV pCM collider in order to accelerate US leadership in this area.

   ii.   Pursue an expanded DOE AS&T initiative to develop foundational technologies for particle physics that can benefit applications across science, medicine, security, and industry.

   iii.   Pursue broad accelerator science and technology development at both DOE and NSF, including partnerships modeled on the plasma science partnership.

b.   Small Projects
   Expand the portfolio of agile experiments to pursue new science, enable discovery across the portfolio of particle physics, and provide significant training and leadership opportunities for early career scientists.

c.   Medium Projects
   i.   Initiate construction of Spec-S5 as the world-leading study of cosmic evolution, with applications to neutrinos and dark matter, once its design matures.

   ii.   Initiate construction of an advanced fourth far detector (FD4) for DUNE that will expand its neutrino oscillation physics and broaden its science program.

   iii.   Initiate construction of a second G3 dark matter experiment to maximize discovery potential when combined with the first one.

d.   Large Projects
   Evolve the infrastructure of the Fermilab accelerator complex to support a future 10 TeV pCM collider as a global facility. A positive review of the design by a targeted panel may expedite its execution (Recommendation 6).

## Figure 1 — Program and Timeline in Baseline Scenario

**Index:** ■ Operation  ■ Construction  ■ R&D, Research  P: Primary  S: Secondary
§ Possible acceleration/expansion in more favorable budget situations

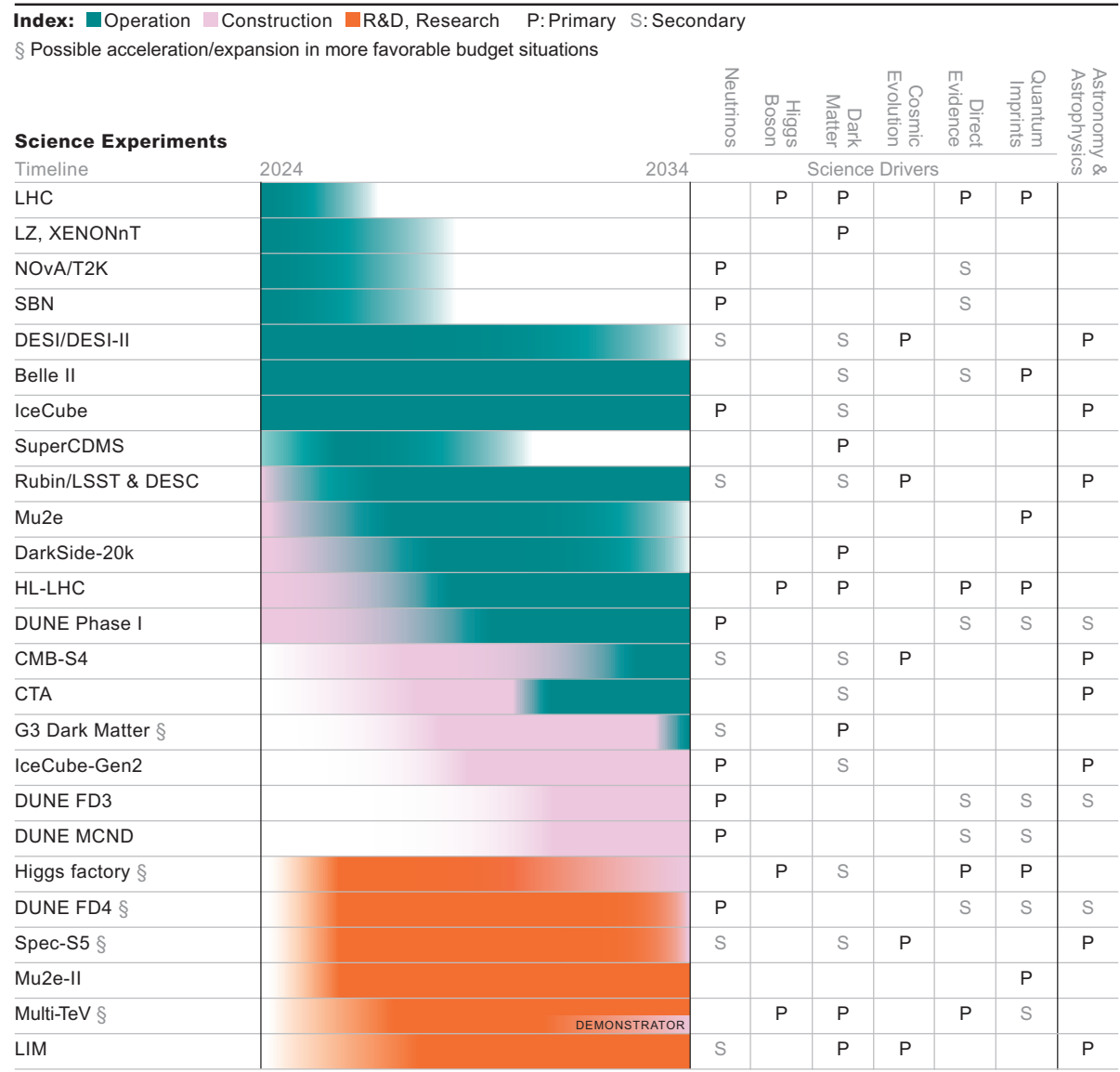

Approximate timeline of the recommended program within the baseline scenario. Projects in each category are in chronological order. For IceCube-Gen2 and CTA, we do not have information on budgetary constraints and hence timelines are only technically limited. The primary/secondary driver designation reflects the panel's understanding of a project's focus, not the relative strength of the science cases. Projects that share a driver, whether primary or secondary, generally address that driver in different and complementary ways.

## Figure 2 — Construction in Various Budget Scenarios

**Index:** Y: Yes  N: No  R&D: Recommend R&D only  C: Conditional yes based on review  P: Primary  S: Secondary
Delayed: Recommend construction but delayed to the next decade
† Recommend infrastructure support to enable international contributions
# Can be considered as part of ASTAE with reduced scope

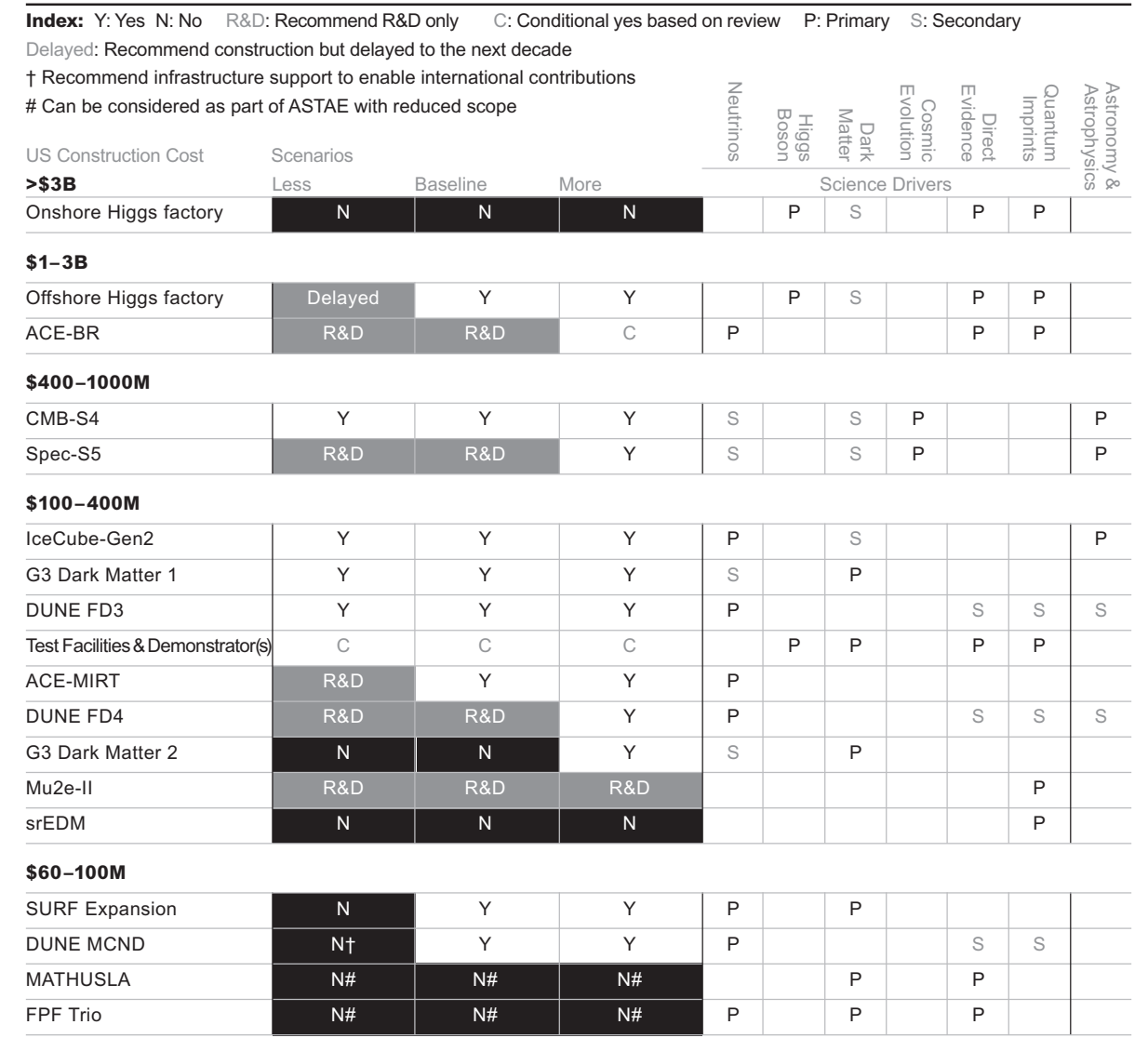

Medium and large-scale US investments in new construction projects for possible budget scenarios. For the three budget scenarios, the projects are ordered in 5 budget brackets according to the number of "N" entries and then by approximate budget sizes. For the offshore Higgs factory, test facilities & demonstrators, see Recommendation 6. See the caption of Figure 1 concerning the science drivers, and section 8 for the rationale behind these choices.

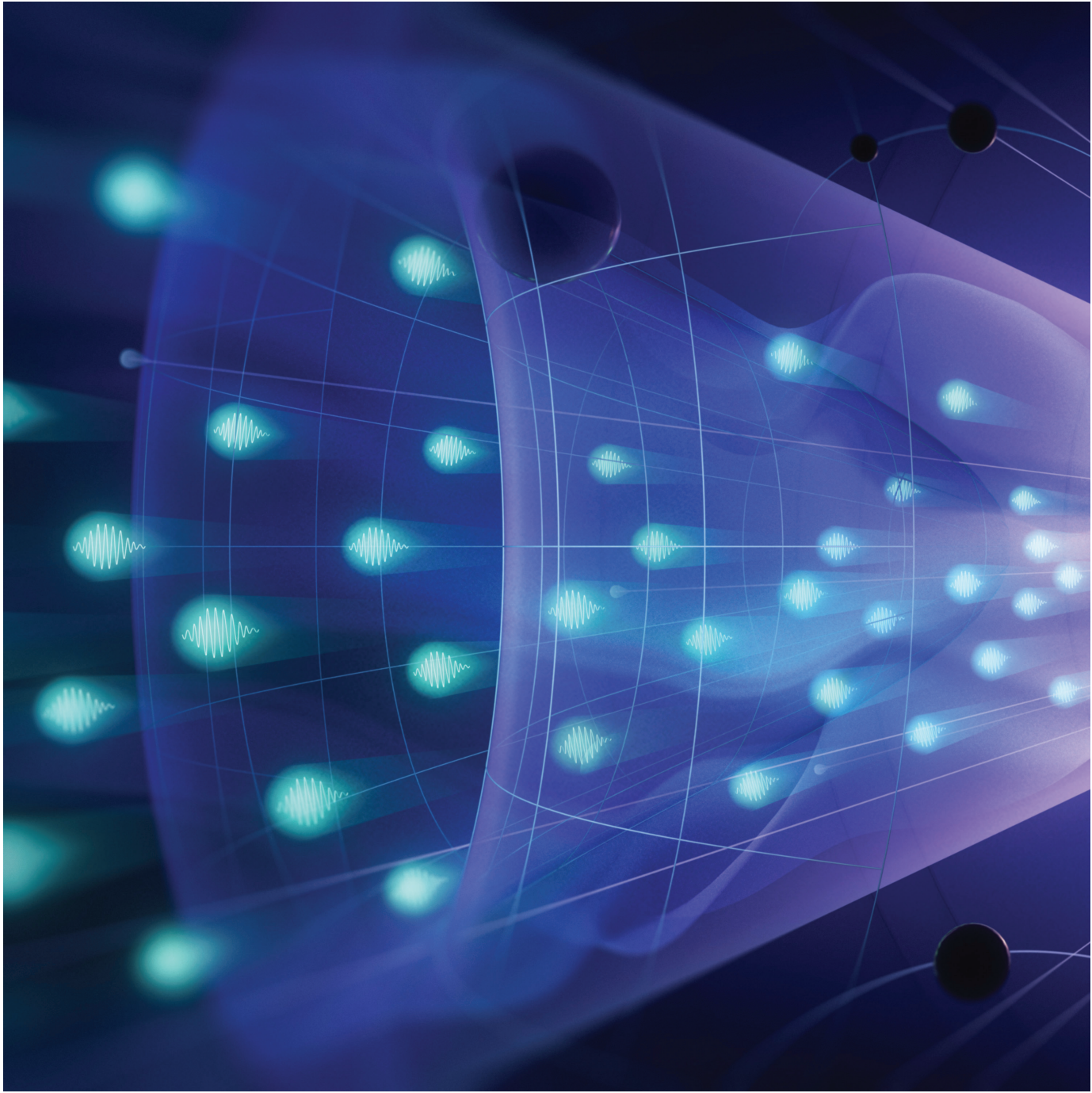

# 3

Decipher
the
Quantum
Realm

Atoms are the building blocks of matter, but what are the building blocks of atoms? For more than a century, we have known that each atom is composed of electrons surrounding a heavy nucleus, bound together by the electromagnetic force. The nucleus is composed of protons and neutrons, which are themselves composed of quarks, bound together by the strong force. The building blocks of atoms are particles, and the forces that bind the building blocks are also described by particles. All known subatomic phenomena can be described by particles and their interactions, an amazing new concept in modern science.

When we look deeper, we see a rich landscape of quantum effects that rule the subatomic realm. The electron appears indivisible, yet its properties are affected by a quantum dance among all subatomic particles. The heaviest known subatomic particle is the top quark, which is surprisingly as massive as a gold atom. While the massless photons and gluons mediate the electromagnetic and strong force respectively, the mediators of the weak force, the W and Z bosons, are massive, blurring the distinction between "force" and "matter." The Higgs boson, the particle associated with the Higgs field that permeates the universe, gives mass to other particles. The Standard Model provides a unified, elegant picture of the subatomic realm that has withstood the most rigorous tests at LHC and KEK. Beautiful as this picture is, it does not yet account for the masses of the mysterious and mutable neutrinos, which oscillate into one another as they travel through the universe.

Is the Standard Model the ultimate description of the quantum realm? Certainly not, and that fact motivates the two science drivers under this theme for the next decade:

*Elucidate the Mysteries of Neutrinos.* Neutrinos come in three types, or "flavors," which undergo quantum oscillations. Although the Standard Model can be augmented to accommodate neutrino mass and oscillations, we do not know in which specific way to extend the model. Moreover, different extensions make vastly different predictions about the birth of the universe. We must further investigate the mysteries of neutrinos in order to explore the deep connections between their physics and the Standard Model.

*Reveal the Secrets of the Higgs Boson.* The discovery of the Higgs boson in 2012 was a major victory for the Standard Model. Subsequent investigations have revealed that the metastable vacuum in the Standard Model puts the universe on a knife-edge of cosmic collapse. Determining the ultimate fate of the universe and looking for physics beyond the Standard Model motivates further scrutiny of the Higgs sector.

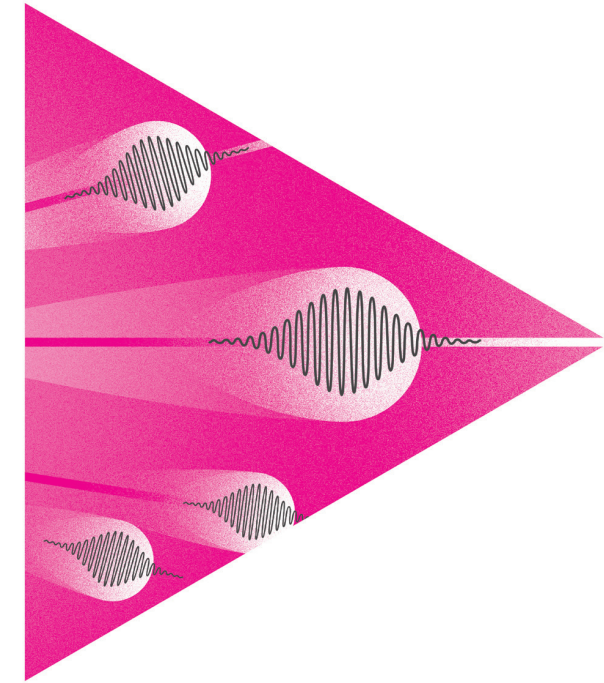

# 3.1

# Elucidate the Mysteries of Neutrinos

## 3.1.1 – Science Overview

Neutrinos are the second most abundant of the known particles in the universe and yet remain an enigma within the framework of the Standard Model. The observation of neutrino oscillations, and the consequent realization that neutrinos have mass, is one of the revolutionary discoveries of recent decades. Despite being the lightest of particles, the tiny mass of neutrinos challenges the standard paradigm of particle physics and has opened a compelling new domain of exploration in the quantum realm.

Nature produces—and we observe—neutrinos in three flavors: electron, muon, and tau. We understand each of these neutrino flavors to be a chimera, a mixture of states with different masses. According to quantum mechanics such a mixture will evolve, or oscillate, into another flavor as it travels. Neutrino oscillation processes have been extensively observed, and the admixture, or mass mixing, of neutrino mass states assigned to each flavor and the differences between the neutrino masses have been measured. Yet the actual values of the neutrino masses remain unknown, and we are not yet sure how they are acquired. Various theoretical frameworks propose diverse methods for neutrino mass generation, often introducing novel particles and interactions. Certain theories even require that neutrinos must be their own antiparticles, which will naturally but surprisingly lead to lepton number violation and provide a necessary condition for the matter-dominant universe we observe now.

Likewise, we have not definitively measured the ordering of the neutrino masses; for example, is the neutrino mass state that has the smallest overlap with the electron neutrino the heaviest or the lightest? We refer to the case where the lightest neutrino has the largest overlap with the electron neutrino as "normal" ordering, because the mass spectrum of the neutrinos in this case follows the familiar mass spectrum of the quarks and charged leptons.

The opposite, or inverted, ordering would be a consequential surprise. For instance, a new symmetry would likely be needed to account for why the two heavier neutrinos are so similar in mass. Neutrino mass ordering impacts efforts that seek to measure the neutrino masses and to understand how neutrinos acquire mass in the first place.

Precision studies of neutrino oscillations can resolve important questions beyond neutrino mass ordering. Is the lightest (or heaviest) neutrino state an equal combination of muon and tau neutrinos, which would hint at new symmetries? Is three-flavor mixing a complete description of all neutrino flavor transitions? Do antineutrinos oscillate differently than neutrinos—namely, is the charge-parity inversion (CP) symmetry violated, and could this difference relate to the origins of matter-antimatter asymmetry in the universe?

We do not fully understand why some fundamental forces respect the CP symmetry, while others violate it. For example, we have long had experimental evidence that interactions involving the weak force do not conserve CP. Conversely, our description of interactions involving the strong force could provide a mechanism for CP violation, yet all measurements so far are consistent with the relevant parameter being zero. Theories that describe particular patterns of CP violation and conservation can be connected to such questions as the nature of dark matter, or why a universe born with equal amounts of matter and antimatter developed into the overwhelmingly matter-dominated universe we observe today. It therefore makes sense to search for and measure CP violation in every context that we can.

Neutrinos have also opened a new view of astrophysical phenomena that can provide unique probes of neutrino physics. The neutrino signal from the core collapse of a massive star in the Milky Way galaxy provides a window into neutrino flavor transitions and transport in a turbulent environment with high neutrino density. Such an environment cannot be simulated on Earth.

Answering the outstanding questions in neutrino physics requires a blend of novel technologies and measurements made with exquisite precision. These may serve as beacons, illuminating the path toward unveiling novel interactions, particles, or symmetries in the universe.

## 3.1.2 – Ongoing Projects: NOvA, T2K, SBN, DUNE Phase I, PIP-II, and IceCube

Ongoing accelerator-based experiments NOvA and T2K have pioneered electron neutrino and antineutrino appearance observations. They have introduced approaches to control systematic uncertainties in combined measurements of mass ordering and CP violation and other mass mixing parameters. A joint T2K and NOvA analysis of the two datasets is ongoing, and results obtained with the complete datasets may provide early indication of future discovery. Ongoing experiments also conduct interesting searches for phenomena beyond the Standard Model.

Over the past decades neutrino oscillation searches at length/distance scales of 1 MeV/m have found some anomalous results: The liquid scintillator neutrino detector (LSND) anomaly, the reactor antineutrino anomaly, the MiniBooNE low-energy excess, and the gallium anomaly. These anomalies have not been confirmed, and the reactor antineutrino anomaly has been recently resolved. The remaining phase space will be conclusively tested by the current short-baseline neutrino program at Fermilab. The SBN program is also crucial in maturing the LAr technology and analysis. SBN, T2K, NOvA, and other ongoing experiments also make measurements of neutrino interactions, which underpin our understanding of neutrino oscillation mixing (Recommendation 1c).

The DUNE experiment consists of three elements: a far detector complex at SURF, a near detector complex hosted at Fermilab, and a neutrino beam sent across the 1300 km distance between the two facilities. The program has achieved significant design and construction milestones, successfully scaling LArTPC technology and preparing the largest underground laboratory in the US, scheduled for completion by 2025.

For the first phase of DUNE, the far detector complex will comprise two 10 kt LArTPCs in an underground area designed to accommodate up to four modules. In this phase these detectors and a near detector facility will be illuminated by the world's brightest neutrino beam, generated by the LBNF at Fermilab. The PIP-II accelerator upgrade currently under construction is central in enabling at least 1.2 MW proton beam operation during Phase I (Recommendation 1b).

DUNE's comprehensive program of neutrino oscillation measurements sets the mass-ordering question as its first goal. Thanks to DUNE's long baseline and broad energy range of neutrinos that result in a strong separation between normal and inverted mass ordering scenarios, DUNE Phase I is expected to achieve a definitive measurement of the mass ordering within its first decade of operation. This result, when combined with measurements made over shorter distances by experiments such as JUNO, probes nonstandard couplings to matter. Together with Hyper-Kamiokande and other ongoing experiments, these measurements may clarify the nature of mixing or uncover where the three-flavor mixing model is incomplete.

DUNE's large, underground LAr detectors also have a unique sensitivity to the electron neutrino component of a supernova neutrino burst. They will therefore complement measurements by other underground neutrino detectors, such as Hyper-Kamiokande, and measurements by the Antarctic neutrino observatory IceCube and future direct dark matter detectors that constrain the overall burst energetics. In addition, IceCube has been able to make measurements of atmospheric neutrino oscillations, and demonstrated that neutrinos are new tools for astrophysics enabling us to peer into the area of supermassive black holes and the disk of our own Milky Way.

Strong software and computing development that includes AI/ML dedicated effort has been integral to the success of the ongoing and planned neutrino experiments. Therefore, it is essential that advances in software, computing, and AI/ML keep up with demands of new neutrino projects, large data rates, and complexity of data analysis as irreplaceable tools for handling high-statistics, rare, and exotic particle searches.

### 3.1.3 — Major Initiatives: Early Implementation of MIRT and DUNE Phase II

Following Recommendation 2b, we envision that DUNE Phase II will include early implementation of an enhanced 2.1-MW beam using MIRT, a third far detector, and an upgraded near detector complex.

Early implementation of MIRT enables beam operations at 2.1 MW to start promptly, permitting DUNE to achieve Phase I design exposure of 120 kt·MW·yr by the mid-2030s, the original planned timescale. MIRT achieves higher beam power prior to a booster replacement by making several changes to the Fermilab accelerator complex. In particular, enhancements of the acceleration and magnet systems reduce the cycle time of the Main Injector and the LBNF target station components, thereby delivering a given beam intensity more frequently. Reliability of the booster synchrotron becomes critical during this period and must be assessed and likely addressed through additional measures as described in section 6.6.2. The changes would enable a robust determination of the neutrino mass ordering by the middle of the next decade (Recommendation 2b).

Phase II completion leaves DUNE poised to deliver the most precise measurement of the CP phase across a range of possible CP phase space. Dune Phase II is the ultimate long-baseline experiment based on a proton-derived, high-intensity neutrino beam and is designed to cover a broad spectrum of neutrino energies, enabling an in-depth exploration of neutrino oscillation quantum mechanics throughout multiple oscillation cycles. Thus, the experiment will comprehensively test the validity of the three-flavor neutrino oscillation framework with best-in-class precision and will search for signatures of unexpected neutrino interactions. In addition, DUNE's long baseline and high neutrino energy provide unique sensitivity to matter effects and new neutrino interactions, and they allow us to study the direct appearance of tau neutrinos.

To fully achieve these goals, DUNE must collect an extremely large sample of neutrinos and gain exquisite control of the relevant systematic uncertainties. The requisite statistics depend on the MIRT upgrade and the expansion of the DUNE Far Detector. The increased detector volume provided by an additional far detector module (FD3) leverages international partnerships and benefits all aspects of the DUNE science program, including those, like supernova neutrino burst studies, that are independent of the neutrino beam. Together these upgrades more than quadruple the DUNE Phase I exposure to achieve 600 kt·MW·yr by the mid-2040s, the originally envisioned timescale. At this integrated exposure we expect statistical and systematic uncertainties to be roughly balanced, giving DUNE significant and unique discovery potential across the neutrino mixing landscape.

With higher statistics, control of systematic uncertainties (such as those arising from the interaction of neutrinos and nuclei) becomes increasingly crucial. An MCND, a gas target combined with a magnetic field and electromagnetic calorimeter, is indispensable for this purpose. In addition, by being exposed to the world's most intense neutrino beam, it will create a unique laboratory for the discovery of novel particles and interactions, many of which could shed light on the nature of dark matter and possible hidden sectors.

The opportunities opened by DUNE Phase II shine brightest when complemented by a strong theory effort. The interaction of neutrinos and nuclei represents a complex many-body quantum problem, and significant theoretical work is required to gain quantitative understanding at a subatomic or nuclear level. This work will further reduce systematic uncertainties. Similarly, theoretical models of new physics will help interpret any anomalies or surprises in DUNE data. In fact, new developments in theory can open up the science opportunities of new physics searches both at the near and far detectors, many of which are not directly related to neutrinos.

### 3.1.4 — Future Opportunities: DUNE FD4, the Module of Opportunity

The advent of MIRT will enable rapid acquisition of beam neutrino statistics and allow DUNE to achieve 600 kt·MW·yr without deploying a fourth detector module (FD4). This paves the way for an expanded physics program, featuring an upgraded, more efficient detector with enhanced charge reconstruction capabilities. Such a detector would allow for full exploitation of the long baseline neutrino program. A more capable detector with significantly improved light collection, charge granularity, and high radiochemical purity would push the detector energy threshold down to MeV, or lower, while improving track and energy reconstruction.

A range of alternative targets, including low radioactivity argon, xenon-doped argon, and novel organic or water-based liquid scintillators, should be considered to maximize the science reach, particularly in the low-energy regime. Increased radiochemical purity would enhance detection sensitivity to low-energy supernova burst neutrinos, and in some cases even to coherent elastic neutrino-nucleus scattering (CEvNS) interactions triggered by a nearby core-collapse supernova. An upgraded detector module will provide excellent prospects for underground physics, including direct dark matter detection, exotic dark matter searches, and expanded sensitivity to solar neutrinos. R&D for advanced detector concepts should be supported.

The plethora of science opportunities has already sparked wide international interest and has been discussed in DUNE-organized workshops featuring presentations of novel technologies and detection approaches to improve DUNE's capabilities. Maximizing the physics potential will require input from all stakeholders: the DUNE collaboration, US funding agencies, and the international community. A decision-making process led by DOE, inclusive of the entire community and driven by all stakeholders, will ensure that the full potential of the FD4 is realized. The timeline should be driven by the most promising scientific opportunities and must include the long-baseline science program (Recommendation 4d).

### 3.1.5 — Interplay with Other Measurements of Neutrino Properties

Understanding the origins of neutrino mass is one of the big questions in physics. However, neutrino masses have not yet been directly measured. There are three approaches to measuring the neutrino mass: direct kinematic mass searches in nuclear beta decay, neutrinoless double beta decay, and cosmology. The first two approaches are under the stewardship of the DOE nuclear science program. Similarly, the question of whether neutrinos are their own antiparticles—Majorana particles—is one of the top science topics highlighted in the recent Nuclear Science Advisory Committee (NSAC) long-range plan via the pursuit of ton-scale neutrinoless double beta decay experiments. Measurements of the mass ordering by the particle physics program set the expected scale for these experiments. Their outcome is one of the most eagerly anticipated pieces to the puzzle of neutrino mass.

Neutrino mass also affects structure formation in the universe. Hence, careful measurements of the distribution of mass in the universe are sensitive to neutrino masses in the range of values indicated by neutrino oscillation. Cosmological surveys DESI and CMB-S4 can probe the sum of neutrino masses, and that information can be directly confronted with the mass ordering measured by DUNE. IceCube and its upgrades test neutrino mixing at high energies and cosmic distances. Any disparities between these two realms would constitute a discovery with profound implications for our understanding of the universe's fundamental properties.

Many models of neutrino mass generation require the neutrino to have heavy partners. In some cases, those partners can be tested in beam dump experiments like the MCND of DUNE. For other mass ranges, those partners can be searched for at the LHC or future 10 TeV colliders.

### 3.1.6 — New Initiative: A Portfolio of Agile Projects for Neutrinos

A healthy portfolio of agile experiments focused on neutrino physics and capable of delivering transformative insights and technological advancements is essential to the future of the field. To advance the understanding of neutrinos, a multifaceted approach must support a versatile and dynamic portfolio of ASTAE experiments, as described in section 6.2.

Key breakthroughs in neutrino physics have been achieved through experiments that shed light on hidden facets of neutrino interactions and resolve outstanding neutrino anomalies. These experiments highlight the potential for discovery science in agile neutrino projects. Accurate measurements are also important for a deeper understanding of neutrinos, their interactions, energy spectra, and flux, and their roles in astrophysical phenomena and long-baseline neutrino oscillation experiments. The adaptability and deployment flexibility of agile experiments, whether near beams or reactors, offer promise for synergistic explorations of hidden-sector particles and other phenomena in the evolving field of physics beyond the Standard Model (BSM). Development of technologies such as innovative materials and unique sensors is critical to shaping the future of neutrino experiments.

The ASTAE portfolio for neutrinos should encompass precise measurements of neutrino interactions, comprehensive neutrino flux assessments, and searches for neutrino BSM physics, coupled with development of cutting-edge technologies for future detectors (Recommendation 3a).

### 3.1.7 — 20-Year Vision

The collaborative efforts of DUNE, Hyper-Kamiokande, and the global oscillation program, including JUNO, could definitively validate the framework of three-flavor neutrino oscillations. The experimental outcomes of the first several years of operation will guide the future vision of DUNE, honing in on the complete picture of neutrino oscillations and even the physics of tau neutrinos, the least explored elementary particle of the Standard Model. Should there emerge indications of a paradigm shift such as CPT violation or of neutrino non-standard interactions, DUNE's long baseline with large matter effects and Hyper-Kamiokande's shorter baseline with small matter effects will be critical in discerning this exciting new physics.

If there are hints of a need for heightened precision, muon-decay based neutrino beams emerge as the logical choice to enhance measurement accuracy. Depending on the nature of the departure from three-flavor oscillations, this could entail the deployment of a low-energy muon storage ring, as exemplified by the Neutrinos from Stored Muons (nuSTORM) experiment. This is certainly the case if novel neutrino types or interactions mediated by light new particles come into play. A facility like nuSTORM also has the potential to significantly refine our understanding of neutrino-nucleus interaction cross sections.

In cases where subtler signs of new particles or interactions surface, a muon storage ring with stored muon energies in the tens of GeV or higher range, commonly known as a neutrino factory, emerges as the most suitable source of neutrinos. Regardless of the specific approach, all muon-decay-based neutrino sources offer unparalleled precision, well-characterized beams, and potent synergies with muon collider research and development efforts.

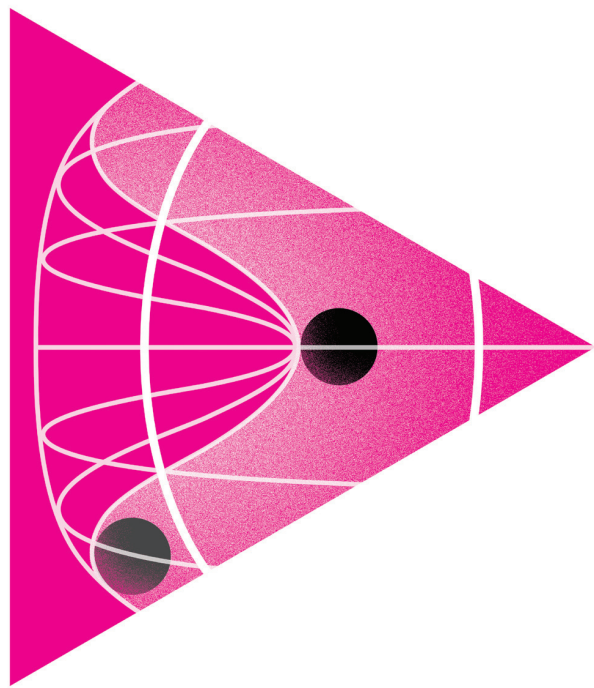

# 3.2

# Reveal the Secrets of the Higgs Boson

### 3.2.1 – Science Overview

The Higgs boson is an extraordinary and unique particle that is connected to the most puzzling questions of particle physics, including the origin of flavor, the matter-antimatter asymmetry, dark matter and dark energy, and inflation. The Higgs boson differs from other particles in that it is a remnant of the Higgs field that is "frozen" and permeates the universe. The field disturbs and slows down the motion of every elementary particle. The Higgs boson slows electrons in atoms so that they stay within the atom instead of flying off into space. Without the Higgs boson, or field, every electron in every atom would move at the speed of light and everything, including us, would evaporate in a nanosecond.

The Higgs boson is the only known fundamental particle that has no spin angular momentum, which permits it to have unique behavior: it can interact with all known matter particles and give them mass, depending on the strength of the interaction. The Higgs boson also provides a novel and distinct gateway to as yet unknown particles, such as dark matter.

Given the unique nature of the Higgs boson and its crucial role in holding together the atoms of the universe, the hunt for it was intense and extensive, beginning in 1964 with both theoretical and experimental efforts. The Higgs boson was discovered to great fanfare in 2012 by an international effort at the LHC at CERN in Switzerland, with crucial contributions from the US community.

Major questions remain about the nature of the Higgs boson. We do not know if the Higgs field is a fundamental field, or if it is actually a composite field made from other constituents. We do not know if there is only one Higgs boson, or if there is a richer sector containing related particles with new dynamics. We do not know why the Higgs boson mass should be as low as it is in the absence of additional particles with similar masses that would stabilize it, or why the mass is not zero in the first place. We do not know if the Higgs boson can decay to non-Standard Model particles. The interactions of the Higgs boson with the matter particles—the generation of fermion masses and mixings—involve the largest number of experimentally measured Standard Model parameters whose values and pattern are not predicted by any theory. Understanding this pattern may shed light on important questions such as the matter-antimatter asymmetry and the origin of neutrino masses.

The properties of the Higgs field play a fundamental role in the evolution of the universe and the attributes of the other Standard Model particles. The Higgs field is unique in that it is the only known fundamental field that has a non-zero value in the vacuum state. In the early universe, all the Standard Model particles were massless and the fundamental forces behaved differently than they do today. As the universe cooled, the Higgs field acquired its current non-zero value. This "electroweak" phase transition, in turn, led to a universe in which the Standard Model particles acquired their current masses and the fundamental forces assumed their current form.

The characteristics of the electroweak phase transition are determined by the interactions of the Higgs field with itself ("Higgs self-coupling") which determines the field's potential energy ("Higgs potential"). How this transition happened and which of the puzzling phenomena in our understanding of the universe are related to this phase transition remain central questions in particle physics. Modifications to the potential and related phase transition, beyond the Standard Model predictions, could provide explanations for the dominance of matter over antimatter in the universe.

The fate of the universe depends on the properties of the Higgs sector. Extrapolations of the currently understood Higgs potential to extremely high energy, using the Standard Model, indicate that the current vacuum state of the Higgs field is not only metastable—not eternally stable—but that the universe is very close to the crossover point between stability and metastability. Further information about the potential will help interpret the meaning of this result. The Higgs boson may even be related to the field that drove the cosmological inflation, called the inflaton field, or to the mysterious dark energy that drives the current accelerated expansion of the universe, both of which require fields of zero spin.

The fact that the properties of the Higgs field are connected to so many of the fundamental questions in particle physics highlights the central role of the Higgs boson and the importance of understanding all aspects of the Higgs field. The quest to reveal the true nature of the Higgs sector is multi-faceted, requiring dedicated experimental and theoretical programs, and it necessitates pushing the frontiers of both precision and energy. In the near term, the LHC and its successor the HL-LHC are crucial for studying the Higgs field. In the longer term, future colliders will be essential for precision measurements of the Higgs sector and for a definitive measurement of the Higgs potential.

### 3.2.2 – Ongoing Projects: ATLAS, CMS, and HL-LHC

The 2012 discovery of the Higgs boson by the ATLAS and CMS experiments at the LHC was a watershed moment in particle physics; it completed the Standard Model and provided the first observation of a fundamental particle with zero spin. The dataset collected since then has provided a wealth of new measurements related to the Higgs sector. ATLAS and CMS have measured the Higgs boson mass to better than 0.2%, have established that it has zero spin, and have made initial measurements of its lifetime using quantum interference effects. The interactions, or couplings, of the Higgs boson with some of the Standard Model particles (W and Z bosons and third-generation charged fermions) have been measured to 5−10% precision.

All the major production modes of the Higgs boson have been observed, with the experimental sensitivity of some modes nearing the precision of state-of-the-art theory predictions, which are at a few percent-level accuracy. This level of precision constrains the scale of new BSM physics to be above a few hundred GeV to a TeV, depending on the model (see section 5 for further discussion). However, the LHC is still far from probing the detailed shape of the Higgs potential to the degree needed to probe the electroweak phase transition described above. Specifically, more precise measurements of the Higgs self-coupling are needed.

The next phase of the LHC, the HL-LHC, will commence in 2029 and dramatically increase the rate of particle collisions that can occur (Recommendation 1a). This challenge will be handled with new, upgraded detectors that build upon innovations in instrumentation and state-of-the-art technology. Advances in software and computing (including AI/ML) will enable experiments to gather more data and detect rare events at a greater rate. About 180 million Higgs bosons are expected to be produced during the HL-LHC run in each experiment, a factor of 10 more than what is projected for the current LHC run. This large dataset is expected to improve our understanding of the Higgs boson in a major way.

Many of the Higgs boson couplings to other Standard Model particles will be measured to a precision within a few percent or lower. Increasing the precision of these measurements to sub-percent level will provide sensitivity to BSM physics above a TeV in mass. The HL-LHC will enable measurements of the rare decay of the Higgs to muon pairs and thus show that the Higgs boson also generates the mass of second-generation fermions. The Higgs couplings will also be tested at the 2% level if the Higgs boson decays to undetected particles lighter than half its mass, such as, for example, dark matter.

For the first time, the Higgs potential will be tested experimentally; the HL-LHC will be able to say if and how strongly the Higgs boson couples to itself, and whether the Higgs field's potential energy has the form predicted by the Standard Model, with precision of around 50%. Deviations of the Higgs potential from the Standard Model predictions can have important implications related to the matter-antimatter asymmetry, as well as the ultimate fate of the universe.

Higgs boson physics can only be studied at high-energy collider experiments, which are currently limited to the LHC and HL-LHC. Longer-term, future colliders, described below, will further our understanding of the Higgs boson by testing its couplings to lighter quarks, by improving the precision of the Higgs couplings, and by measuring the Higgs potential. Advances in theoretical calculations of Higgs properties will be required to fully understand the experimental results.

### 3.2.3 – Major Initiative: Higgs Factory

Beyond the HL-LHC, a Higgs factory will produce large numbers of Higgs bosons with small backgrounds and enable more detailed studies of Higgs boson properties and interactions (Recommendation 2c). Defined as an electron-positron collider that can cover the center-of-momentum energy range of 90 GeV to 350 GeV, a Higgs factory can measure couplings with smaller uncertainties than the HL-LHC due to a combination of more precise knowledge of the momentum of the incoming particles, smaller background environments, and better detector resolutions. Higgs factories offer significant advantages to measuring the production of the Higgs boson. For example, lepton colliders allow us to identify the presence of a Higgs boson independent of how it decays, and hence provide an unbiased sample for a model-independent and high-precision measurement of its properties. This unique feature will also allow a Higgs factory to improve searches for Higgs boson decay to unknown invisible particles, such as dark matter, by an order of magnitude over the HL-LHC, and to improve the sensitivity for unexpected decays into detected particles by up to four orders of magnitude in some cases. Further discussion of these capabilities can be found in section 5.1.

Furthermore, such a collider will enable very strong consistency checks within the electroweak sector of the theory by testing it through quantum loops that relate the masses of the heaviest Standard Model particles: the W and Z bosons, the top quark, and the Higgs boson. The precision of measurements of the masses, the Higgs boson width, and its interactions with other particles will be improved by up to a factor of 10 compared to the HL-LHC. For example, a precision of 0.1−0.2% will be achieved on its coupling to the Z, which will extend the reach for new BSM physics by tens of TeV, well-beyond the HL-LHC reach (section 5.2). A Higgs factory will also significantly improve the knowledge of the coupling to the charm quark, and potentially also the coupling to the strange quark.

### 3.2.4 – Future Opportunities: 10 TeV Parton Center-of-Momentum Colliders

On a longer timescale, a 10 TeV pCM collider—for example, a 10 TeV muon collider, FCC-hh, or possibly a wakefield-based $e^+e^-$ collider—will enable a comprehensive physics portfolio that includes ultimate measurements in the Higgs sector and also a broad search program (Recommendation 4a). A precise measurement of the coupling of the Higgs boson to itself will tell us about the shape of the Higgs potential, which feeds into the behavior of the electroweak phase transition and has consequences related to the matter-antimatter asymmetry and the ultimate fate of the universe. At the HL-LHC a measurement of the Higgs self-coupling with a precision of 50% should be possible. However, a precision of 5%—an order of magnitude improvement—will dramatically enhance our knowledge of the potential and be sufficient to discover or rule out many models which could explain the matter-antimatter asymmetry. This can only be achieved with a collider with 10 TeV or greater pCM, due to the greatly enhanced rate of production of events with multiple Higgs bosons that are needed for measuring the Higgs self-coupling.

At 10 TeV pCM energies, an extremely large number of Higgs bosons will be produced. As a result, these facilities will be able to further improve measurements of Higgs boson couplings, especially for rarer decays such as muon pairs or a Z boson and a photon, which furthers the mass reach to new particles well beyond that of the HL-LHC. Colliders at 10 TeV will also be the first opportunity to improve the measurement of the strength of the top quark-Higgs boson coupling after the HL-LHC, due to the high collision energy required. Overall, the precision on the Higgs couplings increases by an order of magnitude or more at a 10 TeV pCM collider compared to the HL-LHC, achieving sub-percent level precision.

A unique aspect of a 10 TeV pCM collider is its potential to directly probe the causes of possible deviations in Higgs boson properties. At a Higgs factory, a deviation in the measured Higgs couplings would generally point to new physics outside the direct discovery reach of that collider. A 10 TeV pCM collider, on the other hand, would enable both precision measurements that illustrate indirect effects of new physics on Higgs properties and also direct discovery of the particles responsible.

Overall, 10 TeV pCM colliders have a broad search program with a high potential for observing additional Higgs bosons if they exist. They can also directly probe hidden-sector physics through Higgs exotic decays. This and the broader science case for a 10 TeV pCM collider is discussed further in section 5.1.

### 3.2.5 – 20-Year Vision

In 20 years the HL-LHC program will be completed, a Higgs factory will be preparing to take data, and a vigorous R&D program will be paving the path to a 10 TeV pCM collider. Each of these projects will fill in the map of the Higgs boson's behavior in complementary ways: The HL-LHC will deliver the first draft, the Higgs factory will provide incredible detail in key areas of the landscape, and the 10 TeV pCM collider will reveal the challenging heights of the Higgs boson's interaction with itself.

Every refinement will provide an opportunity to test whether the Higgs boson does in fact give masses to other particles as expected, to determine if it is a fundamental object or in fact composed of other particles, to see if it has unexpected interactions, and to verify that it bootstraps its own mass as predicted by the Standard Model. These studies, propelled by advances in theory, and software and computing, will enable us to obtain a much clearer picture of the Higgs boson and a better understanding of how it has shaped our universe.

This roadmap relies on the design and construction of accelerators and detectors at the forefront of particle physics. The technology choices for achieving the Higgs factory and 10 TeV pCM collider goals need to be determined based on technical feasibility, cost-effectiveness, host site capacity, sustainability, and synergies with the demands of other science topics. The European Strategy for Particle Physics Update, which typically includes scientists and funding agency representatives from the US, is planned for later this decade and will be a milestone in the decision process. In this context, a separate panel (Recommendation 6) is recommended to provide additional guidance to the accelerator program.

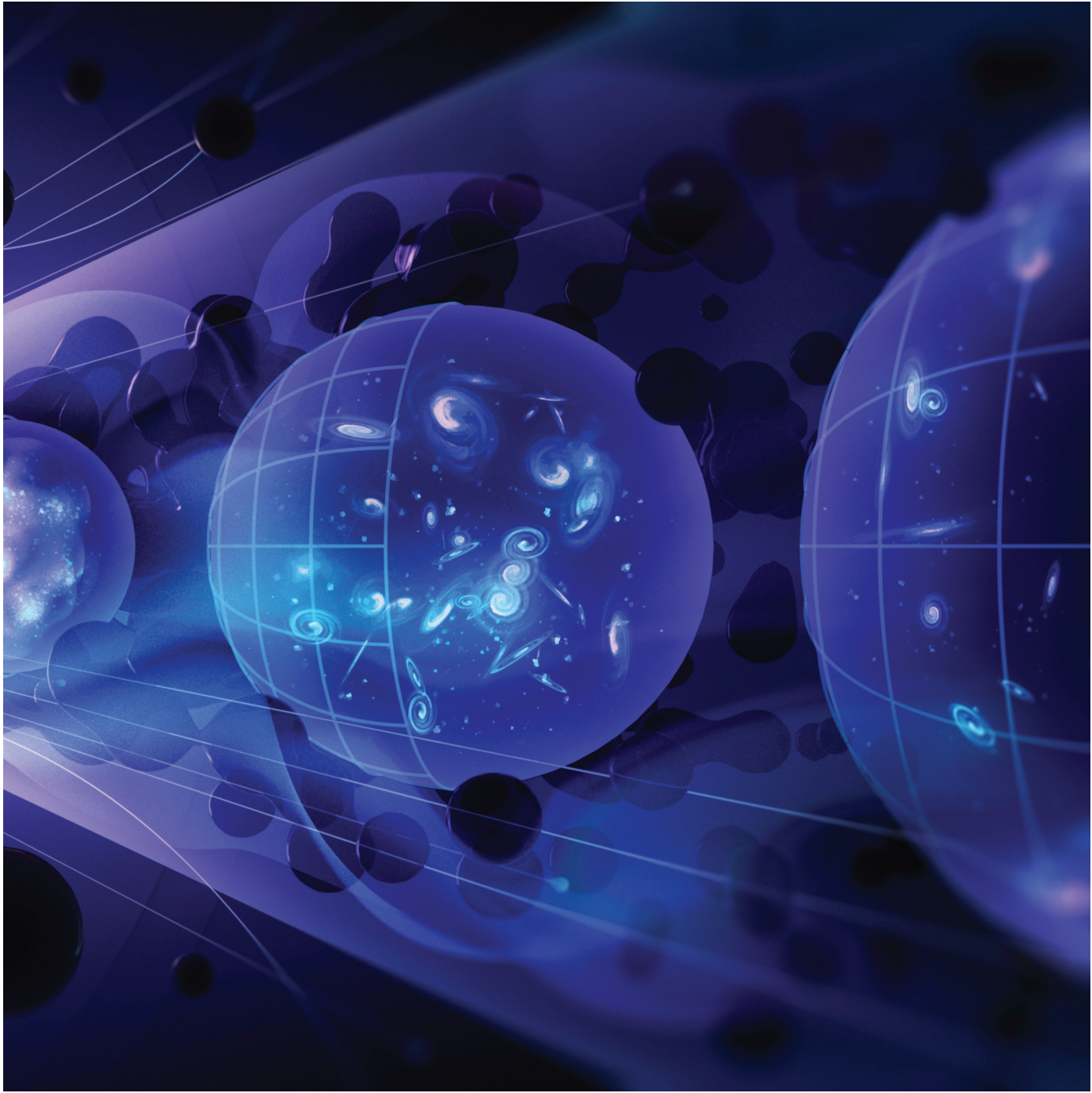

# 4

Illuminate
the
Hidden
Universe

Looking out into the night sky, we see pinpricks of light surrounded by darkness. Everything we see in the universe, including stars, planets, and plumes of interstellar gas, makes up only 5% of the universe. There is a vast hidden universe of dark matter and dark energy that is responsible for guiding the evolution of the cosmos itself, from forming the structures of matter to driving an accelerated expansion that will determine the ultimate fate of the universe.

The ΛCDM cosmological paradigm captures the physics that describes how the universe has changed over cosmic time. In the first tiny fraction of a second after the Big Bang, the universe underwent rapid, accelerated expansion called cosmic inflation, during which the seeds of the structure that was to come were created. The universe continued to expand and cool, evolving from its earliest moments of being filled with light, to the epoch of galaxy formation driven by dark matter, to the current era of accelerated expansion driven by dark energy.

Although we do not yet understand the underlying nature of dark matter and dark energy, or the specific physical processes driving cosmic inflation, we do know that they are not described by the subatomic structures in the Standard Model. Answering some of the deepest questions about particle physics itself requires detailed studies of cosmic evolution to reveal the underlying nature of dark matter, dark energy, inflation, and other particles in the universe that might have played a role in driving cosmic evolution.

To illuminate the hidden universe, our efforts are focused on two main science drivers:

*Determine the Nature of Dark Matter.* The gravitational evidence for dark matter is overwhelming. We have many ideas for what dark matter could be, with a handful of particularly compelling candidates having viable cosmological histories. The number of strong candidates inspires a multifaceted campaign to determine the nature of dark matter by leveraging underground facilities, quantum sensors, telescopes, and accelerator-based probes.

*Understand What Drives Cosmic Evolution.* The evolution of the universe has been determined by physical processes not described by the Standard Model, from the exponential expansion called inflation during the first moments of time, to intermediate periods dominated by radiation (potentially including unknown light species) and dark matter, to the cosmic acceleration of today. Measuring the growth of cosmic structure and the expansion history of the universe through multiple complementary methods offers unique explorations into inflation and dark energy, while also yielding insights into neutrino properties and the possible existence of cosmic relic particles from the earliest moments of the cosmos.

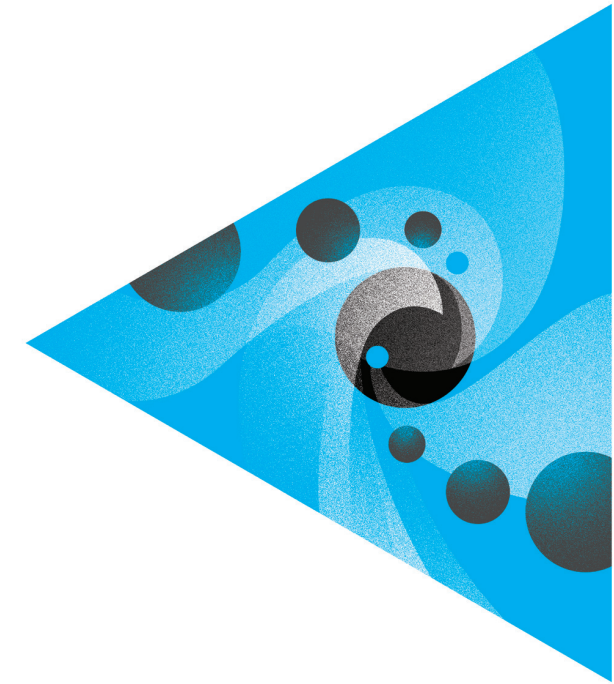

# 4.1

# Determine the Nature of Dark Matter

## 4.1.1 — Science Overview

Dark matter constitutes the vast majority of the universe's mass, influencing its structure and galaxy formation through gravity. Despite this dominant influence, its particle composition and interactions beyond gravity remain unknown. This profound mystery drives research across all frontiers of our field.

Unraveling the dark matter enigma requires a diverse approach that falls into four main categories: cosmic surveys, accelerator-based experiments, indirect detection experiments, and direct detection experiments. Cosmic surveys probe the distribution of dark matter on a variety of length scales, yielding essential data about its properties and guiding the other approaches. The other three categories search for the particles themselves. Accelerator-based experiments attempt to produce the dark matter particles, while indirect detection experiments look for the cosmic messengers resulting from dark matter interactions. Finally, direct detection experiments focus on detecting dark matter's interactions here on Earth. Each approach, guided by theoretical work, plays a vital role in increasing understanding of the nature of dark matter.

The WIMP has been the primary candidate for dark matter for three decades. With a mass between a few GeV and about a hundred TeV, the particle would have been created in great abundance at early cosmological times in thermal equilibrium with the universe. As the cosmos expanded and cooled, the WIMP would fall out of equilibrium, which would result in the formation of the cold dark matter structures we observe today.

The US has led significant efforts, using tools from all areas of particle physics, to seek evidence of the WIMP through its interactions with Standard Model particles. These efforts have shown that if WIMPs are the dark matter, they must couple to the Standard Model very feebly. Detecting these extremely rare interactions requires large detectors. Because of strong motivation from theory for WIMPs as a dark matter candidate, the pursuit of these feeble couplings remains an important benchmark.

Since the 2014 P5 report, progress in theory has expanded our understanding of other plausible dark matter candidates that have both compelling implications for particle physics and viable cosmological histories. These developments have been accompanied by significant advances in direct-detection technologies, particularly in quantum sensing, that have enabled discernment of the most minute signals. This convergence of theoretical developments and cutting-edge detection capabilities holds tremendous potential for discovery.

One theoretical approach leads to models of dark matter interacting with the Standard Model through hidden-sector particles beyond the dark matter candidate. These new interactions allow for the dark matter to be produced through mechanisms distinct from those of the WIMP, as well as permitting new dynamics within the hidden sector, such as dark matter self-interactions. Hidden-sector particles can be produced in accelerators, whereas innovative techniques can search for the cosmic dark matter particles themselves. This synergy of accelerator experiments and advanced detection methods could shed light on the nature of hidden-sector models and their crucial role in the dark matter puzzle.

Another theoretical approach has led to wave-like dark matter candidates. These candidates possess masses less than 1 eV, making them so light that they behave more like waves than particles. As a result, detection techniques are inherently quantum in nature, pushing the boundaries of quantum sensing. Within this category are the highly compelling QCD axion, which provides a solution to why interactions involving the strong force do not appear to show CP violation, and the related axion-like particles (ALPs). These particles would have been abundantly produced during the early universe, but in contrast to the WIMP, they would not have been in thermal equilibrium due to their light mass and small couplings. Instead, their abundance would have been dictated by the initial conditions of the universe. They can be searched for directly, and cosmology and astrophysics measurements are pivotal in constraining the mass range of particles of interest.

Dark matter experiments currently taking data are venturing into unexplored territories and hold the potential for groundbreaking discoveries. Developments in detector instrumentation lay the foundation for future campaigns to identify dark matter in new scenarios (Recommendation 4d). Our recommendations provide a portfolio of projects, research, and tools, including theory and computational work (Recommendations 4b and 4f), that can comprehensively target and characterize the dark matter model benchmarks.

### 4.1.2  –  Ongoing Projects: Direct Detection, Indirect Detection, and Collider Searches

Ongoing projects probe dark matter using a complementary suite of techniques and technologies. The larger ones include: the LHC, which can produce electroweak-scale dark matter candidates in a controlled environment; IceCube, which has sensitivity to spin-dependent and ultra-heavy particle dark matter candidates; and the second-generation (G2) direct detection experiments, such as the DOE-funded LZ and Axion Dark Matter Experiment (ADMX) G2, the NSF-funded DarkSide-20k and XENONnT, and the jointly funded SuperCDMS SNOLAB.

In the coming decade, the HL-LHC will be in a unique position to explore whether dark matter couples to the Higgs boson and to test weak scale supersymmetry and many other theories with dark matter candidates. The high energy and intensity of LHC collisions also enable auxiliary experiments searching for long-lived or feebly interacting hidden-sector particles. Both LZ and XENONnT started data-taking in 2021; SuperCDMS SNOLAB and DarkSide-20k are scheduled to begin data-taking in 2025 and 2026, respectively.

These experiments will improve the sensitivity to a wide range of models by more than an order of magnitude. The ADMX has already excluded the QCD axion for masses between 2.66 $\mu$eV and 4.2 $\mu$eV and is currently working to push sensitivity to higher masses. These experiments should be supported to achieve their maximum sensitivity and potential (Recommendations 1a and 1e). This support should include the necessary theory and simulation work, as well as background modeling (Recommendation 4b).

### 4.1.3  –  New Initiative: A Portfolio of Agile Projects for Dark Matter

In pursuit of understanding dark matter, a diverse and agile portfolio of ASTAE experiments, as described in section 6.2, offers the potential for significant discoveries and technological advancements. Small but sensitive detectors are ideal for studying low-mass dark matter since the needed size of the detectors scales roughly with the dark matter mass. This strategic approach focuses on two promising areas: hidden-sector models and QCD axions, both of which boast high-priority benchmark models that can best be addressed by this scale of experiment.

Accelerator-based searches for the production of hidden-sector particles leverage beam dumps at existing beamlines and have sensitivity to thermal benchmark models in the MeV–GeV mass range. The direct searches for these hidden-sector particles utilize innovative materials and ultra-low-noise detectors with the ability to detect down to sub-eV energy depositions to reach the benchmarks. This synergistic combination of approaches is necessary to understand and unlock the secrets of hidden-sector dark matter.

The search for axions and ALPs is also well-suited for this agile portfolio. Specific QCD axion models provide definitive benchmarks, and through a series of carefully designed experiments, masses from 40 neV to 1 eV can be thoroughly explored. Additionally, these endeavors lay a foundation for even more ambitious projects that target the lightest masses falling within the range of 1 peV to 40 neV.

This multi-faceted approach maximizes the potential for seminal discoveries and pushes the boundaries of what is measurable in the realm of dark matter. Notably, this portfolio has already been set in motion by the Dark Matter New Initiatives (DMNI) experiments, which have completed their design phases and now await construction funding. These initiatives are integral components of the broader portfolio of ASTAE experiments (Recommendation 3a; section 6.2).

### 4.1.4 – Major Initiative: G3, the Ultimate WIMP Dark Matter Search

The next phase of the search for WIMP dark matter requires experiments capable of reaching roughly order-of-magnitude weaker interaction strengths than current experiments. A large Generation-3 (G3) WIMP dark matter search would build on the most successful designs of the current G2 experiments, providing sensitivity to dark matter-Standard Model interactions that are small enough that neutrinos become an irreducible background (the "neutrino fog").

This improvement in reach would provide significant coverage of important benchmark WIMP models, such as the constrained minimal supersymmetric extension to the Standard Model. Such a G3 experiment would also perform important measurements of solar and possibly supernova neutrinos. A G3 direct detection experiment would be the ultimate WIMP search within the current approach; moving past the reach of the G3 experiment and deeper into the neutrino fog would require significant changes in method and technology.

Although supporting more than one G3 experiment would be beneficial, expected costs are high enough, especially compared to the costs of the portfolio of smaller dark matter projects, that funding two does not appear feasible. Our recommendation supports one G3 experiment, preferably sited on US soil to help maintain US leadership (Recommendation 2d). Investment in the expansion of SURF, taking advantage of the DUNE excavation infrastructure and potential private funding, would enable such siting. Continued support by both DOE and NSF is needed to maximize the science and US leadership. A second, complementary G3 experiment would maximize the discovery potential and would teach us more about dark matter if one of the G2 experiments has promising results.

### 4.1.5 – New Initiatives: IceCube-Gen2 & CTA

In the next decade, NSF-funded astrophysical gamma-ray and neutrino observatories will provide unprecedented views of the high-energy universe. These observatories will look for cosmic messenger particles that are made in dark matter interactions (Recommendations 2e and 3c). In addition, observations of photons, cosmic rays, neutrinos, and gravitational waves can give a more complete picture of the physics that drives the most energetic sources in the universe.

The IceCube-Gen2 Observatory will provide a tenfold improvement in sensitivity to astrophysical neutrinos over the IceCube Observatory, and will be the most sensitive probe of heavy decaying dark matter. IceCube-Gen2's wide-ranging particle physics portfolio also includes searching for signatures of neutrino physics beyond the Standard Model. In addition, IceCube-Gen2 has a wide-ranging multi-messenger astrophysics portfolio, which gives us a more complete picture of the physics that drives the most energetic sources in the universe that produce the highest energy neutrinos and cosmic rays.

The CTA offers a parallel improvement in sensitivity to very-high-energy gamma rays, along with refined energy resolution over an expanded energy range and a more sharply resolved picture of the gamma-ray sky. The CTA and the Southern Wide-field Gamma-ray Observatory (SWGO) provide sensitivity to WIMP thermal targets that lie beyond the reach of the G3 direct detection experiments. The CTA's excellent energy and angular resolution play a key role in disentangling a dark matter signal from conventional astrophysical backgrounds. Beyond dark matter, the CTA's broad astrophysics portfolio will provide insights into the most extreme environments in the universe and the origin and role of relativistic cosmic particles.

### 4.1.6 – Complementarity: Astrophysical and Cosmological Probes

Astrophysical and cosmological probes, such as observations of the Milky Way satellite galaxies, stellar streams, strong lensing systems, and the CMB, can map the distribution of dark matter on small length scales where the standard cold and collisionless nature of dark matter may break down. On these scales new interactions among dark matter particles, which are predicted in many hidden-sector models, can lead to structure-formation phenomena such as halo core formation or gravo-thermal collapse that would be absent in WIMP or QCD axion models. The insights gained could guide terrestrial experiments.

Similarly, a dearth of dark matter structure on small scales could be indicative of dark matter with a significant thermal velocity, the quantum pressure from an ultra-light axion, or dark matter particles that had significant interactions with relativistic species in the recent past. CMB-S4 will have exquisite sensitivity to such light species that reside in hidden-sector models. Over the next decade Rubin/LSST is expected to discover a large number of new Milky Way satellites, stellar streams, and strong lensing systems. Follow-up observations of these objects with existing ancillary telescope resources or future observatories, such as DESI-II and Spec-S5, combined with state-of-the-art cosmological simulations and analyses, will access parameter space inaccessible to laboratory experiments.

Thus, developing a comprehensive understanding of the nature of dark matter requires the complementary support of both astrophysical and terrestrial probes. In practice this can be done within the DOE HEP Cosmic Frontier model by supporting individuals contributing to the primary HEP science goals of ongoing projects (DESI, Rubin/LSST, and DESC) to also carry out complementary work on dark matter as a secondary science goal. In addition, computational and modeling work relevant to astrophysical probes can be performed with theory support.

### 4.1.7 — 20-Year Vision

The program outlined above encompasses a series of experiments planned for this decade that hold immense potential for discovery. Simultaneously, it nurtures the growth of next-generation experiments. Over a 20-year timeframe, this carefully curated portfolio of experiments will conduct targeted searches encompassing WIMP, hidden-sector, and QCD axion dark matter models. The investments in the construction of a Higgs factory and in the R&D and technology tests and demonstrators for a 10 TeV pCM collider will be essential steps toward achieving unprecedented sensitivity to feebly coupled particles (see section 5.1). Reaching the 10 TeV scale is needed to achieve definitive covering of the thermal targets for minimal WIMP candidates.

Meanwhile, astrophysical probes will provide complementary insights into the nature of dark matter. Discoveries would be followed up with studies by multiple means with improved sensitivity, informed by this proposed portfolio of experiments. These endeavors not only foster the development of novel technologies but also challenge the limits of what can be effectively measured. Acting as a vital bridge, theory interconnects diverse measurements and guides us toward uncharted avenues of exploration. This multi-pronged strategy, harmonizing experiments of varying scales, theoretical frameworks, and technological advancements, provides a coherent roadmap that optimizes the potential for seminal discoveries.

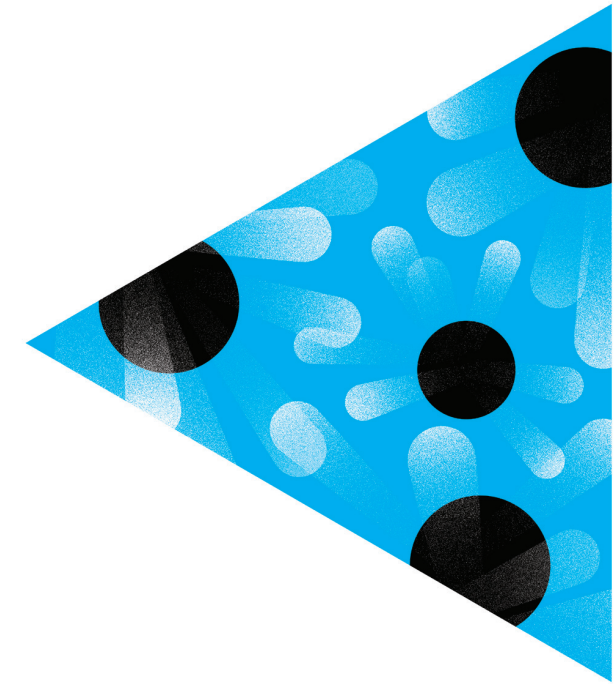

# 4.2

# Understand What Drives Cosmic Evolution

### 4.2.1 — Science Overview

The dynamical evolution of the universe is deeply connected to its energy content. At its earliest moments, the universe was sensitive to particle physics processes at energies far beyond what can be probed even in the LHC, today's most powerful particle accelerator. As the universe expanded and cooled, clues about those early high-energy phenomena were imprinted on the CMB, the distribution of cosmic matter and light. Those clues allow us to probe new physics that is inaccessible by other means, playing an essential role in advancing our knowledge of particle physics.

As we peer into the night sky, we see the cumulative effect of the multiple eras that the universe has gone through to reach its present rich structure. The earliest epoch was an era of apparent accelerated expansion, referred to as inflation, during which the initial seeds of structure were created. Once inflation ended, the universe transitioned to a hot radiation era in which ultra-relativistic particles dominated its energy density. During that era, new light species predicted by many promising theories beyond the Standard Model could leave subtle signatures on the evolution of the universe and give insight to the nature of dark matter. The matter era that followed allowed the universe to mature under the dominant influence of dark matter and form the stars, galaxies, and clusters now populating the cosmos. Most recently, the universe entered another era of accelerated expansion, requiring dark energy to form the majority of the energy budget of the universe today.

The key particle physics questions about the universe's evolution that cosmic surveys seek to answer are the following: What physics is responsible for the rapid, accelerated expansion during the early inflationary era? Were there extra light species beyond photons and neutrinos present in the universe during the radiation-dominated era? What is driving the current accelerated expansion of the universe?

Answering these questions requires scientists to develop a detailed understanding of (i) the nature, properties, and type of primordial fluctuations created during the inflationary era, (ii) the evolution and growth of these initial fluctuations into the visible objects we observe today, and (iii) the cosmic expansion history of the universe. The current paradigm for addressing these questions attributes the recent cosmic acceleration to a cosmological constant, a uniform repulsive energy throughout the universe. However, some tensions arise when attempting to consistently interpret both early and recent expansion data within this framework. These tensions may hint at physics that requires a significant paradigm shift.

The early universe's primordial fluctuations, resulting from inflation, reveal critical insights into the physics of that era. These primordial fluctuations comprise energy density variations shaping the universe's structure and gravitational waves indicating space-time's response to high inflation energies. Analyzing the statistical properties of density fluctuations through the distribution of galaxies and CMB anisotropies uncovers the inflationary dynamics. Primordial gravitational waves leave a unique signature in CMB polarization; that signature offers clues about fundamental physics at high energies. Post-inflation, the universe transitioned into a hot radiation phase that allowed investigation of new light particle species beyond the Standard Model. CMB measurements provide essential constraints on these species. Future measurements may probe relics present during the quark-hadron transition and, eventually, light species present at temperatures above the electroweak scale.

The matter-dominated epoch that followed the radiation-dominated era was recently interrupted by a burst of accelerated expansion driven by what is termed dark energy. The simplest explanation for dark energy, a cosmological constant ($\Lambda$), is a pillar of the $\Lambda$CDM-cosmological paradigm, but confirming the nature of dark energy as either $\Lambda$ or something else is a key science driver for our field. Investigating dark energy's impact on cosmic structure growth and expansion history is crucial, with both observational and theoretical advances needed to distinguish between the cosmological constant of $\Lambda$CDM, more complex dark energy models, and alternative cosmological models with modified gravity.

Since the last P5 report, advances in our understanding of the early universe have been made through precise observations of the CMB's temperature and polarization fluctuations. The US-led ground-based observing program, including the BICEP program, the South Pole Telescope (SPT), the Atacama Cosmology Telescope (ACT), and POLARBEAR, combined with the European Planck Satellite, have provided us with crucial bounds on the energy scale of inflation, the abundance of light relics in the early universe, and the sum of neutrino masses, in addition to shedding light on the fundamental nature of dark matter and dark energy.

Significant progress has also been made in building statistically consistent portfolios of observations from galaxy surveys with well-understood systematics. On the growth of structure, a combination of (i) gravitational lensing, (ii) the distribution of galaxies across time and distances, (iii) the abundance of galaxy clusters, and (iv) the velocities of galaxies near over-dense regions can precisely probe the parameters characterizing dark energy, determining whether it is consistent with a cosmological constant or has a dynamic of its own. For the cosmic expansion history, the transition from the past decelerating phase to the current era of accelerated expansion can be probed primarily through four types of measurements: the brightness of distant supernovae, strong lensing cosmography, the evolution in distance of a distinctive pattern in the distribution of galaxies known as the baryon acoustic oscillation signature, and gravitational wave standard sirens, which indicate cosmic distance. The major advances made by the Dark Energy Survey (DES) and the extended Baryon Oscillation Spectroscopic Survey (eBOSS) in development and application of these measurement methods, and in theoretical modeling, have directly influenced the ongoing and new projects recommended in this P5 report.

Our recommended portfolio will enable seminal discoveries connected with these crucial scientific questions along with other secondary science cases noted below. It is important to note that support for computational work and theory efforts (Recommendations 4b and 4f, as well as a robust workforce (Recommendation 5), are essential for realizing the scientific potential of the experiments described below.

## 4.2.2 – Major Initiative: CMB-S4

CMB-S4 is the transformative next-generation CMB experiment, with the ambitious primary science goals of constraining the energy scale of inflation and determining the abundance of light relic particles in the early universe. CMB-S4 will also measure the sum of neutrino masses and probe the physics of dark matter and dark energy, in addition to a rich astrophysics program. This project will provide a significant advancement in sensitivity to gravitational waves produced by inflation and to light relics. Importantly, this sensitivity to light relics will extend beyond the quark-hadron transition, which is an important benchmark in searches for new physics. CMB-S4 construction is planned to begin in Chile and at the South Pole late in this decade (Recommendation 2a).

The current generation of ground-based CMB experiments includes the South Pole Observatory (BICEP Array and the South Pole Telescope, both currently operating), and Simons Observatory. CMB-S4 builds on decades of experience from US-led ground-based CMB experiments, but with increased sensitivity achieved by scaling up to nearly 500,000 detectors.

CMB-S4 presents an important opportunity for the field of particle physics: the discovery of gravitational waves produced by inflation in the extremely early universe would provide a direct window to that previously inaccessible epoch in cosmic history and to the highest energy scales in the universe. CMB-S4 is designed to cross critical thresholds in the search for gravitational waves from inflation. Even a non-detection from CMB-S4 would rule out large classes of inflation models, placing interesting constraints on the theoretical landscape.

Achieving CMB-S4's ambitious science goals requires installing telescopes at and observing from both the South Pole and Chile, which are proven sites with good observing conditions and infrastructure. For ground-based CMB experiments, the unprecedented sensitivity to the physics of inflation that probes the highest energy scales in the universe is possible only at the South Pole. The site at the South Pole is unique in that it provides a dry, stable atmosphere with continuous observation of the same patch of sky. The site in Chile is complementary because Earth's rotation leads to the ability to observe large portions of the sky, which is important for constraining the abundance of light relic particles. Coordination between DOE-HEP, NSF-AST, and NSF-OPP is critical for the success of CMB-S4. NSF-OPP and CMB-S4 should continue to work closely together to ensure that the logistics footprint of the project at the South Pole is consistent with site capabilities (see 6.6.4).

Given the planned landscape of ground- and space-based CMB experiments, CMB-S4 plays a unique role in using demonstrated technology with a two-site survey design that is crucial for addressing the key science goals. The two-site design ensures that the whole is far greater than the sum of its parts; it enables important crosschecks on systematics that would otherwise be impossible. CMB-S4 also provides important synergies with the space-based LiteBIRD instrument, which aims to launch in the next decade and constrain the energy scale of inflation through a complementary technique.

### 4.2.3 – Ongoing Projects: Rubin/LSST and DESC, DESI

This decade will see tremendous advances in our understanding through the galaxy survey program established by the last P5 report: DESI (a spectroscopic survey), and Rubin/LSST. Rubin Observatory will carry out the LSST, a 10-year imaging survey, and the LSST Dark Energy Science Collaboration (DESC) will carry out the fundamental physics analyses of LSST. DESI and the LSST will enable analyses with multiple complementary methods of both structure growth and expansion history of the universe, with extensive programs to control systematics (Recommendation 1e). In particular, these experiments will provide unprecedented constraints on cosmic acceleration using several probes of structure growth and expansion rate during the entire time period of cosmic acceleration. The experiments will also reach back into the weakly matter-dominated era, when the expansion was still decelerating. Strong support for operations and data analysis will ensure a return on the investment in these experiments. This program will stress-test the standard cosmological paradigm and is particularly powerful when combined with space-based datasets (for example, from Euclid or the Nancy Grace Roman Space Telescope), with current CMB surveys, and even more so with CMB-S4. These combinations would benefit from dedicated efforts towards joint analysis including N-body simulations and simulated survey products.

As these surveys yield discoveries about cosmic evolution and improve our understanding of how to robustly constrain the cosmological model despite astrophysical and observational systematics, the particle physics community should use that new understanding to formulate future galaxy surveys. This includes, during the first five years of LSST, engagement with discussions of the post-LSST future of Rubin Observatory; and during DESI/DESI-II, refinement of Spec-S5 survey concepts.

These cosmic surveys also have important secondary science goals, such as constraining the sum of the neutrino masses (providing complementary information with the measurements of the mass ordering by DUNE) and dark matter (where astrophysical probes have an important place among other types of measurements).

### 4.2.4 – New Initiative: DESI-II upgrade

The DESI-II program (Recommendation 3c) is a first step at going beyond the galaxy surveys constructed as a result of the previous P5 report. Besides providing an opportunity for testing technology for next-generation spectroscopic surveys, its scientific goals include constraining cosmic acceleration by extending DESI dark energy constraints deeper into the matter-dominated regime, and complementing/enhancing dark energy and dark matter science with Rubin/LSST by leveraging the power of overlapping spectroscopic and imaging surveys.

For example, the program could provide ~5%-level constraints on the dark energy density at a time when the standard cosmological paradigm predicts that it is only a few percent of the energy density of the universe. The program also provides opportunities for spectroscopic observations that would reduce key systematic uncertainties in Rubin/LSST measurements of structure growth and cosmic expansion. Given the low construction cost for this extension to the DESI project, executing DESI-II provides a high scientific return on the existing investments in both DESI and Rubin/LSST, especially as DESI-II will serve as a pathfinder for the next proposed major initiative, Spec-S5.

### 4.2.5 – Future Opportunity: Spec-S5

The proposed next-generation spectroscopic survey, Spec-S5, holds great promise to advance our understanding and reach key theoretical benchmarks in several areas: inflationary physics via the statistical properties of primordial fluctuations, late-time cosmic acceleration, light relics, neutrino masses, and dark matter. The balance between these scientific goals, which affects survey design, should be refined in light of early DESI and Rubin/LSST results.

The coming years will see important preparations for Spec-S5—site selection and other crucial decision points—with support for necessary instrumentation R&D and refinement of the survey concept. Going beyond the capabilities of DESI-II and Rubin/LSST, Spec-S5 will permit us to map cosmic expansion deep into the matter-dominated regime, while also enabling order-of-magnitude improvements in our understanding of the early era of cosmic inflation. Carrying out these preparations during the 2020s, including the computational and theory work necessary to interpret the data, is essential to continuing the robust program of spectroscopic surveys.

Spec-S5 could be ready for construction at the end of this decade if key decisions regarding siting and other issues are resolved. With limited funds under the baseline budget scenario, the difficult choice was made to support Spec-S5 R&D (Recommendation 4e) but not construction; that support would result in a significantly more mature survey concept for consideration for immediate construction by the next P5. However, in the event of exciting discoveries in DESI and/or Rubin/LSST, and in better funding scenarios, a more mature Spec-S5 concept should be considered for construction at the end of this decade (section 2.6.2).

## 4.2.6 — Future Opportunities: Line Intensity Mapping and Gravitational Waves

Line intensity mapping (LIM) techniques are potentially a valuable future method to address key particle physics science issues during the next 20 years by probing the expansion history and the growth of structure deep in the matter-dominated era when the first galaxies were forming. LIM observations of that era could enable tests of the theory of inflation by providing a precise map of tracers of structure formation, such as emission from primordial hydrogen gas or other atomic or molecular lines. This technique has the potential to access an earlier epoch in the universe than Spec-S5. Work in both analysis and instrumentation to prove the viability of this method should continue with multi-agency support (Recommendation 4e), including low-cost instrumentation development competed through the DOE R&D program. DOE has already partnered with NASA to construct one pathfinder LIM experiment, LuSEE-Night, and there are exciting opportunities for investment in ground-based activities in the coming decade.

Gravitational waves are a powerful new tool for exploring a range of astronomical and particle physics topics, including probing the expansion history of the universe using standard sirens. NSF has been an excellent steward of this program and should support the development of new capabilities and a next-generation project. The particle physics case for studying gravitational waves at all frequencies should be explored by expanded theory support.

## 4.2.7 — 20-Year Vision

We are entering an exciting era in our study of cosmic evolution. The projects recommended by the last P5 report that are now beginning operations, the project portfolio recommended by this P5 report, and the future projects for which R&D and project definition will occur in this decade will allow for great progress in our knowledge of the entirety of our cosmic history, from the inflationary era, through the radiation and then matter dominated eras, to the dark energy era. Together with strong theory and computational support, that progress lays the foundation for the next generation of projects.

To support the success of this portfolio of cosmic surveys at a range of wavelengths, continued work and advocacy will be important to prevent or mitigate the effects of human-produced nuisances, including light pollution, satellite constellations in low-earth orbit, and radio-frequency interference.

The knowledge gained from CMB-S4 and eventually from Spec-S5 will enlighten us about the nature of inflation at the earliest cosmic times, both in terms of the energy scale and the inflationary dynamics. We recommend pathfinding works in the next decade, specifically LIM R&D and research, that will allow us to follow up any detected primordial signal from the inflationary era. Moving forward in cosmic time to the radiation and matter eras, we will have a window to new relics during the quark-hadron transition, and lay the groundwork for future projects that can push down to the electroweak scale.

In the event of a discovery beyond the standard cosmological paradigm, LIM and high-resolution CMB experiments could be formulated to confirm and characterize the discovery. Future gravitational wave experiments could provide complementary means to probe the expansion history deeper in the matter era. And at late times, our recommended portfolio sets us up with multiple complementary means to rigorously test the cosmological constant hypothesis and discover the time evolution of dark energy.

The flexibility of Spec-S5 to address multiple scientific goals (inflation, late-time cosmic acceleration, dark matter) depending on the priorities that emerge from DESI, early DESI-II, and Rubin/LSST results makes it a crucial part of this 20-year vision. Similarly, future survey concepts for Rubin Observatory, to be developed later this decade after early LSST science results are available, could address key questions that come to the forefront of particle physics studies of cosmic evolution in five to ten years.

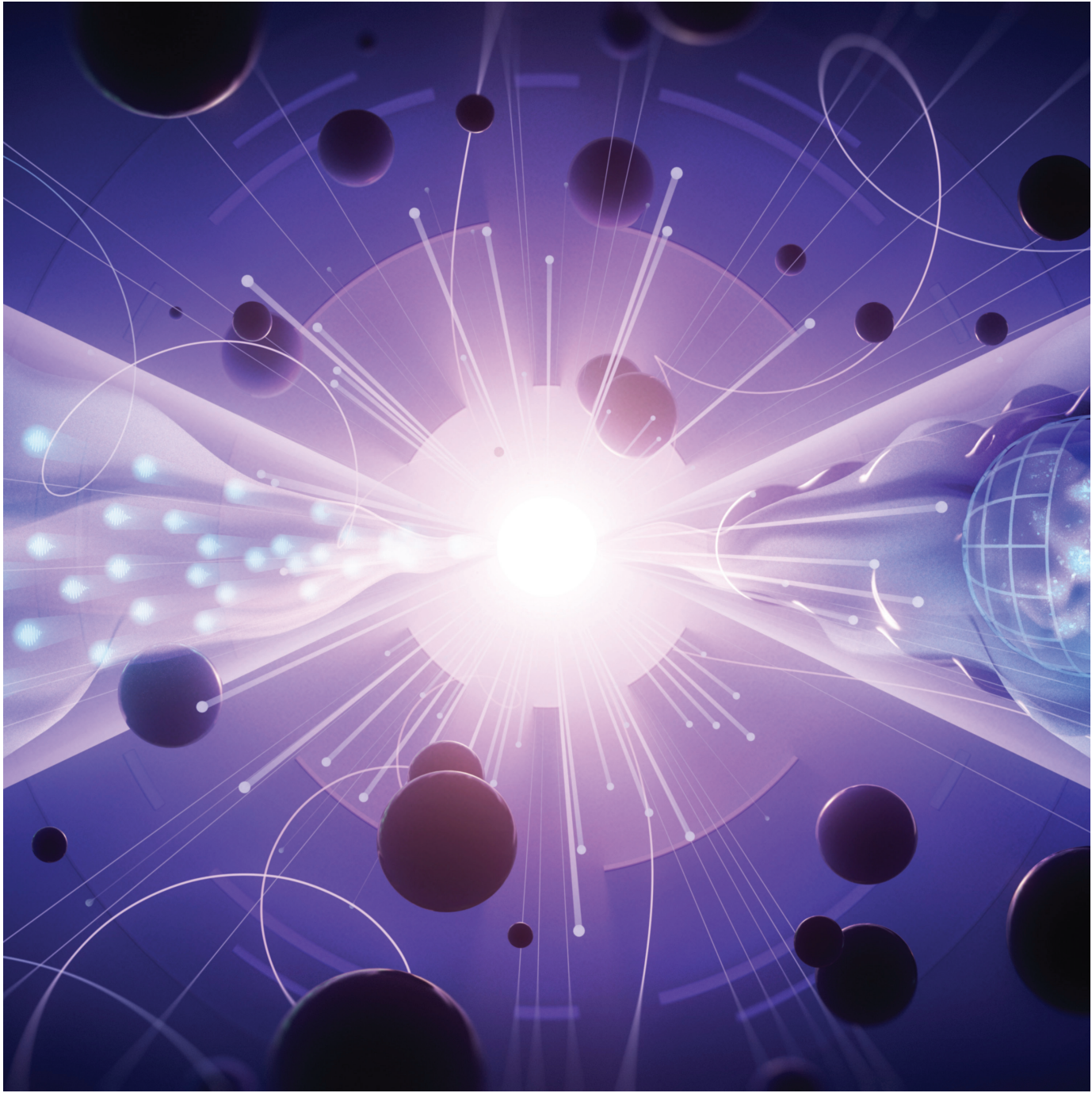

# 5

---

Explore
New
Paradigms
in Physics

The prevailing paradigms of particle physics and cosmology—the Standard Model and ΛCDM, respectively—are triumphs of experimental ingenuity, theoretical creativity, and human curiosity. Together, these pillars of fundamental physics explain a vast range of phenomena, from the gravitational scaffolding of the universe due to dark matter, to the complex structure of nuclei due to quarks and gluons. Particle physics lays the conceptual foundations for modern science, and it exemplifies the power of multigenerational and international collaborations to tackle grand scientific challenges.

The quest to understand the fundamental structure of the universe is far from over and mysteries remain. Our existence can be traced to a tiny difference between the amount of matter and antimatter in the early universe, which can be accommodated but not explained by the ΛCDM model. The structure of atoms only requires one generation of matter particles, but the Standard Model has three generations, with no obvious pattern or logic to the triplication. A repeated theme in the history of physics is unification, where seemingly disparate phenomena turn out to be manifestations of a common structure. Although there are tantalizing opportunities for unification within these prevailing paradigms, none has thus far withstood experimental scrutiny.

Given these open questions, we explore new paradigms that might yield transformational insights into our universe. There are two broad strategies for venturing into the unknown:

*Search for Direct Evidence of New Particles.* Experiments that seek direct evidence for new particles set the gold standard for discovery. Heavy particles can be produced at colliders with sufficiently high energies, whereas light but elusive particles can be produced at accelerator-based experiments with sufficiently high intensity. The discovery of new particles, or definitive evidence for their absence, would ignite major paradigmatic shifts and determine the direction of future research.

*Pursue Quantum Imprints of New Phenomena.* Even if new particles cannot be produced directly, they can still leave clues to their existence via quantum imprints on known particles. This is especially true if the new particles break a fundamental symmetry of the Standard Model. Many known particles were first detected indirectly through their quantum imprints, with follow-up direct experiments providing definitive evidence. This motivates continued investments in a broad search program for possible quantum imprints of new phenomena.

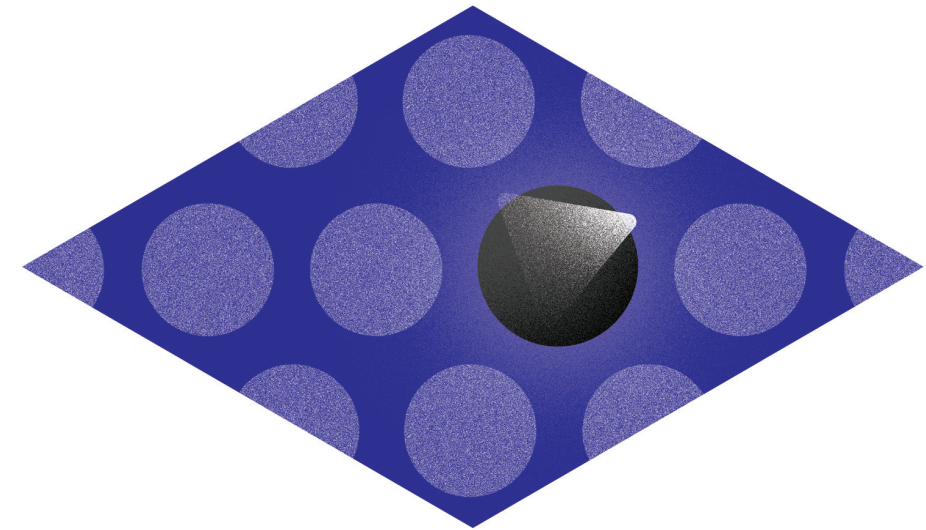

# 5.1

---

# Search for Direct Evidence of New Particles

## 5.1.1 – Science Overview

Particle physics is a field that earned its name through the discovery of many particles that were thought to be the fundamental building blocks of matter. This began with the extraordinary discovery that every object in our world is made of scores of chemical elements. Each of those chemical elements is made of a particular kind of atom, with an atomic nucleus of specific electric charge surrounded by orbiting electrons that negate that electric charge. Although the electrons are understood to be point-like elementary particles with no internal structure, the atomic nuclei were discovered to consist of protons and neutrons. The number of protons in an atom's nucleus determines the nature of the chemical elements that, in turn, determine the nature of every object we see. This was a unifying paradigm that explained the physical world around us.

But nature is complicated, and further discoveries revealed that the proton and neutron were made of fundamental particles called quarks bound together by a force particle called the gluon, carrier of the "strong force." It was also discovered that many particles disintegrate over time into lighter ones. This led to the discovery of the "weak force," which was later determined to be similar to electromagnetism. The weak force is mediated by massive W and Z bosons, as opposed to the massless photon. Through the process of direct observation of new particles, the community of particle physicists arrived at the current paradigm: the Standard Model of particle physics.

However, we also learned that the Standard Model is far from complete. From the establishment of the ΛCDM paradigm of cosmology, we need dark matter, inflation, and dark energy, none of which are part of the Standard Model. For the moment, we do not know what led to the development of our matter-dominated universe which is suitable for life. When inflation started, the universe was much smaller than an atomic nucleus. Inflation rapidly turned energy into matter and expanded it to create the immense universe that exists today. Our current understanding is that this expansion must have resulted in an equal number of matter and antimatter particles. If it had stayed that way, however, all of the matter and antimatter particles would have annihilated each other, reverting back into pure energy.

Some physical process must have turned an extremely small fraction of the produced antimatter into matter, but the Standard Model does not explain this physical process. The model also lacks the quantum description of gravity consistent with the general theory of relativity developed by Albert Einstein early in the 20th century. It is increasingly clear that discovery of new particles and their interactions is awaiting us and is likely to come from the exploration of the energy frontier.

The answers to the current mysteries about the universe are believed to be related to not-yet-discovered physics at the electroweak energy scale, a fundamental scale of nature at about 100 GeV. The apparently arbitrary size of this scale affects parameters as diverse as the dimensions of atoms and the half-lives of radioactive nuclei. The mechanism that sets this fundamental energy scale, or temperature, is unknown, and almost all theoretical explanations require new particles with masses of the same order of magnitude as this scale. Cooling through this temperature is known as the electroweak phase transition and results in many fundamental particles acquiring masses. The strength of this phase transition may have implications for the origin of matter, or baryogenesis. Understanding the origin of this energy scale and the impact that the associated physics has on our universe is a major quest of particle physics today.

The most direct way of answering these questions is by discovering new fundamental particles. If these are very massive, they can only be produced directly in high-energy colliders, as the higher the collider energy the higher the mass that can be produced. Another possibility is that these particles are produced at lower energy but very rarely—for example, in decays of known particles such as the Higgs boson. Discovering these rare, lower energy particles requires accelerators that produce very high numbers of particles, including neutrino experiments with their high intensity beams and massive detectors.

These complementary approaches provide access to an extensive theoretical parameter space that covers both higher mass scales and new physics that is weakly coupled to the Standard Model. Overall, these searches can be broadly categorized into those that are guided by specific theoretical ideas, searches driven by questions resulting from experimental data (for example, dark matter), and searches that are model-agnostic and perform a general exploration of the unknown. Together, these approaches provide comprehensive coverage of the landscape beyond the Standard Model (BSM) and have the potential to yield groundbreaking insights into the universe.

A broad set of theoretical ideas guide BSM searches at colliders. These include, for example, supersymmetry, a well-justified mathematical theory that predicts partner particles with a broad range of properties, and theories with composite particles or extra dimensions. A new experimental focus centers on maximizing the ability to identify exotic phenomena predicted by specific theories that are harder to find. Looking beyond specific models, experimental searches target exotic BSM states, including new gauge bosons or Higgs bosons, fermions, and other resonances that are a feature of many Standard Model extensions, including the ones motivated by the smallness of the neutrino masses.

Searches for dark matter are particularly well-motivated by many astrophysical and cosmological measurements that require BSM explanations. Weakly interacting massive particles (WIMPs) continue to be an important target for particle experiments, and high-energy colliders are well-suited to search for them. In a widely studied and simple WIMP scenario, the dark matter particle is identified as a new heavy stable particle in the TeV range. In many scenarios the dark matter is only a part of the dark (or hidden) sector. Some of the hidden-sector particles may be directly observed through their tiny couplings to the Standard Model particles. Yet another class of candidates includes axion and axion-like particles. These typically couple very weakly to Standard Model particles and can be searched for at high-intensity beam-dump experiments, colliders, and dedicated experiments (see also section 4.1).

The Higgs boson is connected to most of the fundamental questions about the universe and is an integral part of the collider search program. The question of whether the Higgs boson is an elementary particle or a composite particle built of more fundamental constituents remains an important mystery. Current constraints from resonance searches suggest that if the Higgs is composite, the relevant energy scale is beyond the TeV scale. Future colliders that explore this energy scale could reveal the composite structure.

Another important question is whether there are additional Higgs bosons as predicted in supersymmetry models and elsewhere. The Higgs mechanism, as postulated in the Standard Model, is the most mathematically simple method to realize the electroweak symmetry breaking and give particles their masses. However, this simplicity causes serious problems when extrapolating the Standard Model to high energies; for example, a fantastical accident is required to accommodate the observed mass of the Higgs boson. These problems are alleviated if additional particles related to the Higgs are introduced, which is why it is important to search for those. The search must be very wide, since even the simplest extensions to the Higgs sector display a broad range of phenomenology and connections to fundamental physics questions like the origins of the electroweak phase transition and baryogenesis.

The small natural width of the Higgs boson presents an opportunity for observing its decays into new particles even if the couplings between those particles and the Higgs are exceedingly small. Searches for exotic Higgs boson decays remain highly motivated even though the currently measured Higgs boson properties match the Standard

Model expectations. Indeed, in many plausible scenarios these new particles could be our only direct window into physics beyond the Standard Model. These scenarios include decays into invisible particles, a mix of invisible and Standard Model particles, and complicated cascades with many-body final states. Long lifetimes are a generic feature of BSM particles in these cascades, yielding hard-to-detect but extremely low background signatures. The rate of such decays can be very small, so large samples of Higgs bosons are needed.

Specific examples of theoretical ideas and experimental searches that address some of the open questions in particle physics have been described above. However, the exploration of the energy frontier will allow us to observe exotic particles and interactions irrespective of whether or not they have been predicted by current theoretical understanding. High-energy colliders enable us to explore the unknown with the potential for discoveries beyond our current imagination.

## 5.1.2 — Ongoing Projects: ATLAS, CMS, LHCb, and HL-LHC

The ATLAS and CMS experiments at the LHC have explored an enormous amount of BSM parameter space and dramatically changed theoretical perspectives on the most pressing questions of high energy physics. The results from these studies have tightly constrained minimal models of supersymmetry, while dark matter searches have ruled out large chunks of the theoretical WIMP territory, and have set strong constraints on the allowed values of the properties of both the dark matter and mediator particles. There have been major improvements in sensitivity to new, heavy gauge bosons and new fermions with the reach going beyond five times what was achieved prior to the LHC. Innovative experimental techniques, propelled by novel theoretical insights, have started to explore challenging signatures such as compressed spectra, boosted topologies, and long-lived particles.

In the immediate term, the HL-LHC will expand on this BSM program with a factor of 20 enhancement of the LHC Run 2 luminosity (Recommendation 1a). The reach for new, heavy particles will be extended significantly and the HL-LHC will produce approximately 180 million Higgs bosons in each of ATLAS and CMS, enabling a robust program of searches for exotic Higgs decays. Searches for dark matter will further scrutinize additional WIMP parameter space and target hidden sectors and particles interacting more feebly than WIMPs.

New detectors, such as picosecond-precision timing detectors, and forward tracking and extended trigger systems, will enable searches to better target new physics with challenging signatures. Alternative data-taking strategies and novel analysis techniques leveraging advances in AI/ML (for example, anomaly detection) will provide access to parameter space that is currently unexplored.

The LHCb experiment, despite smaller luminosity and detector coverage, has a unique design that covers particles produced at small angles to the beam, which allows it to be competitive with the general-purpose detectors for some new particle searches. This is particularly true for Higgs decays into long-lived particles, where LHCb can leverage its advanced tracking and vertex detection capabilities along with real-time data processing. Its upgrade for HL-LHC will allow for significant increase in instantaneous luminosity and sensitivity to the hidden sectors (Recommendation 3c).

## 5.1.3 — New Initiative: A Portfolio of Agile Projects to Search for Direct Evidence of New Particles

Another strategy to look for long-lived particles at colliders is to construct auxiliary experiments that are placed far away from the primary collision points. Proposed auxiliary experiments like CODEX-b and MATHUSLA can extend the sensitivity to BSM particle lifetimes in Higgs decays by several orders of magnitude. Experiments like FASER2 and FORMOSA at the proposed Forward Physics Facility at CERN would be sensitive to the hidden sectors through the vector and heavy neutral lepton portals. At Fermilab, PIP-II is expected to make many more protons than needed for DUNE, and we anticipate proposals for experiments using the excess protons. These experiments should compete in the portfolio for agile projects (see Recommendation 3a and section 6.2).

## 5.1.4 — Major Initiative: Higgs Factory

Beyond HL-LHC, an electron-positron Higgs factory (Recommendation 2c) will provide a very large sample of Higgs bosons with small backgrounds, as well as unique access to exotic Higgs decays, which a hadron collider may find challenging to identify. Such a machine will provide access to new direct production processes below the Higgs boson mass and will have indirect sensitivity to higher energy scales, as described in section 5.2. Results from searches related to extended Higgs sectors are expected to improve upon the HL-LHC results by an order of magnitude, covering a wide range of plausible parameter space where a strong electroweak phase transition enabling baryogenesis is allowed. Accumulation of large luminosities will enable the exploration of uncharted territories in direct searches for feebly coupled light states, such as heavy neutral leptons and axion-like particles. Since it will also produce very large numbers of Z bosons, new particles will also be searched for in Z-boson decays.

## 5.1.5 — 20-Year Vision and Future Opportunities

The program described in this section consists of a combination of large and small projects and holds great promise for discovery. By the end of this 20-year period we will have ultimate LHC results from the general-purpose experiments and a constellation of agile auxiliary experiments. We will also be in the final stages of construction of a Higgs factory and will have made progress on the high-field magnets, multi-MW proton driver, wakefield accelerator technology, and muon cooling, including operation of several technology tests and demonstrators (see sections 6.4 and 6.5). All of this progress will enable us to move forward with a 10 TeV pCM collider.

A 10 TeV pCM collider (muon collider, FCC-hh, or possible wakefield collider) will provide the most comprehensive increase in BSM discovery potential (Recommendation 4a). Dramatic increases in sensitivity are expected for both model-dependent and model-independent searches. Such a collider will be able to reach the thermal WIMP target for minimal WIMP candidates and hence will play a critical role in providing a definitive test for this class of models.

In many cases, the sensitivity for new gauge bosons, fermions, or other resonances will be extended by an order of magnitude beyond the HL-LHC. For example, for a universal Z′ benchmark scenario, direct searches at a 100 TeV proton collider provide extensive coverage up to masses of about 45 TeV for a range of couplings. Furthermore, a 10 TeV muon collider can uniquely probe Z′ masses around 100 TeV using indirect effects.

A 10 TeV pCM collider will provide access to new hidden sectors by producing a substantially higher mediator mass or probing even smaller couplings. It will provide opportunities to produce new states with masses of order 10 TeV and directly address the open question related to the composite nature of the Higgs boson. It can also serve as a giga-Higgs factory and provide the ultimate reach for Higgs-like scalars. For example, direct searches for an extra scalar at a 10 TeV muon collider, taking advantage of vector boson production enhancement, could probe most of the parameter space corresponding to percent level deviations in Higgs couplings, and even explore regions with smaller deviations that will be difficult to observe in precision measurements.

Overall, 10 TeV pCM colliders are a unique tool to directly produce and study new particles and their properties. Their comprehensive coverage of the BSM parameter space enables us to explore the unknown for potential discoveries that may address some of the most fundamental questions about the universe.

For example, a muon collider, if technologically achievable and affordable, presents a great opportunity to bring a new collider to US soil. A 10 TeV collider fits on the Fermilab site and is a good match with Fermilab's strengths. Its development has synergies with the neutrino program beyond DUNE, and the required upgrades to the accelerator complex would also enable fixed target and beam dump experiments to look for new particles directly or via their quantum imprints.

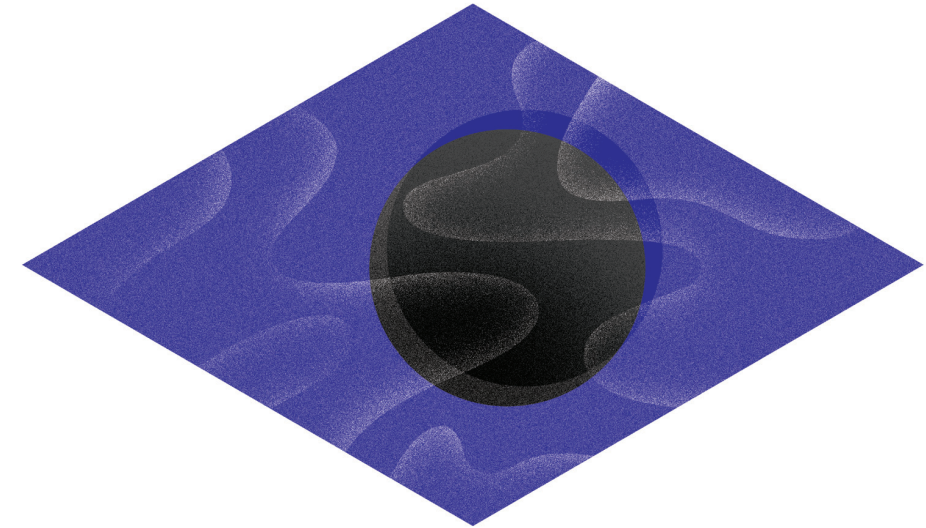

# 5.2

# Pursue Quantum Imprints of New Phenomena

## 5.2.1 — Science Overview

The direct search for new phenomena in particle physics has often been accomplished by going to higher energies with accelerators. This is simply because of Einstein's famous equation $E=mc^2$, which allows for producing heavier particles (large $m$) once there is higher energy (higher $E$). Here, $c$ is the speed of light, a constant of nature that converts the unit of mass to energy. This is how new heavy particles have been discovered, and it clearly demonstrates new phenomena that lead to new theories, and eventually, new paradigms.

However, in the quantum world, another window for discovery is available. Even when new particles are beyond the reach of accelerators, their *quantum imprints* can be searched for. The quantum effects are fuzzy, and just like quantum bits can be 0 and 1 at the same time, particles can exist and not exist at the same time. They are called "virtual" particles. But they can leave clear imprints on the behavior of particles we observe.

There is a long history in particle physics of unexpected discoveries through quantum imprints and their theoretical interpretation. The study of radioactive beta decay led to the prediction of the neutrino and the properties of the W boson; the explanation of matter-antimatter asymmetry in kaons led to the prediction of the third quark family; the observation of neutral current weak interaction events in neutrino scattering resulted in the indirect discovery of the Z boson; measurements of B meson mixing predicted the high mass of the top quark. The characteristics of the observed quantum imprints often led to the definition of the accelerators required to directly produce the relevant new particles successfully.

Before the Higgs particle was discovered by the LHC, its mass had been constrained around the actual measured value from precision measurements of the top quark and the W boson. Measurements of the decays of the Higgs boson already rule out the existence of a fourth generation of particles that follow the same template as the known three.

The physics of flavor is particularly sensitive to quantum imprints of particles that are not present in either the initial or the final state of interactions. The existence of three families of elementary particles of matter and their mass and mixing patterns are a key feature, and a puzzle, in the Standard Model. Rare quark flavor transitions have unparalleled sensitivity to the existence of new physics many orders of magnitude beyond the reach of direct searches in current and planned energy frontier accelerators.

For electrons, muons, and tau leptons, Standard Model mixing is so rare that any observation of a flavor-changing signal called charged lepton flavor violation in any current or planned experiment would be an important discovery and an unambiguous signature of new physics. Collider experiments can search for subtle effects that would be caused by particles with masses much too heavy to be produced directly.

A discovery made by any current or planned experiments would indicate specific directions in which to focus subsequent experiments to directly observe the underlying new physics and potentially suggest the new energy scale to be probed. Progress necessitates clean theoretical predictions and high precision experiments with huge data samples and excellent control of systematic uncertainties. Theoretical and experimental progress go hand in hand, with advances in one side motivating further activity and progress in the other side, in a continuous synergistic interplay.

Currently several intriguing experimental deviations from theoretical Standard Model predictions have been observed. The most significant of these anomalies are related to g-2, the magnetic moment of the muon, which appears larger than anticipated; decays of the bottom quark to a strange quark and a pair of charged leptons, which may be less frequent than expected; and a possibility that bottom quark decays to tau leptons and muons do not have the same strength, pointing to lepton flavor universality violation. All these anomalies might be resolved by improved experimental precision or theoretical predictions. They also might be the first signs of new discoveries.

## 5.2.2 — Ongoing Projects: Mu2e, Belle II, LHCb, ATLAS, and CMS

The largest US experimental efforts currently dedicated to indirect probes of new physics are the Mu2e charged lepton flavor violation experiment, hosted by Fermilab, and the Belle-II and LHCb programs, hosted by KEK in Japan and CERN, respectively. These programs will continue over the time frame of this report with alternating periods of data taking and upgrades (Recommendation 1). The other HL-LHC experiments also present many opportunities for indirect searches, as will experiments at future colliders.

Mu2e, which searches for the conversion of a muon captured by a nucleus into an electron with no emitted neutrinos, will extend our sensitivity to charged lepton flavor violation in this sector by a remarkable four orders of magnitude. Our access to new muon-electron-quark interactions will increase by an order of magnitude in energy scale, up to $10^4$ TeV. Any observed signal at Mu2e would indicate new physics and should be followed up by further experiments for its full characterization.

The Belle-II and LHCb programs focus on decays of bottom and charm quarks and of tau leptons. The clean, controlled environment of electron-positron collisions (for Belle-II) and the extremely high rates and broad-spectrum production in proton-proton collisions (for LHCb) make the two experiments complementary in many ways. These experiments will further probe, with unprecedented precision, Standard Model predictions for quark behavior and lepton physics including leptoquark searches and flavor changing neutral currents, and will conduct wide hidden-sector searches. They will reduce the experimental uncertainties in the measurements in tension with the Standard Model by an order of magnitude or more and introduce qualitatively new tests made possible by the extremely large number of particle decays available for study. The experiments will continue the established program of searches for potential new signs of unexpected matter-antimatter differences, and of measurements that overconstrain and challenge Standard Model parameters to search for inconsistencies between different processes.

The quantum imprints of particles too heavy to be produced in significant rates at the LHC can still result in observable departures from the predictions of the Standard Model. ATLAS and CMS at LHC currently have the unique capability of studying the interactions of directly produced top quarks, Higgs bosons, and W and Z bosons. The data from the ATLAS and CMS experiments, in conjunction with other results, are being used to assemble a comprehensive picture of potential deviations from the Standard Model caused by massive new particles in a largely model-independent manner. The sensitivity of HL-LHC for certain new physics models can reach scales as high as 20 TeV. These measurements complement those at lower energies to provide a comprehensive view of the possible existence of new particles.

## 5.2.3 — New Initiatives: Belle II and LHCb upgrades

The upgraded Belle II experiment will record 25 times more electron-positron collisions by 2035 at the SuperKEKB accelerator at KEK in Japan than Belle did previously. The facility produces world-record luminosity using the most advanced nano-size beams. The unique environment of the SuperKEKB offers access to decays with multiple undetected particles in the final state, such as hadron and tau decays that produce more than one neutrino. The experiment will also further constrain hadronic vacuum polarization, which is important for the precise comparison of theoretical and experimental results on the muon g-2. Quark mixing parameters will be measured with ultimate precision (Recommendation 3c).

The US community has extensive experience with this science at the BaBar experiment at the SLAC National Accelerator Laboratory from the early 2000s, as well as participation in the original Belle experiment, and has a lot to contribute to the operation of the detector and analyses of data.

To achieve such an unprecedentedly powerful beam, there are many challenges in the accelerator. A major technological challenge is producing a nanobeam, namely the ability to focus the beam down to the nanometer scale just before collisions. Another is to maintain a very high degree of vacuum in the beam pipe, as a small amount of residual gas can interact with the beam and cause problems. Finally, the intense beams can also cause a high level of background at the collision point that needs to be mitigated. These technologies are important for all future electron-positron colliders, and therefore it is crucial that the US is involved in their development for the future of the field.

LHCb after its second upgrade will produce huge B hadron samples and will reach unprecedented precision in a large number of observables in the time period from 2035 to 2041, or earlier. The scientists in the LHCb collaboration will explore extremely rare flavor processes, including matter-antimatter asymmetries in the charm sector, and will also search for and study hidden-sector particles, anomalous B meson decays, and more. LHCb upgrade II will be a major project that opens a new era of precision in the rare phenomena explored by the experiment (Recommendation 3c).

The LHCb experiment, together with BaBar, Belle, and Belle II, has also produced a new type of matter composed of quarks. It has been known that most matter around us is either made of three quarks, such as protons and neutrons (collectively called baryons), or of a quark and an antiquark, such as pions and kaons (called mesons). However, these experiments discovered a type of matter made of two quarks and two antiquarks (tetraquarks) or four quarks and an antiquark (pentaquarks). The study of these exotic particles helps us understand the forces that bind quarks together. Such novel states of matter are of strong interest not only in particle physics but also in nuclear physics and might also exist at the interior of very dense stars such as neutron stars.

These experiments also produce large samples of tau particles. In addition to charged lepton flavor violation searches at Mu2e from muon decays, they can search for the same violation in tau decays, a complementary probe to quantum imprints of particles well beyond the reach of current collider experiments.

All experiments mentioned here are major international projects with small but important US contributions and demonstrate the good investment value of international engagements.

## 5.2.4 — Major Initiative: Higgs Factory

One of our recommendations for major initiatives is the US involvement in a Higgs factory (Recommendation 2c). The main purpose of the factory is to reveal the secrets of the Higgs boson (section 3.2). However, the Higgs boson is also a sensitive probe of the quantum imprints of new phenomena. For instance, it is possible that there is more than one type of Higgs boson, and the discovered particle is only the first one in a new family. Even when the additional Higgs bosons are well beyond the reach of HL-LHC, their existence affects the interaction of the first one with various particles. Another possibility is that the Higgs boson is not an elementary particle but is a composite consisting of smaller constituents, and has a non-zero size. Both these cases can be inferred from precise measurements of the Higgs couplings.

The Higgs factory we recommend can be run at the Z pole. Its high luminosity could produce of the order of $10^9$–$10^{12}$ Z bosons and a large sample of WW events. These abilities would enable an exceptional program of precision studies of electroweak interactions, extending the probed energy scale by a factor of 3–10 beyond the HL-LHC. In fact, a similar program was conducted at the Large Electron Positron Collider (LEP) and the Stanford Linear Collider (SLC) in the 1990s that resulted in the prediction of masses of the top quark and the Higgs boson, the exclusion of a fourth family of light neutrinos, and other important results without ever reaching the relevant energy scales. With a much bigger sample of Z and W bosons, we will obtain an unprecedented reach to quantum imprints of new phenomena. A successful Z pole program will involve challenging, high-collision-rate environments that will necessitate advances in accelerator and detector design, as well as imposing computing requirements an order of magnitude beyond those of the HL-LHC.

In addition, the Z bosons would then produce large samples of bottom and charm hadrons and tau leptons in their decays, and at the $10^{12}$ Z boson scale these will become extremely useful for studying their properties. For example, the FCC-ee circular collider is expected to produce a sample of bottom mesons 20 times larger than that of Belle-II, enabling a strong indirect search program that will complement its Higgs boson and electroweak parameter measurements. That search program is unfeasible at LHCb-II or Belle II.

Precision measurement of the top quark mass is an indirect measure of its interaction with the Higgs boson, which controls the quantum mechanical evolution of the Standard Model at high energies; a 350 GeV Higgs factory stage will reduce the uncertainty in this crucial parameter by a factor of 10. Comparing the direct measurements of the top quark and Higgs boson masses at a Higgs factory to the precision measurements of Z and W boson properties can reveal hidden quantum imprints of new particles and phenomena at 10 TeV energy scale.

## 5.2.4 — 20-Year Vision and Future Opportunities

The intense proton beams provided by the Fermilab accelerator complex have the potential to further extend searches for quantum imprints of new phenomena. There is ongoing discussion of an advanced muon facility using the beams of the PIP-II accelerator currently under construction. We encourage further development of this concept, which would increase sensitivity to multiple possible charged lepton flavor violation processes by about a factor of 100 (Recommendation 4e).

The possible future replacement of the Fermilab booster with a new accelerator producing higher intensity proton beams can support a comprehensive facility for experiments on flavor physics, searches for extremely rare phenomena, and CP violation. With careful planning the new accelerator can also become a key enabler for future neutrino experiments and/or a muon collider, in line with the long-term vision of our report.

# 6

Investing in the Future of Science and Technology

Flagship projects often start as ideas on a blackboard. Future innovations require that the entire lifecycle of scientific discovery be supported, from initial concept to construction, operations, and data analysis. R&D is critical at multiple stages along the way.

To promote robust R&D efforts across a range of enabling technologies, we recommend sustained investments in key areas essential to the future of particle physics: theory, an agile project program, detector instrumentation, particle accelerators, collider R&D, facilities and infrastructure, software and computing, and data science. We also discuss the ways that particle physics leads to technology innovations for society and make recommendations for a more sustainable future for the field. Note that collider R&D includes both accelerator and detector R&D targeting specific future colliders.

# 6.1

# Theory

Particle physics theory lies at the heart of our understanding of the universe at both the smallest and largest scales. Theory plays an essential role in guiding which new experiments should be pursued, informing experimental design, interpreting experimental measurements and observations, and exploring uncharted regimes.

Theoretical developments have far-reaching implications and connect particle physics to other areas of science, including nuclear physics; astrophysics; condensed matter physics; and atomic, molecular, and optical physics. In addition, theoretical work closely connects physicists with mathematics and computer science. Theoretical insights also incorporate new perspectives, from quantum information to AI/ML, that propel innovative experimental techniques.

Theorists have opened new avenues of research and proposed new experiments; for example, theoretical investigations into possible hidden sectors led directly to the portfolio of DMNI projects. Theorists support the experimental portfolio of particle physics, providing the predictions necessary to analyze, interpret, and understand experimental results. Examples of theoretical calculations underpinning the extraction of physical quantities from experimental data are neutrino-nucleus interactions for DUNE, particle collisions at the LHC, and the detailed structure of the CMB power spectra for CMB-S4.

Theorists uncover the mathematical patterns that describe the universe and explore alternate mathematical universes to deepen our understanding of nature. Theoretical investigations into quantum gravity and the mysteries of black holes have unlocked connections between extreme space-time geometries and information theory. The perspectives theorists bring to particle physics are important in inspiring young scientists.

Computational physics is a vital component of particle theory. Powerful computers simulate the behavior of fundamental particles of the Standard Model, enabling meaningful comparisons between theoretical expectations and experimental measurements. For example, computational simulations of the strong nuclear force have been critical to precision tests of the Standard Model at the LHC and other colliders; they provide new insight into the origin of matter in the early universe. Theoretical techniques are also used to create "digital twins" of the universe to test different scenarios for cosmic evolution.

Maximizing the physics reach of experiments relies on the collective effort of the particle physics community, encompassing theoretical and computational research. Maintaining a balanced program, with healthy support for theoretical research, will magnify the impact of the entire field and enable future discoveries.

Universities are home to the vast majority of theoretical research and nearly all student training, so it is critical that university support be strengthened to ensure equitable access to particle physics. The level of support for theoretical research has eroded over the past decade. The cuts were primarily absorbed by universities, to the detriment of the field. To catalyze tomorrow's theoretical breakthroughs and to retain the best talent, this trend needs to be reversed. Increasing support (Recommendation 4b) will enable particle physics to thrive in a robust manner, bring diverse perspectives, strengthen talent identification, and create a more inclusive and effective workforce.

**Area Recommendation 1: Increase DOE HEP-funded university-based theory research by $15 million per year in 2023 dollars (or about 30% of the theory program), to propel innovation and ensure international competitiveness. Such an increase would bring theory support back to 2010 levels. Maintain DOE lab-based theory groups as an essential component of the theory community.**

# 6.2

# Advancing Science and Technology through Agile Experiments (ASTAE)

Experiments at multiple scales in cost, size, and duration are critical to unlocking the mysteries of the universe. We recommend DOE create a portfolio of experiments that are small in scale and quick to execute, yet significant in their impact. This complements the Mid-Scale Research Infrastructure (MSRI) program, both within the Division of Physics and across NSF. It enables DOE to strategically target emerging opportunities within its program while supporting the overall DOE mission. The program for these agile DOE experiments is ASTAE, which is Latin for "javelins," reflecting the experiments' focused nature. Across particle physics there are important experiments that meet these criteria, experiments capable of bringing forward discovery science, advancing technology, developing a workforce, and furthering US leadership.

These experiments are key to paradigm-shifting discoveries, both in their own right and as incubators for new technologies and physics directions. The recent DOE-HEP DMNI program exemplifies the opportunity for discovery that small-scale experiments can provide, while incubating key quantum technologies. Similarly, the recent first observation of coherent elastic neutrino-nucleus scattering (CEvNS) demonstrates the potential for discovery science for experiments of this scale. In addition to discovery science, this proposed portfolio can enable small-scale supporting experiments that simplify the interpretation of data from flagship measurements. They can also act as critical demonstrators for innovative instrumentation that will enable the next generation of experiments.

ASTAE will provide an essential training ground for a workforce capable of innovation and leadership to sustain the vitality of the field. These experiments train the full HEP workforce (students, postdoctoral fellows, scientists, engineers, technicians, and project managers) in the complete life-cycle of an experiment: design, construction, commissioning, operations, and the publication of the results. The ASTAE portfolio will provide the hands-on experience needed to conceive the technological breakthroughs that will enable the scientific discoveries of the future and foster scientific ingenuity among the next generation of researchers.

Over time, this portfolio should maintain a balance of experiments across the science driver focus areas. The possibility of projects that push beyond these focus areas should be left open for compelling scientific or technological cases. The projects should be reviewed for their potential for discovery, technology innovation, and ability to provide critical inputs to the success of the greater HEP mission.

**Area Recommendation 2: For the ASTAE program to be agile, we recommend a broad, predictable, recurring, and preferably annual call for proposals. This ensures the flexibility to target emerging opportunities and fields. A program on the scale of $35 million per year in 2023 dollars is needed to ensure a healthy pipeline of projects.**

**Area Recommendation 3: To preserve the agility of the ASTAE program, project management requirements should be outlined for the portfolio and should be adjusted to be commensurate with the scale of the experiment.**

**Area Recommendation 4: A successful ASTAE experiment involves 3 phases: design, construction, and operations. A design phase proposal should precede a construction proposal, and construction proposals are considered from projects within the group that have successfully completed their design phase.**

**Area Recommendation 5: The DMNI projects that have successfully completed their design phase and are ready to be reviewed for construction should form the first set of construction proposals for ASTAE. The corresponding design phase call would be open to proposals from all areas of particle physics.**

# 6.3
# Detector Instrumentation

The field of particle physics is at an exciting juncture where current detector instrumentation technologies have pushed the boundaries of sensitivity and scalability. Whereas modest advancements in some cases can pave the way for the next generation of experiments, numerous research areas eagerly await new approaches, innovative materials, and cutting-edge technologies to overcome existing technological challenges. A vital particle physics program requires a balance of investments in both evolutionary and transformational "blue sky" detector development to achieve paradigm-shifting ideas. Many of the R&D needs for the next decade and beyond are outlined in the report DOE Basic Research Needs for High Energy Physics—Detector Research and Development.These advances can be achieved if we support careers in instrumentation, including research scientists at universities who can support local detector innovation, provide continuity, and educate the next generation of experts (Recommendations 4d and 5d).

Purely instrumentation-based and interdisciplinary graduate programs should be empowered, and the lateral move of scientists between universities and laboratories, including joint positions, should be made easier. Career paths and recognition in detector instrumentation should be extended to all scientists, including chemists, material scientists, engineers, and technicians. Cross-disciplinary opportunities of collaboration and fertilization at universities and laboratories are essential.

The particle physics community has identified the need for stronger coordination between the different groups carrying out detector R&D in the US. We strongly support the R&D collaborations (RDCs) that are being established and will be stewarded by the Coordinating Panel for Advanced Detectors, overseen by the APS Division of Particles and Fields. The RDCs are organized along specific technology directions or common challenges and aim to define and follow roadmaps to achieve specific R&D goals. This coordination will help to achieve a more coherent detector instrumentation program in the US and will help to avoid duplication while addressing common challenges. International collaboration is also crucial, especially in cases where we want to have technological leadership roles. Involvement in the newly established detector R&D groups at CERN is encouraged, as are contributions to the design and planning for the next generation of international or global projects. Targeted future collider detector R&D, such as for Higgs factories or a muon collider, is covered in section 6.5.

To enable groundbreaking detector innovation and US leadership in this field, we need to invest in a coherent set of modernized facilities with enhanced capabilities. These include test beam and irradiation facilities with beam properties and intensities appropriate for future experimental demands, low-background and underground facilities, cleanroom space, access to nano-fabrication facilities, and microelectronics foundries.

To stimulate transformational breakthroughs, we need to pursue synergies with other disciplines outside of particle physics, as well as close collaborations with industry. Recent examples of this are quantum sensors and microelectronics, where communication and exchange between particle physics and other communities led to enhanced funding for technology development and the addition of external experts to the particle physics community. Over the past few years this collaboration has led to the development of a new suite of detector technologies and methods; these in turn have established a new set of small-scale experiments that enhance the particle physics portfolio.

The use of quantum sensors in particle physics experiments has grown extensively, and the developments within particle physics have benefitted communications, computing, and many other areas. Quantum sensors have been used in searches for dark matter, fifth forces, dark photons, permanent electric dipole moment (EDM), variations in fundamental constants, and gravitational waves, among others, and they come in a wide range of technologies: atom interferometers and atomic clocks, magnetometers, quantum calorimeters and superconducting sensors. Quantum sensors are being deployed in all areas of particle physics. To further advance these technologies, we need to continue strong support for a broad range of quantum sensors.

Much of the growth in quantum sensors over the past decade has occurred in small, laboratory-based experiments. Support for such experiments should continue as a way to rapidly develop sensor technologies and help determine the areas where quantum sensors can have the greatest impact. Several recent developments have reached the point where plans for larger-scale, longer-term experiments can and should be conceptualized. These concepts can evaluate the potential reach that can be achieved in a larger effort and the scale of required technological development. Here, potential reach is meant in terms of our physics goals: enhanced sensitivities to the phenomena we are trying to measure compared to traditional non-quantum technologies. We recommend mechanisms to support interactions outside the particle physics program to enable collaborations with experts in other fields who have experience with quantum sensors, including QIS theory.

In microelectronics, particle physics specializes in extreme environments such as high radiation, ultra-high vacuum, and cryogenic operations. These areas have been spearheaded by the particle physics community over the past several years and show benefits to fields outside particle physics, such as nuclear physics, space exploration, nonproliferation, and homeland security.

Related innovations and proof of concepts from within particle physics are benefitting the microelectronics industry, with specialized cryogenic and energy-efficient designs and integration techniques being adopted by industry. In addition, the facilities established at the national laboratories are of growing interest to commercial partners. Close collaboration between particle physics microelectronics experts and CAD tool companies is needed to provide feedback on novel applications of needed advanced tools. Close collaboration between microelectronics teams at the national laboratories and engineers and scientists at universities helps cultivate talent to advance the workforce for the microelectronics industry.

Particle physics benefits from developments in other fields; for example, atomic clocks developed decades ago for precision timing standards are now being proposed for the search for variations of fundamental constants and gravitational waves. We should strive to collaborate with other scientific communities to maximize mutual benefits, such as gaining access to new sensor technologies or providing access to our development of large magnets, vacuum systems, and superconducting radio-frequency cavities for outside fields. To facilitate this exchange, we need to move beyond traditional funding boundaries and allow scientists and engineers working with quantum sensors, whatever their specific field, to tackle the most interesting and challenging problems. Focused support for the development of quantum materials, including theoretical explorations, should be provided.

Workforce development is needed to encourage workers with the needed skills to engage with particle physics and stay in the field for our long-term success in the face of increasing competition from industrial quantum computing. The particle physics community needs to invest now to train and retain the next generation of quantum scientists.

**Area Recommendation 6: Increase the yearly budget for generic Detector R&D by $20 million in 2023 dollars. This should be supplemented by additional funds for the collider R&D program.**

**Area Recommendation 7: The detector R&D program should continue to leverage national initiatives such as QIS, microelectronics, and AI/ML.**

# 6.4

# Particle Accelerators and R&D

Particle accelerators play an essential role in high energy physics research. They deliver the unique beams that enable the majority of the P5 science drivers for the next decade (section 6.4.1). Physics goals beyond this decade place radical new demands on accelerator capabilities. To achieve these demanding performance requirements while reducing costs and minimizing environmental impacts requires focused investment in generic R&D (section 6.4.2) and collider R&D (section 6.5) along with strategic investment in the existing accelerator complex at Fermilab (section 6.6.2). In this context, DOE-HEP, in partnership with NSF and other offices and agencies, has the opportunity to lead accelerator development into the future while simultaneously delivering broader benefits across science and society.

Particle accelerators rely on a broad suite of technologies. This starts with the source that injects particles into the accelerator. When the particles involved are rare or unstable, or the required intensities are very high, these injector systems require significant technology development. At the highest intensities and brightness, techniques to increase the density (in energy and space) of particles in the beams—for example, beam cooling—must often be applied. The particles must be accelerated to the required energy, typically with metallic radio-frequency structures, dielectric structures, or, most recently, plasma-based wakefields. Particle accelerators generally utilize advanced magnets to guide and focus the beams. At the highest energies, manipulating particle beams often requires very-high-field magnets that rely on developing technologies such as high-temperature superconductors. The resulting beams are directed to an experimental target or brought into collisions with another beam inside a collider detector. To manage and understand the behavior of the beams in each of these subsystems, sophisticated beam physics theory and computation, as well as dedicated instrumentation and control systems are required.

High-energy colliders represent perhaps the most formidable and extensive integration of particle accelerator technology. The challenge comes not only in achieving the requisite beam parameters, but in managing the complex interactions between the multiple systems that make up the machine. Incorporating the results of technology R&D to meet these challenges drives the need for sustained investment in the R&D and design of future colliders. Validation of collider designs through simulations, practical tests of individual design elements, and ultimately demonstrator facilities that test integrated segments of the design are essential steps toward transforming the results of R&D into realistic prospects for the future.

It is vital to build on the investments of the past while also developing innovative solutions for the future. Fermilab, which currently supports a strong neutrino and muon flavor physics program with its proton accelerator complex, represents one of those investments. In the near term, upgrades to that complex will allow Fermilab to deliver high-power, high-intensity neutrino beams to DUNE. Developing a path for the evolution of the Fermilab accelerator complex beyond that goal supports a multi-decadal vision for Fermilab as a world-leading facility and lays the foundation for a 20-plus-year vision for particle physics.

## 6.4.1  –  Particle Physics Accelerator Roadmap

Major advances in accelerator design are central to realizing some of our most ambitious scientific goals. The vision laid out in this report foresees decision points on the 5-, 10- and 20-year timescales where accelerator technology choices will need to be made. Informed decisions at each of these junctures will not be possible without a robust and responsive R&D program to deliver crucial but as yet unknown information about the capabilities, cost, and risks of promising technologies. While it will take time to assemble the teams required to inform these decisions, it is imperative that this R&D is pursued aggressively if we hope to act on our most ambitious goal of initiating a 10 TeV pCM collider shortly after the conclusion of the HL-LHC program.

Accelerator design is strongly influenced by the type of particle being accelerated. Electron (and positron) accelerators are constrained by the fact that beams of these very light leptons quickly lose energy to synchrotron radiation, which must be accommodated in the accelerator design. This has limited the energies of practical electron-positron circular machines to a fraction of a TeV. Linear accelerator designs aim towards the TeV beam energy scale or, using wakefield acceleration methods, potentially reach beyond it. Synchrotron radiation is strongly suppressed for heavier particles such as muons and protons, so very high energies can be achieved with circular accelerators. In the case of muons, the short lifetime of the particle means that the acceleration process must occur very rapidly.

Non-collider experiments in domains such as neutrino and flavor physics also benefit from ongoing R&D to deliver beams with increased intensity and specialized beam formatting, while carefully managing the beam physics that drives the onset of instabilities and loss of performance.

We anticipate that a concrete plan to build an offshore Higgs factory will take shape on roughly a five-year timescale. There are reasonably mature linear collider and circular collider designs that can achieve our science goals. These designs are still utilizing R&D to further optimize and reduce risk in the designs. The US accelerator community is well-positioned to engage in this final pre-project stage of R&D to ensure a cost-effective machine that meets all performance requirements.

Beyond the Higgs factory, the physics landscape that has shaped the science drivers points to still higher energy scales, the 10 or more TeV pCM scale. Three technological approaches are under development that have the potential to enable physics exploration at this scale. They are a proton-proton collider based on very-high-field magnets, a muon collider, and possibly a linear collider based on advanced wakefield technology. All three of these technologies have different appealing features and must be developed further. To make a confident, informed decision on the path forward—a decision that we hope to make within the next 20 years—one or more of these technologies must reach technical maturity, allowing us to reliably estimate both cost and technical risk. Reaching that point is only possible with strategic, intensive, and focused development of both the foundational technologies and the resulting collider designs.

The proposal for an affordable proton-proton machine rests on a very plausible extrapolation of the parameters for the proton beams. However, that design requires magnet technology that is currently beyond the state of the art. A multi-decade, international R&D program is essential to produce magnets meeting the necessary specifications. The US is actively engaged in this effort through the US Magnet Development Program.

In the case of the muon collider, concepts and preliminary specifications exist for each of the required subsystems of the collider complex. That complex would be most readily built on an existing proton accelerator complex. Fermilab in the US emerges as a premier candidate for such a facility (section 6.6.2). A muon collider baseline design that will support a detailed cost estimate is in preparation. Significant system R&D—for example, developing the prototype cryomodules that integrate the high-field magnets and radio-frequency accelerating cavities for the muon cooling system—along with engineering design work before the end of the decade would produce a fully costed conceptual design for a demonstrator facility, which a future panel can consider for construction.

Wakefield concepts for a collider are in the early stages of development. A critical next step is the delivery of an end-to-end design concept, including cost scales, with self-consistent parameters throughout. This will provide an important yardstick against which to measure progress along this emerging technology path.

In addition to developing the technologies and expertise to build future colliders,

maintaining US expertise in their operation and optimization is crucial. Engagement with the LHC and its high-luminosity upgrade at the energy frontier, with SuperKEKB for flavor physics, and with the Relativistic Heavy Ion Collider (RHIC) and its successor, the Electron-Ion Collider (EIC), for nuclear physics can provide key pathways to maintaining a vibrant US accelerator and collider workforce.

The US accelerator program stewards not only accelerator technology but also the workforce that adapts that technology to particle physics science goals. The R&D efforts outlined in the following sections will inform our future decisions and should not be deferred. Acting now capitalizes on the current energy and enthusiasm of the community and maintains and motivates the necessary workforce. These are essential to effective US participation in a Higgs factory and they maximize the likelihood of realizing a 10 TeV pCM collider within a realistic timeframe.

## 6.4.2 — Particle Accelerator R&D

Accelerator R&D drives the innovation required to meet the increasing demands of particle physics on accelerator capacity, performance, and cost. These demands span improving the performance of operating accelerators such as the Fermilab Accelerator Complex (section 6.6.2), advancing the technologies to deliver a Higgs factory, and developing the designs and technologies needed for colliders capable of exploring the 10 TeV pCM scale (section 6.5). Development for muon, proton, and advanced accelerators and superconducting magnets will support collider R&D and drive recruiting and workforce development.

Over the last decade, sustained R&D has created capabilities that are driving this decade's research (Recommendations 1a and 1b). The HL-LHC upgrade leveraged generic accelerator R&D to produce Nb3Sn conductor and cable, which were utilized by the targeted LHC Accelerator Research Program (LARP) to deliver high-field magnet prototypes. The Proton Improvement Plan-II (PIP-II) project is in progress to provide high intensity beams to DUNE and other experiments based on state-of-the-art superconducting radio-frequency (RF) cavities.

The US accelerator R&D program has also made substantial progress in the past decade that affects the future design and development of a Higgs factory and of high intensity accelerators for neutrino and flavor physics. For example, in collaboration with international partners, the US program has set new records in normal and superconducting RF gradients, in the context of an ILC module. Tests of integrable optics and optical stochastic cooling demonstrated key steps toward the future of intense beams. High-intensity beam experiments also validated new accelerator target materials.

At the same time, advances in R&D now allow us to consider technology paths that can enable scientific discovery at the 10 TeV pCM scale (Recommendation 4a). New magnetic field records were set toward the needs for FCC-hh and muon colliders. Tests also demonstrated muon cooling and thus enabled a new international design effort in which US participation is essential. High-gradient plasma-wakefield-based advanced accelerators demonstrated advances toward future colliders including 8 GeV energy gain over just 20 cm (approximately a thousandth of the distance conventionally required) and positron acceleration.

Advances in accelerator R&D have a profound effect on both particle physics and a range of other disciplines. Accelerator structures developed for the ILC enable a new generation of light sources important to basic energy science (BES), including the Linac Coherent Light Source-II (LCLS-II) and the high-energy LCLS-II-HE. The Cryomodule Test Facility and associated test stands support both the LCLS-II (BES) and PIP-II (HEP) accelerator projects. Investments in high-field magnets by the DOE Magnet Development Program and NSF's MagLab have advanced the state of the art in conductors and magnet design to the benefit of particle physics, materials science, fusion energy research, and commercial development.

Broad generic R&D with a long-term focus is critical to extending the reach of accelerators to meet future physics needs. Technical breakthroughs are required to enable accelerators to meet the field's science drivers, to push costs lower than estimates based on current technology, and to reduce environmental impact. There are exciting opportunities in the development of (i) new high average power and efficient drivers (RF, lasers, and electron beams); (ii) accelerating structures that can sustain high average power and gradient (metallic, plasma and dielectric); (iii) high-temperature superconducting magnets; and (iv) computing, instrumentation, and controls. Normal conducting radio frequency (RF), superconducting RF, superconducting magnets, targets, and advanced acceleration concepts are essential to develop the next generation of accelerators for particle physics. The normal conducting RF program should incorporate innovative concepts such as cryogenic cool copper and distributed coupling. Accelerator and beam physics research is also critical, including large-scale computation as machines become more complex. Superconducting high field magnet R&D is essential to future proton (FCC-hh) and muon collider options; timely execution of magnet R&D would leverage expertise becoming available with the completion of the HL-LHC Accelerator Upgrade Project.

Stronger investment in accelerator R&D is needed across the program (Recommendation 4c). The DOE HEP GARD program sponsors crucial work at both laboratories and universities across the program areas.

**Area Recommendation 8: Increase annual funding to the General Accelerator R&D program by $10M in 2023 dollars to ensure US leadership in key areas.**

Test facilities are ever more important to develop the advanced technology for future machines (Recommendations 4a and 4c). The need is magnified by the small number of training opportunities on operating machines and the significant timescales and technical demands of the next colliders. Use of the existing test facilities should be maximized. The development of high-intensity beams, supported by proton and superconducting RF accelerators, can be pursued using FAST and IOTA at Fermilab. Key acceleration and beam requirements of a stage for a future collider based on wakefield technology, including energy gain with high brightness beams at high efficiency, can be developed using FACET-II at SLAC, BELLA at Lawrence Berkeley National Laboratory (LBNL), AWA at Argonne National Laboratory (ANL), ATF at Brookhaven National Laboratory (BNL), ZEUS at the University of Michigan under NSF, and other facilities. In addition to demonstrating a single self-contained stage, some of these facilities, particularly BELLA and potentially ZEUS, can demonstrate the next step toward a plasma wakefield collider—operation with two linked stages.

Future test facilities would typically be mid-scale projects. Technical and scientific plans should be developed for test facility projects that could be launched within the next 5–10 years. These could include the second stage cool copper test, which could develop high-gradient normal conducting RF technology. Advanced accelerator test facilities can explore technology and concepts that could significantly reduce cost and risks associated with a 10 TeV pCM collider.

An upgrade for FACET-II $e^+$ is uniquely positioned to enable study of positron acceleration in high-gradient plasmas. New kW-class efficient lasers, and use of their kilohertz repetition rate for active feedback at kBELLA, will advance stage performance and enable beam tests. An AWA upgrade would support GeV advanced structures. These, together with muon collider development, will advance the technology and feed into a future demonstrator facility to make possible a 10 TeV pCM collider (see section 6.5). Many of these projects may be ready for scientific, technical, and cost reviews within the context of the HEP program toward the middle to end of this decade (Recommendation 6).

**Area Recommendation 9: Support generic accelerator R&D with the construction of small-scale test facilities. Initiate construction of larger test facilities based on project review and informed by the collider R&D program.**

Robust growth in the field requires strengthened investment in education, training, and retention to renew the workforce and develop expertise in accelerator disciplines. A strong and creative workforce is necessary to develop and build new accelerators and colliders. Creating such a workforce is driven by state-of-the-art R&D that attracts high-level talent that can execute the machine design, development, and research needed for major new accelerator projects. Such growth is seeded by development of university groups and targeted training, such as the curriculum at the US Particle Accelerator School, structured to provide opportunities to a broad base of talent (Recommendation 5).

NSF support is important to the unique role of universities, including exploration of novel ideas at the forefront of accelerator science, and educating the next generation. NSF has an effective program in plasma-wakefield accelerators, in magnet science, and in focused conventional accelerator centers. Building a general program, potentially modeled on the successful NSF-DOE partnership in plasma science, would be of strong benefit.

Investments in AS&T drive innovation that benefits extending well beyond particle physics. AS&T efforts that directly support particle physics goals also meet critical needs in other offices, agencies, and organizations. For example, major US facilities such as LCLS-II, the future LCLS-II-HE, and the EIC rely on past and ongoing AS&T investments and provide workforce development opportunities that should be encouraged.

More broadly, these investments lead to valuable partnerships with other DOE Science offices, other US agencies such as the Defense Advanced Research Projects Agency (DARPA) and the National Nuclear Security Agency (NNSA), academia, and industry. These partnerships are dynamic drivers of innovation and progress in accelerators that benefit both particle physics and the nation. They help build a national workforce that can develop and execute major accelerator projects (section 6.8).

In the global context, investment in accelerator-related technologies has increased dramatically over the last decade. This reflects a broad consensus about the importance of advanced technology development and scientific advances for the health of societies—from supporting a strong technology workforce to providing critical scientific and technical capabilities. Continued US investment is needed to maintain leadership and a robust workforce. Sustained investment and continued development of strong AS&T partnerships are key to maintaining a US leadership role (Recommendation 4c).

# 6.5

---

# Collider R&D

Targeted collider R&D is required to translate advancements in detector and accelerator technology into the experimental facilities that shape our understanding of the universe. The development of these future colliders requires broad R&D programs (sections 6.3 and 6.4). It also requires increased investment in promising directions for specific future colliders. In the near term, an offshore Higgs factory and the exploration of 10 TeV pCM collider technologies present exciting possibilities, each requiring further development (section 6.4.1). Involvement in these efforts will reinvigorate accelerator physics research in the US and guide the direction of future R&D.

Design choices for the detector and accelerator elements of a collider are inextricably intertwined. In this section, collider R&D therefore refers to both elements.

Targeted R&D investments are crucial for developing comprehensive designs with cost models, guiding technology advancements and collider pathways, establishing advanced performance benchmarks for detectors and accelerators, and training the next generation of experts. This increased investment complements general detector and accelerator R&D (sections 6.3 and 6.4), which focuses on developing the necessary infrastructure and technologies. Such a synergistic approach is essential for positioning the US as a leader in projects outlined in our 20-year vision. The approach includes robust participation in an offshore Higgs factory and a pivotal role in shaping the path towards a future 10 TeV pCM machine, potentially on US soil.

Decisions related to construction of an offshore Higgs factory are anticipated to be made later this decade. The current designs of both FCC-ee and the ILC satisfy our scientific requirements. To secure a prominent role in a future Higgs factory project, the US should actively engage in feasibility and design studies (Recommendation 2c). Engagement with FCC-ee specifically should include design and modeling to advance the feasibility study, as well as R&D on superconducting radio frequency cavities designed for the ring and superconducting magnets designed for the interaction region. These efforts benefit from synergies in workforce development through participation in SuperKEKB and the EIC.

Maintaining engagement with ILC accelerators through the ILC Technology Network can include design updates and cryomodule construction, which will support significant US contributions to potential projects. A global framework for future collider development, such as the ILC International Development Team as implemented by ICFA for the ILC, is relevant for all future colliders.

For Higgs factory detectors, a concerted effort of targeted R&D synchronized with the targeted accelerator R&D program is needed. The US should engage in international design efforts for specific collider detectors. To achieve the scientific goals, several common requirements apply to the detectors of the various collider options, including vertexing, tracking, timing, particle identification, calorimetry, muon detection, and triggering. Central coordination of these requirements is crucial.

Major international decisions on the route to a Higgs factory are anticipated later this decade. Supported by the International Committee for Future Accelerators (ICFA), the Japanese HEP community remains committed to hosting the ILC in Japan as a global project. The FCC-ee feasibility study is scheduled for completion by 2025, followed by an update by the European Strategy Group and a decision by the CERN Council. Once a specific project is deemed feasible and well-defined, the US should focus efforts toward that technology. A separate panel should determine the level and nature of US contribution while maintaining a healthy US onshore program in particle physics (Recommendation 6). In the scenario in which a global consensus to move forward with the Higgs factory is not reached, the next P5 should reevaluate.

Parallel to the R&D for a Higgs factory, the US should pursue a 10 TeV pCM collider. Designs addressing technical and cost feasibility for the future energy frontier at the 10 TeV pCM are required and build on technical progress in accelerator and detector technologies (Recommendation 4a).

End-to-end designs are needed well before a decision can be made on a project in order to understand potential performance parameters and costs. These will guide research priorities, technology development, and demonstrator facilities. Such early designs will also play a critical role in creating and sustaining the expertise to design such machines. Progress on these end-to-end designs should be evaluated (Recommendation 6).

The 10 TeV pCM energy scale and potential performance benefits motivate muon collider development and ongoing work to advance proton and possible advanced wakefield accelerator paths (section 6.4.1). The US should pursue a leading role in the muon collider design effort, in concert with the International Muon Collider Collaboration (IMCC). The US role should include R&D on relevant technologies and preparations for a demonstrator facility. Delivery of a baseline design later this decade is also a crucial milestone. R&D of accelerator technologies (section 6.4)—including higher-field superconducting magnets crucial to both proton (FCC-hh) and muon colliders, and high-temperature superconductors suitable for high field and temperature—are essential to this effort. Similarly, progress in advanced wakefield accelerators motivates efforts to develop a self-consistent design to understand feasibility and costs. Each of these research areas will benefit from international engagement to enable timely progress.

Targeted detector R&D for 10 TeV pCM machines is needed to address challenges specific to these high-energy machines, such as ultrafast timing, radiation hardness, and high rate capabilities. In particular, detector R&D for a muon collider is needed to address challenges related to the unstable nature of muons. Beam-induced backgrounds due to the in-flight muon decays from the beam line can potentially inhibit the detectors' ability to successfully reconstruct collision products. While many aspects of detector design and optimization are common to all future collider detectors, this unique feature requires dedicated study and R&D.

Overall, an iterative co-design process that integrates accelerator, detector, and simulation expertise is crucial for addressing challenges specific to 10 TeV pCM machines and for demonstrating their technical and costing feasibility. R&D efforts in the next five years will inform test facilities as discussed in section 6.4 for the mid- to late-decade time period, and collider design results will set the stage for initiating a demonstrator facility (Recommendation 6) that would feed into future decisions on a potential collider project.

**Area Recommendation 10: To enable targeted R&D before specific collider projects are established in the US, an investment in collider detector R&D of $20M per year and collider accelerator R&D of $35M per year in 2023 dollars is warranted.**

For the targeted detector R&D, we suggest initially allocating 70% of the funds for Higgs factory detector R&D, with about 30% reserved for 10 TeV pCM detector R&D. Once detector R&D for a Higgs factory is funded and coordinated as a project, targeted detector R&D for a 10 TeV pCM machine should be ramped up.

# 6.6

# Facilities and Infrastructure

Experimental particle physics in the US relies on strategic facilities and infrastructure that enable exploration of the unknown. Stewardship of these facilities through maintenance, planning, and development is central to the future efforts recommended in this report. Maintaining existing capability is specifically outside the scope of this report. The sections below describe important enhancements to enable the science in this report.

## 6.6.1 – Role of National Facilities

National laboratories and facilities are central to the major initiatives slated for the next decade, providing scientific and technical expertise, as well as the management, support, and site infrastructure essential to successful projects. Leading these global projects requires seamless collaboration between international and domestic partners, which in turn requires frequent access to the laboratory sites. To meet this demand, national laboratories and funding agencies should pursue streamlined access policies that will enable researchers to work both on-site and remotely. In addition, facilitating procurement processes and ensuring technical support for experimenters are essential to timely project completion and delivery of scientific results. Equally vital is the creation of a safe, inclusive, and welcoming culture within these facilities; diversity and openness are essential for the flourishing of innovation and scientific discovery on a global scale. In this pivotal role, US national laboratories and facilities are poised to drive the future of scientific exploration and collaboration.

**Area Recommendation 11: To successfully deliver major initiatives and leading global projects, we recommend that:**

a. **National laboratories and facilities should work with funding agencies to establish and maintain streamlined access policies enabling efficient remote and on-site collaboration by international and domestic partners.**

b. **National laboratories should prioritize the facilitation of procurement processes and ensure robust technical support for experimenters.**

c. **National laboratories and facilities should prioritize the creation and maintenance of a supportive, inclusive, and welcoming culture.**

## 6.6.2 — Fermilab Accelerator Complex

The Fermilab Accelerator Complex began operations in 1967 and has become the center of accelerator-based particle physics in the US. Over the decades Fermilab has been modernized and upgraded based on necessities, and that continues today. Ongoing major upgrades of the proton complex by the PIP-II project (Recommendation 1c) will enable production of the world's most intense high-energy beam of neutrinos for the DUNE experiment and will provide a variety of beams to other experiments. The ACORN project is modernizing portions of the electronic controls and power systems. Planned improvements to the MIRT will allow timely delivery of 2.1 MW proton beams on target to ensure high-impact initial data sets from the DUNE detectors (Recommendation 2b).

Beyond these upgrades, the Fermilab Accelerator Complex must continue to evolve to support the next generations of experiments based on a high-performance proton complex. Preliminary concepts for the evolution of the accelerator complex have been explored; further development of a plan that accounts for changes in the science landscape and supports a multi-decadal vision for Fermilab is needed. We recommend that Fermilab work with the US and international communities to deliver a compelling long-term plan that identifies required R&D and defines a roadmap for upgrade efforts over the next two decades (Recommendations 4g and 6; section 6.6.2).

**Area Recommendation 12: Form a dedicated task force, to be led by Fermilab with broad community membership. This task force is to be charged with**

**defining a roadmap for upgrade efforts and delivering a strategic 20-year plan for the Fermilab Accelerator Complex within the next five years for consideration (Recommendation 6). Direct task force funding of up to $10M should be provided.**

Optimal performance and reliability of the accelerator complex must be ensured to capitalize on the investments in the accelerators and the experiments, including DUNE. In particular, the reliability of the booster synchrotron was identified as a central factor in maximizing the DUNE physics production, providing partial justification for a replacement. Even in the case of a rapid replacement, the DUNE experiment will operate for several years at up to 2.1 MW, with MIRT, prior to a complete booster replacement. This period will be the most crucial for delivering the physics results. The reliability of the complex in this mode must be critically analyzed; toward that end, Fermilab should devise a modest set of mid-term investments to the booster and other existing infrastructure.

**Area Recommendation 13: Assess the booster synchrotron and related systems for reliability risks through the first decade of DUNE operation, and take measures to preemptively address these risks.**

## 6.6.3 — Sanford Underground Research Facility

The Sanford Underground Research Facility (SURF) is the nation's premier site for particle physics research deep underground, enabling detection of rare interactions that would otherwise be swamped by the far more numerous interactions from particles bombarding Earth's surface from space. Excavation for the Long Baseline Neutrino Facility (LBNF) will provide the shielded space for the DUNE far detectors. Additional expansion of SURF using state and private funds, at a greatly reduced cost due to the mobilization for the DUNE excavation, is currently planned. DOE support for outfitting the resulting cavern(s) into a laboratory space would provide a US home for a G3 WIMP dark matter search (as described in section 4.1), as well as space for other future particle or nuclear physics experiments for neutrinos or dark matter.

**Area Recommendation 14: To provide infrastructure for neutrino and/or dark matter experiments, we recommend DOE fund the cavern outfitting of the SURF expansion.**

## 6.6.4 — South Pole Station

The South Pole, the site of the NSF-operated Amundsen-Scott South Pole Station, is a unique site that enables groundbreaking scientific discoveries in particle physics and astrophysics. The world-leading infrastructure and logistics capabilities of the US Antarctic Program and South Pole Station in particular are a unique national resource that enables US scientific leadership. We commend the NSF Office of Polar Programs (OPP) for their leadership and vision in constructing and maintaining this facility, which is a monumental effort in such a remote location and a harsh environment. In the next decade, the South Pole can continue to be a place of US scientific leadership, pushing the boundaries of our understanding of fundamental physics in a way that can only be done at this site.

US-led experiments sited at the South Pole—the South Pole Telescope, the BICEP suite of telescopes, and the IceCube Observatory—have produced important discoveries in fundamental physics. For experiments that observe the CMB, the South Pole provides a high, dry, stable atmosphere with continuous access year-round for deep observations of the same low-foreground patch of sky. These observing conditions are particularly important for measuring the signal from inflation in the early universe that is imprinted on the CMB. For IceCube, the South Pole provides an unmatched volume of deep, clear glacial ice that allows for precision observations of optical signals made by astrophysical neutrino interactions in the ice.

The significant advancements in our understanding of inflation and the early universe by CMB-S4 and the wide range of exciting science enabled by neutrino astrophysics in IceCube-Gen2 will be made possible through continued NSF investment in infrastructure at the South Pole. NSF has begun a South Pole Master Planning Process to create a plan for maintaining critical infrastructure and capabilities with a 30- to 500-year horizon. Beyond maintenance, further expansion of the capacity to support projects at the South Pole would allow for more future scientific opportunities. Renewable energy infrastructure at the South Pole is an exciting opportunity to streamline and expand the capabilities for supporting science.

**Area Recommendation 15: Maintaining the capabilities of NSF's infrastructure at the South Pole, focused on enabling future world-leading scientific discoveries, is essential. We recommend continued and critically important direct coordination and planning between NSF-OPP and the CMB-S4 and IceCube-Gen2 projects.**

# 6.7

# Software, Computing, and Data Science

Software and computing play a critical role in maximizing the science output of particle physics experiments. They are an integral component of experimental design, trigger and data acquisition, simulation, reconstruction, and data analysis. They also underlie simulation and design of particle accelerators. The complexity, computational needs and data volumes of particle physics experiments are expected to increase dramatically in the next decade or so. Advances in software and computing, including AI/ML, will be key for solving the challenges associated with the data deluge and for enhancing the sensitivity of the experimental results.

## 6.7.1 – Software, Computing, and Cyberinfrastructure

A hallmark of particle physics is scientific instruments that produce extremely large datasets. For example, at the ATLAS and CMS experiments at the LHC, sophisticated real-time algorithms implemented in hardware and software discard all but the most interesting 0.01% of collisions, with a final recorded output that still consists of tens of petabytes of raw data per year. These datasets are then processed to translate detector signals into a coherent picture of the particles created in the collisions.

Collecting, managing, and analyzing these datasets is a key challenge and represents a significant portion of the total operations cost for many projects. At the same time, simulating the quantum mechanics of particle interactions and the corresponding detector signals at the necessary level of precision to interpret the data poses processing and storage challenges of equivalent or larger complexity to analyzing raw data. Software, computing, and the overarching cyberinfrastructure are essential parts of project portfolios.

The enormous computational demands of particle physics call for expanded investments in R&D to maximize the physics reach of the programs. These R&D efforts have been synergistic with national initiatives in AI/ML, quantum computing, and microelectronics. Funding agencies should continue to leverage the resources made available through those national initiatives.

**Area Recommendation 16: Resources for national initiatives in AI/ML, quantum computing, and microelectronics should be leveraged and incorporated into research and R&D efforts to maximize the physics reach of the program.**

Constantly emerging advances in computing can reduce costs and expand capabilities that enable new science. These advances are powered by new computing architectures, software paradigms, network capabilities, and access to expanded computing resources. Leveraging these advances, however, requires significant effort.

Based on experience from the past decade, we recommend adding support for a sustained R&D effort—at the level of $9M per year in 2023 dollars—to adapt to emerging hardware and other computing technologies. This should include efforts to transition the products of software R&D into production. We also suggest that DOE's Office of High Energy Physics and NSF's Directorates for Mathematical and Physical Sciences and Astronomy coordinate with other programs within the Office of Science and NSF to ensure that the profile of computing resources available matches the needs of particle physics experiments.

**Area Recommendation 17: Add yearly support by $9M per year in 2023 dollars for a sustained R&D effort to adapt software and computing systems to emerging hardware, incorporate other advances in computing technologies,**

and fund directed efforts to transition those developments into systems used for operations of experiments and facilities.

The potential of the quantum computing paradigm and QIS has given rise to a National Quantum Initiative. The initiative encourages transformative scientific discoveries through investments in core QIS research programs. The national strategy includes investments that target the discovery of quantum applications and foster quantum-relevant skills in the next generation of scientists and engineers. The quantum nature of particle physics phenomena provides a rich set of problems in which quantum computing is expected to have an inherent advantage over classical computing. For example, quantum computing approaches can resolve fundamental challenges to studying the structure of nuclear matter that are encountered with the classical computing approaches.

To adapt to the quickly changing landscape, we support the creation of a group focused on software and computing, modeled after the Coordinating Panel for Advanced Detectors (CPAD). Such a group would serve as a valuable resource for DOE and NSF to consult in order to ensure a coordinated strategy for computing that maximizes the impact of investments in computational infrastructure.

We must ensure sustained development, maintenance, and user support for key cyberinfrastructure components, including widely used software packages, simulation tools, and information resources, such as the Particle Data Group and INSPIRE. Although most of these shared cyberinfrastructure components are not specifically tied to projects, nearly all scientists in the field rely on them. A significant investment—at the level of $4M per year in 2023 dollars—for this type of shared cyberinfrastructure with dedicated personnel is appropriate.

As the nation and the world increasingly embrace open science, now is the time for a paradigm shift in the long-term preservation, dissemination, and analysis of the unique data collected by various experiments and surveys in order to realize their full scientific impact. An investment at the level of $4M per year in 2023 dollars is needed to establish these new forms of shared infrastructure and the dedicated personnel. The infrastructure should support the requisite theoretical inputs and computational requirements for analysis.

**Area Recommendation 18: Increase targeted investments that ensure sustained support for key cyberinfrastructure components by $8M per year in 2023 dollars. This includes widely-used software packages, simulation tools, information resources such as the Particle Data Group and INSPIRE, as well as the shared infrastructure for preservation, dissemination, and analysis of the unique data collected by various experiments and surveys in order to realize their full scientific impact.**

Research scientists and research software engineers at universities and labs are key to realizing the vision of the field. They possess highly specialized knowledge and are critical for maintaining a technologically advanced workforce. Strong investments in career development, including recruiting, training, and retention, will ensure future success.

**Area Recommendation 19: Research software engineers and other professionals at universities and labs are key to realizing the vision of the field and are critical for maintaining a technologically advanced workforce. We recommend that the funding agencies embrace these roles as a critical component of the workforce when investing in software, computing, and cyberinfrastructure.**

## 6.7.2 – Artificial Intelligence, Machine Learning, and Data Science

The particle physics community benefits from and contributes to the rapid advances in artificial intelligence (AI) and machine learning (ML). Our field has used various forms of ML for decades to enhance the sensitivity of experiments through more efficient accelerator control, data collection, and data analysis. The rise of deep learning has dramatically expanded the potential for these approaches by bringing qualitatively new capabilities, including the ability to work with low-level sensor data, to generate synthetic data, and to identify anomalies in data.

These capabilities can have a comparable impact on the physics reach of experiments as do enhancements to accelerator facilities, instrumentation, and detector design. Performing AI inference in reprogrammable hardware, closer to the detectors in the readout path, will unlock new capabilities in data collection. Emerging techniques in generative artificial intelligence are leading to productivity enhancements, particularly in writing code for software; these technologies may dramatically alter the way physicists develop code to analyze data or accelerate the process of porting code to new computing architectures. Advances are also being made in formal mathematics and mathematical physics where generative AI can serve as a creative assistant or be paired with a verification system.

The transformative potential of AI should drive a robust investment targeted at individual projects and cross-cutting capacity to accelerate technology transfer. Resources for the national initiatives in AI should be incorporated into research and R&D efforts to maximize the physics reach of the program.

Specific skills are needed to deploy AI and advanced statistics techniques to datasets at the scale found in particle physics, both in real time during data collection and during the subsequent data analysis. Because particle physicists must understand analysis techniques, data management, and software and computing infrastructure, the field of particle physics has proven to be an excellent training ground for data scientists. By aligning the tools and techniques used in particle physics with those used in industry, the field can more quickly benefit from advances driven by industrial data science applications and ensure that the training experiences translate outside of particle physics.

# 6.8

# Technology Innovations and Impact on Society

Particle physicists have a long history of recognizing, embracing, and fostering emerging technologies. The devices and tools that particle physics develops enable applications beyond fundamental physics in fields as different as medicine and aerospace. This section outlines the many ways particle physics spurs innovation, with example impacts.

## 6.8.1 — Computing

Particle physicists are the stewards of some of the world's largest datasets and therefore have a special connection to computing and data applications. Particle physicists were among the first to transition from analog to digitized data systems and were early adopters of the use of machine learning algorithms. The world wide web and the web browser were invented at CERN as an efficient information sharing system for particle physicists. To cope with enormous datasets, particle physicists played leading roles in the development of global-scale distributed computing technologies and high-performance networking. The ambitious portfolio of projects proposed for the next 10 years will drive innovation that addresses the scientific needs of the particle physics community and will broadly impact society.

## 6.8.2 — Artificial Intelligence and Machine Learning

The unique challenges encountered in particle physics have proven to be fertile ground for innovation of AI/ML. Importing new techniques into the particle physics context requires a process of translation, through which those techniques are stress-tested, generalized, and improved. Those improvements feed back into the wider AI/ML research community, completing a cycle of use-inspired research.

For example, when particle physicists began incorporating advances in deep learning for images, it revealed that these techniques were not well suited to nonuniform detector geometries or sparse data with high dynamic range. Coping with these unique challenges led to rapid innovations in graph neural networks and geometric deep learning. These developments were effective at building cross-disciplinary collaborations including universities, national labs, and the private sector and led to innovations that have been incorporated into the mainstream of AI/ML research.

We are witnessing an unprecedented level of cross-pollination among scientific disciplines as AI/ML has become a common language and vehicle to transfer innovations. AI-enhanced statistical inference techniques developed for particle physics are now being used in fields as disparate as materials science, neuroscience, systems biology, epidemiology, and genetics. Similarly, AI-enhanced Monte Carlo sampling techniques developed for lattice QCD are co-evolving with the techniques used to study molecular dynamics and gravitational waves. In addition, many of the technical developments at the core of modern machine learning frameworks, such as automatic differentiation, are being incorporated into traditional scientific computing. These advances have involved partnerships among scientists working closely with AI researchers and applied mathematicians. Furthermore, they impact a wide range of industrial applications.

## 6.8.3 — Quantum Information Science

Particle physics is intrinsically rooted in the realm of quantum mechanics. This unique intersection of particle physics and quantum science positions the field as a vital contributor to the National Quantum Initiative. Particle physics is deeply involved in all five National Quantum Information Science Centers and actively engages in various activities sponsored by DOE and NSF.

The precision and sensitivity of particle detectors directly hinge on the capabilities of the sensors and the noise levels in the associated electronics. To meet the stringent demands of the field, particle physics has long used cutting-edge quantum sensing techniques and ultra-low-noise electronics to meet the stringent requirements of the science. Hence particle physics is an ideal testing ground for these technologies. Several of the DMNI experiments are perfect examples of this paradigm. They are pushing beyond the standard quantum limit for electronic noise and scale the number of channels deployed.

A strong connection exists between quantum information science and particle theory. A wide variety of problems in high energy physics cannot be addressed using classical computation. There is a strong effort to explore the potential of quantum simulation to render complex and previously insurmountable problems into manageable ones. There is significant effort to use the profound connections between quantum information theory and quantum gravity to form a better understanding of these complex fundamental problems. In addition, next-generation experiments are beginning to explore technologically and theoretically what could be measured by actively preparing entangled quantum systems.

## 6.8.4 — Microelectronics

The particle physics community has developed application-specific integrated circuits (ASICs) to cope with extreme environments, such as high levels of ionizing radiation or cryogenic temperatures. These developments are benefitting areas outside of particle physics, such as space exploration and defense needs. Specialized cryogenic and energy efficient microelectronics designs and integration techniques are now adopted by the microelectronics industry. In addition, the microelectronics facilities established at the national laboratories are of growing interest to commercial partners.

## 6.8.5 — Detectors and Instrumentation

Detectors for particle physics have found their way into many applications that benefit society. One of the most profound is in the area of medical imaging. Detectors developed for vertexing, tracking, and photon detection in particle physics can also be used to minimize exposure times for patients. Over time, technology advancements in particle detectors have lowered detection thresholds and have thus allowed reduction of the dose needed for medical imaging applications. Single photon detectors that were developed for low-mass dark matter searches, such as Skipper-CCDs, are being used in nuclear non-proliferation applications. Silicon-based tracking detectors have also found application in muon tomography of ancient structures like pyramids, enabling scans for hidden caverns. Development in tracking detectors of cosmic-ray muons made the recent archeological discovery of hidden chambers in the Great Pyramid of Giza possible. In addition to these examples, the development and construction of advanced detectors and instrumentation is critical for training the technical and scientific workforce. Many detector physicists move into careers that focus on the application of these technologies.

## 6.8.6 — Accelerators

Particle accelerators exemplify how development of frontier machines for high-energy physics enables broad science and societal applications, including x-ray light sources, material science, medical therapy, and particle sources for national security and industry. As an example of such broad science, accelerator structures developed for the ILC are enabling a new generation of light sources including LCLS-II and LCLS-II-HE. Light sources and other new capabilities have been made possible through decades of development of transformational accelerator techniques by the particle physics community.

Accelerator technology development enables further applications. Magnets have benefitted applications including medicine and fusion development. Similarly, lasers have greater impacts including advanced manufacturing, medicine and security applications. Future advances, such as those needed to reach higher beam powers and energies, will inspire yet more research on very-high-field magnets, extraordinarily intense beams, and accelerator technologies with the potential to be radically more compact (Recommendations 4a and 4c; sections 2.3.2, 6.3, 6.4, and 6.5). In turn, these developments may lead to improved efficacy in cancer treatment, more compact fusion devices, and compact sources of particles and radiation for industry and security, both benefiting from and re-inforcing particle physics investment.

# 6.9

# Sustainability and the Environment

Commitment to sustainability is a high priority for particle physics activities. This includes energy and carbon management, energy efficiency and savings, and environmental impact. It concerns present and future accelerators as well as testing and computing facilities.

In the study of future accelerator projects, it is important to establish and launch at an early stage a full lifecycle sustainability effort. Because future accelerators, conceived for higher beam energies and intensities, will have higher energy demands, the promotion of energy efficient accelerator concepts and identification and development of energy-saving accelerator technology are critical. These considerations will affect the affordability of new accelerators and demonstrate the responsible role of the HEP community in society.

Accelerator technologies play a key role in sustainability. Investments in high field magnets by the DOE Magnet Development Program and NSF's MagLab have advanced the state of the art superconductors and magnet design to the benefit of particle physics, but also materials science, fusion energy research, and commercial development. In this context, high temperature superconducting materials, operated at temperatures higher than superfluid or liquid helium, can be significant in reducing energy consumption—for example, for future colliders including muons or FCC-hh. Accelerator structure improvements could also be significant, including higher quality factor cavities, and concepts like the Cool Copper Collider.

Other new technologies such as muon and possibly wakefield colliders have potential to improve power required for a given luminosity and reduce the size and environmental impact of future facilities. Innovation in electric power generation, management, and distribution also contribute to sustainable development and will be encouraged.

Upgrade, construction, and operation of accelerator complex (section 4.2) and test facilities are part of the global effort to advance particle physics. Defining sustainable requirements on industrially procured technology, including construction, electrical, and cooling equipment, should be included in the project development.

To contribute to the global decarbonization effort, future projects will aim to reduce the $CO_2$ footprint of their civil engineering components. Research and adoption of sustainable and eco-friendly materials will be encouraged, and this applies to gases with high global warming impact used in some types of detectors. Study of alternative solutions will be investigated and implemented, in addition to the adoption of stringent recirculation requirements when those gases are used. Reuse of materials should be encouraged to foster sustainable development and to limit the use of natural resources and raw materials, including critical minerals that are subject to potential shortages.

The international nature of particle physics activities calls for extensive travel to participate in meetings and conferences and to carry out experiments. The dramatic increase in the use of remote conferencing mandated by the COVID-19 pandemic spurred an evolution of practices and technologies that allow researchers to increase participation and inclusivity. With this new paradigm in place, laboratories should facilitate hybrid meetings and invest in upgrades to video conferencing equipment as necessary. Meeting organizers should establish protocols that ensure people connecting from remote locations can participate fully, including in discussions. At the same time, in-person meetings remain strongly valuable and should not be discouraged; they facilitate open and effective communication and build trust in international partnership. Assessing and implementing sustainability strategies require R&D and investment of appropriate resources. The field would benefit from development of consistent metrics for sustainability of research, construction, and operations.

**Area Recommendation 20: Charge HEPAP, potentially in collaboration with international partners, to conduct a dedicated study aiming at developing a sustainability strategy for particle physics.**

# 7

## A Technologically Advanced Workforce for Particle Physics and the Nation

Addressing the profound scientific inquiries within particle physics, from understanding the fundamental building blocks of nature to mapping out the evolution of the universe, requires a creative and technologically advanced workforce operating in an environment of mutual trust. The inherent curiosity driving our exploration of the natural world is a universal aspect of human nature. This shared curiosity serves as the driving force behind our commitment to strengthening and expanding this workforce and prompts us to actively seek talent from all corners of society and all regions of the country, and the world.

The stewardship of our field demands active engagement with diverse perspectives that provide innovative strategies and solutions to complex problems. Achieving increased participation and diversity of perspective entails implementing practices that support ethical conduct of research, dismantling existing barriers, recruiting broadly, and forging new pathways of opportunity to ensure that particle physics thrives as an inclusive and dynamic discipline.

Moreover, the design, construction, operation, and analysis work of high energy physics not only advances our scientific endeavors but also trains the researchers, engineers, and technicians urgently needed in mission-critical areas for the US. These areas include hardware and chip development, large-scale computing, QIS, and AI/ML. It is imperative that we provide exciting career pathways within our field to retain the necessary talent. Simultaneously, we must continue to educate the many technologically adept researchers and technical personnel who robustly contribute to society outside particle physics. Additionally, engaging with the public by sharing our scientific discoveries and underscoring how our work contributes to industries beyond particle physics further enhances the impact and relevance of our pursuit.

# 7.1

# Ethical Research in Particle Physics

Science is built on trust. When we work ethically, we maintain confidence in our scientific results within our community and with the public. Trust within our community is developed through transparency and an open, constructive exchange of views. The two key principles for ethical behavior are telling the truth and treating others with respect. Telling the truth means our science is robust and transparent: our data is presented accurately, we present original work, we make our data publicly available whenever possible, and we fairly allocate credit. Treating others with respect requires maintaining a professional work environment, free from harassment and abuse. Discrimination, harassment, or bullying in a scientific collaboration harms individuals, disrupts scientific progress, and is therefore scientific misconduct.

We must collaborate to do our science, which by necessity brings together people at various career stages and from a diverse set of backgrounds. Within these collaborations, ethical standards must be stated clearly and upheld in order for our community members and our science to thrive. Establishing and maintaining these standards includes setting professional expectations, reporting violations, and having a process to rectify violations. Such infrastructure should include ombudspersons, independent investigators in cases of egregious alleged violations, training for those in leadership positions, and training to encourage collaborators to create an environment in which people can thrive and scientific output can reach its potential. Laboratories and funding agencies are charged to support such infrastructure (Recommendation 5a).

# 7.2

# Recruiting, Training, and Retaining

Particle accelerators and detectors are among the most complex instruments ever created by human ingenuity. Electronics, cryogenics, magnets, vacuum systems, and a complex array of other hardware must work together to deliver signals that probe the fundamental nature of the universe. These instruments, boasting billions of electronic channels and systems sensitive to rare processes that operate at the quantum noise threshold, create intricate datasets. Harnessing this wealth of data requires advanced computing techniques and mathematics, sophisticated statistical tools, and the transformative power of AI/ML. At the core of this endeavor lies a critical component: an exceptionally talented workforce.

To nurture this workforce, it is imperative to recruit individuals with a broad range of backgrounds and provide them with the skills and resources needed to advance the field. To retain this workforce we must provide career pathways and an environment that encourages and rewards exemplary work.

Some particle physics-trained researchers will chart career paths in other areas of academia, and in government and industry. Their multidimensional and technical expertise in areas such as accelerators, quantum sensors, and AI/ML, will make them sought after for jobs in quantum computing, medical physics, and other data-rich environments. Their particle physics training will equip them to drive innovation and will have a profound impact on the nation's scientific and technology landscape, as well as society at large.

*Recruiting:* We should draw research and technical personnel from both a national and international talent pool. Recruitment should address historic and ongoing patterns of underrepresentation in particle physics by engaging broadly in educational efforts and dissemination of scientific findings to the public. Pathways to cutting-edge research at the frontiers of our field should be expanded to reach a broader cross section of students. Strategic academic partnerships connecting different types of academic and research institutions are important to this effort, and continued support is recommended (Recommendation 5b). There should be a concerted effort to include institutions which reach historically underrepresented groups such as minority-serving and rural institutions.

*Training:* Given the technologically advanced skill sets needed to ensure a robust future for particle physics, we must have strong programs that pass knowledge from experienced experts to new participants in the field. Imparting knowledge to the next generation is crucial to keeping our field at the forefront of AI/ML, quantum sensors, accelerators, and a host of other technologies needed for current and future research. Programs like the US Particle Accelerator School and the Theoretical Advanced Study Institute in Elementary Particle Physics (TASI) are essential ingredients in equipping the next generation with key skills and require sustained support (Recommendation 5b).

*Retaining:* People are our most precious resource. Scientific advances rely on the co-ordinated efforts of experimental and theoretical physicists, technicians, engineers, and administrative support staff. An infrastructure that facilitates individual participation in these efforts will allow us to meet the community's ethical standards, ensure good stewardship of human resources, and help retain talent. This infrastructure must ensure accessibility, including living wages and sufficient support for those with family and caregiver responsibilities (Recommendation 5b). Studies of the work climate of the field, informed by experts in sociology and organizational psychology, can allow us to identify barriers to participation and gauge our progress (Recommendation 5c).

In addition to creating an environment conducive to good science, we must maintain a critical mass of technical expertise. Researchers applying for funding should be required to have mentoring plans in place for all group members, with an emphasis on early career researchers and those without permanent positions. Strategic support for research scientists, engineers, and other technical experts at universities and national labs also plays a key role by providing alternative pathways to retain key expertise (Recommendation 5d).

Requirements from funding agencies via proposal guidelines, project evaluations, and reviews of operating experiments, are critical to ensuring that the particle physics community will successfully recruit, train, and retain a technologically advanced workforce with the diverse talents to meet our scientific needs. NSF and DOE have provided significant leadership, accelerated in recent years, that includes funding opportunities that expand the number of institutions and individuals engaging in particle physics. The agencies have also strengthened the requirements surrounding mentoring plans for grant-funded personnel, increased the focus on dissemination of results, and paid closer attention to improving workplace climate in the national laboratories and experimental collaborations. Efforts on these topics are rapidly evolving, and a systematic review of the existing framework, with an eye toward additional opportunities, would advance the development of needed expertise in particle physics.

# 7.3

# Engaging With the Public

The dissemination of results to the public is a crucial ingredient of successful scientific projects. The big questions—uncovering the dynamics of the early universe and investigating the nature of matter, energy, space, and time—can inspire the public and contribute to long-term workforce development by nurturing interest and attracting future STEM talent. In addition to physics results, there are exciting stories to tell about how large international teams work together to invent new technologies, solve hard problems, and achieve common goals.

In many cases, technological advances in particle physics lead to progress in industries beyond scientific research. The public, which funds the science, deserves to learn about the discoveries of particle physics and understand the broader impact the field has on society. Dissemination of that information, in turn, is critical in engaging the workforce needed and enabling future science discoveries and societal benefits.

It is the duty of the particle physics community to recognize the value of public engagement. Opportunities to learn science communication skills are available through the funding agencies and professional societies, but they should be more broadly advertised. Additional recognition of the value of public engagement is essential to building the necessary culture. A plan for dissemination of scientific results to the public should be included in the proposed operations and research budgets of projects and experiments, and such a plan should be supported by the agencies (Recommendation 5e).

"Science is built on trust. When we work ethically, we maintain confidence in our scientific results within our community and with the public."

# 8

## Budgetary Considerations

A clear challenge to this P5 was to realize our vision within the provided budget scenarios. We carried this out with clear principles of prioritization. Below are the rationales for the choices we made for the baseline scenario and for the less favorable scenario.

# 8.1
# Prioritization Principles

In the process of prioritization, we considered scientific opportunities, budgetary realism, and a balanced portfolio as major decision drivers.

Clear principles for the prioritization of scientific opportunities guided the deliberations of P5, as outlined in section 1.6. In addition, depending on project scale, our minimum requirements were:

Large projects (> $250M)
- Paradigm-changing discovery potential
- World-leading
- Unique

Medium projects ($50M–$250M)
- Excellent discovery potential or development of major tools
- World-class
- Competitive

Small projects (< $50M)
- Discovery potential, well-defined measurements, or outstanding technology development
- World-class
- Excellent training grounds

To provide us with a fair and dependable basis for evaluating big projects, we set up a subcommittee with project experts from the physics community, which reviewed the maturity of cost estimates, the risk analysis and mitigation plans, and the proposed schedule. The subcommittee provided an expected cost range for each project and comments on project risks, budget maturity, and schedule. We also considered the uncertainties in the costs, risks, and schedule as part of our prioritization exercise. The prioritized project portfolios were chosen to fit within a few percent of the budget scenarios and to ensure a reasonable outlook for continuation into the second decade, even though that is beyond the purview of this panel.

Finally, we paid careful attention to the balance in the program in terms of:

*Size and timescale of projects.* Large projects push the boundaries of our capabilities and bring together broad communities of people. Small projects operate on shorter time scales, which provide crucial training and leadership opportunities for young scientists and have the potential for transformative breakthroughs.

*Onshore vs. offshore.* Onshore projects can leverage the unique capabilities of US facilities and ensure our leadership. The global nature of our field necessarily means that we would like to participate in offshore projects for important physics experiments, with the benefits of lower cost and cooperation with international partners. Yet there are logistical and financial issues associated with the need to travel abroad frequently.

*Project vs. research.* Projects are meant to produce cutting-edge data, which need to be analyzed and interpreted by activities in research. However, research is where new ideas and concepts are conceived of in order to guide the design of projects. In addition, research supports people, namely graduate students, postdocs, and scientists, and therefore requires an adequate level of funding to ensure the future health of the field.

*Current vs. future investment.* The field relies on current projects to produce science, train people, and advance the field. Without careful investment in the future, our capabilities will diminish and the US will lose its global leadership. Future investments include technology, such as accelerators, detectors, and computing, as well as the theory that guides the field. The portfolio, like all investments, requires regular rebalancing.

# 8.2
# Hard Choices for Baseline Budget Scenario

The overall cost of the projects proposed during the Snowmass community study exceeded the budget scenario by many times. As a result, we had to make tough choices based on the principles stated above. We carefully selected projects that have the biggest scientific impact while maintaining the balance among science areas and satisfying the budgetary constraints, resulting in an exciting program for the next ten years. The baseline budget scenario assumes an initial bump in the budget thanks to the CHIPS and Science Act, and an annual 3% increase to keep pace with inflation.

The following project list reflects the sacrifices made in P5's prioritization process, starting with the largest and moving down to the funding level of approximately $100M.

*Onshore Higgs factory.* We could not identify room in the budget executable in the next 20 years for an onshore Higgs factory unless the overall budget is several times higher. On the other hand, there is an ongoing process in Europe to see if FCC-ee is feasible. The Japanese HEP community has been making an effort to realize ILC as a global project hosted in Japan. We therefore recommend exploring offshore options and vigorously pursuing international collaborations so the US can play a major role when one of those projects becomes reality. If FCC-ee and ILC are judged to be not feasible, a new panel should revisit the possibility of bidding to host a Higgs factory in the US, potentially as a global project and including advanced technology options.

*Fermilab Accelerator Complex Extension Booster Replacement (ACE-BR).* Extending the capability of the accelerator complex is important to secure the future of particle physics. Unfortunately, the cost of the proposed options was prohibitive for the given time period. Additionally, it is not yet clear how the extension will enhance Fermilab's capability to support future high-energy colliders, which are key to this report's long-term vision. We therefore recommend MIRT, a less expensive option that achieves 2.1 MW of beam power (instead of 2.4 MW) earlier than envisioned by ACE-BR, augmented by a reliability upgrade (section 6.6.2) of the current booster to provide the required beam for DUNE.

*DUNE FD4.* As described in sections 3.1.3 and 3.1.4, the re-envisioned plan for DUNE Phase 2 enables DUNE's science goals to be achieved with only three LAr far detector modules in the 2040s time frame. It also provides an opportunity for the FD4 to host a different type of detector or enhanced LAr detector to expand the science program of DUNE, leveraging the deep underground site and powerful beam. This budget scenario can accommodate only R&D for FD4 during this period.

*One of two G3 experiments for direct detection of WIMP dark matter.* There were two major proposals for dark matter experiments with sensitivity to reach the neutrino fog, envisioned to be the ultimate sensitivity achievable with the current technology. Even though the two proposals are attractive, we cannot accommodate both in this budget scenario. We recommend that the US bids to host one G3 experiment at a new cavity at SURF. If the bid fails, the US should become a leading partner in one outside the US. If the funding situation is better than in the baseline scenario, or if NSF or other partners can make a substantial contribution, or if an experiment shows a hint of a detection, a second experiment may be considered, possibly outside the US.

*Spec-S5.* This project follows the current world-leading project DESI and the follow-up DESI-II program to significantly enhance the sensitivity to the physics behind dark energy, dark matter, and inflation. We endorse the scientific objectives of the proposal, but it, including the cost estimate, was not sufficiently mature to recommend it. Instead, we recommend support for R&D on smaller fiber positioners to take spectra of a much larger number of objects simultaneously and to define the survey concept and required telescopes. In a more favorable budget scenario, the project may be considered once its scope is well-defined and technical feasibility has been demonstrated.

*Mu2e-II.* Mu2e extends the discovery reach for charged lepton flavor violation by four orders of magnitude, while Mu2e-II is proposed to further extend it by another order of magnitude. Mu2e, once in operation, will take data until 2034. Considering budget reality, scientific case, and the need for informed design choices, we recommend R&D and await the outcome and performance of Mu2e before starting to fund Mu2e-II.

*srEDM.* Electric dipole moment experiments are powerful searches for quantum imprints at higher energy scales. A rich program of EDM searches exists within other DOE programs. We judge that the science case for HEP to enter this area needs to mature, especially relative to the cost of such a program.

*MATHUSLA.* By exploiting the HL-LHC beam by placing a detector away from the collision point, MATHUSLA has a large discovery potential. It does require commitment from CERN for civil engineering, which is yet to be decided. If downsized, it would qualify as a candidate in the ASTAE portfolio.

*Forward Physics Facility.* This item has a broad physics program utilizing the beam of HL-LHC by placing detectors at the forward direction from the collision point. The facility does require commitment from CERN for civil engineering, which is yet to be decided. The bundled request for the US funding of the FPF trio of experiments does not fit into our budget. However, individual experiments in the facility can be candidates in the ASTAE portfolio.

Some of the tough choices about which projects to include were based on the need to rebalance the portfolio to allow for research in underfunded areas of theory, instrumentation, computing, GARD, a new program for agile experiments, and a virtually non-existent targeted collider R&D program.

# 8.3

# Difficult Choices for Less Favorable Budget Scenario

The less favorable budget scenario assumes a 2% increase in budget per year, which does not keep pace with the assumed 3% annual inflation. Actual inflation may be even higher. In this scenario, we had to make harder choices than those made for the baseline budget scenario. We can maintain a minimum portfolio to continue some scientific progress, although US leadership will begin to erode in much of the field, jeopardizing our 20-year vision for the US particle physics program. Impacts on some of the major projects are described below:

*Reduced contribution to an offshore Higgs factory.* The US contribution will be reduced and the US cannot play a commensurate role as an international partner in the project.

*DUNE FD3 with deferred ACE-MIRT.* This scenario explicitly forces a delay in the DUNE Phase 2 timeline of execution, making the project less competitive and hurting the US reputation as a host for large international projects. In a technically limited schedule, the order of phase II elements is MIRT, FD3, and MCND. There is a compelling science case for these three components and hence when budgets allow this would be the preferred order of construction. If, on the other hand, budgets are more constrained, trade-offs also have to consider the science lost. In the long run the same statistics for the beam physics would be obtained by the addition of either MIRT or FD3. However, FD3 offers a broader set of science topics related to non-beam physics like supernovae and also has a better potential of attracting international support than MIRT. Therefore, in a budget-constrained scenario FD3 is prioritized over MIRT. MCND requires the combined statistics of FD3 and MIRT to accomplish its main goals and hence in all scenarios is the third priority. This scenario is also based on the understanding that MIRT and MCND can be added at any time to the program should additional funds become available. In all scenarios we preserve long-lead-time MIRT elements to enable staging of the beam.

*G3 experiment outside the US for direct detection of WIMP dark matter.* One G3 experiment for direct detection of WIMP dark matter can be supported at less than 50% level and only outside the US so that the SURF expansion is not needed. The US will cede its leadership in this area.

*Reduced increase in support for research.* We identified four areas in which the current support for research requires reinforcement to regain or sustain US leadership: theory, general accelerator R&D, instrumentation, and computing. This scenario would reduce the level of this critically needed reinforcement.

# 8.4

# US Support for Scientific Discovery

The field of particle physics entails curiosity-driven research that relies heavily on federal funding, which comes from taxpayers. We are grateful for that support.

This report discusses the difficult choices we made in our recommendations to maximize science output and make necessary investments in the future of the field with a long-term vision. Note that even a modest increase in the particle physics budget will allow for additional pathways to discovery by expanding the scope of science areas, accelerating projects, and investing more vigorously in the future. We listed additional opportunities in section 2.6.2.

Research in particle physics has an excellent track record of producing revolutionary technologies that enrich society as a whole (section 6.8). Pushing the boundaries of human knowledge requires bringing technology to the next level and training the next generation of scientists to become a technologically advanced workforce. Cutting-edge quantum technologies and AI advances are commonplace in particle physics. People trained in the field become a technologically advanced workforce in many areas of society. Knowledge is its crucial element. Investment in all areas of curiosity-driven research is critical for this foundation.

We are keen to bring knowledge and excitement of major innovations and discoveries to the people who paid for them. Join us as we explore the quantum universe.

# Appendix

# Charge to P5

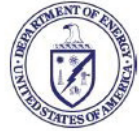

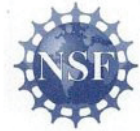

U.S. Department of Energy
and the
National Science Foundation
November 2, 2022

Dr. JoAnne Hewett
Chair, High Energy Physics Advisory Panel
Theory Group, Mail Station 81
SLAC National Accelerator Laboratory
2575 Sand Hill Road
Menlo Park, California  94403

Dear Dr. Hewett:

The 2014 report of the Particle Physics Project Prioritization Panel (P5), developed under the auspices of the High Energy Physics Advisory Panel (HEPAP), successfully laid out a compelling scientific program that recommended world-leading facilities with exciting new capabilities, as well as a robust scientific research program.  That report was well received by the community, the U.S. Department of Energy (DOE) and the National Science Foundation (NSF), and Congress as a well-thought-out and strategic plan that could be successfully implemented.  HEPAP's 2019 review of the implementation of this plan demonstrated that many of the report's recommendations are being realized, and the community has made excellent progress on the P5 science drivers.

As the landscape of high-energy physics continues to evolve and the decadal timeframe addressed in the 2014 P5 report nears its end, we believe it is timely to initiate the next long-range planning guidance to the DOE and NSF.  To that end, we ask that you constitute a new P5 panel to develop an updated strategic plan for U.S. high-energy physics that can be executed over a 10-year timeframe in the context of a 20-year, globally aware strategy for the field.

A critical element of this charge is to assess the continued importance of the science drivers identified by the 2014 P5 report and, if necessary, to identify new science drivers that have the potential to enable compelling new avenues of pursuit for particle physics.  Specifically, we request that HEPAP 1) evaluate ongoing projects and identify potential new projects to address these science drivers; 2) make the science case for new facilities and capabilities that will advance the field and enhance U.S. leadership and global partnership roles; and 3) recommend a program portfolio that the agencies should pursue in this timeframe, along with any other strategic actions needed to ensure the broad success of the program in the coming decades.

In developing the plan, we would like the panel to take into consideration several particularly relevant aspects of constructing a compelling and well-balanced portfolio:

- A core tenet of the 2014 P5 Report is that particle physics is fundamentally a global enterprise.  Thus far, the U.S. program has achieved high impact through U.S. researchers participating in the programs at world-class facilities outside the U.S. and international researchers working at world-class U.S. facilities.  The recommendations developed for this report should carefully consider the current and future international landscape for particle physics.  The panel's report should include an explicit discussion of this context, including the extent to which it is necessary to construct, maintain, and/or upgrade leading U.S.hosted high-energy physics facilities so that our leadership position in the global scientific arena continues, while at the same time preserving the essential roles of, and contributions by, the National Laboratories and universities to global collaboration on large-scale initiatives.

- A number of the projects recommended by the 2014 P5 report are still being built, and the agencies take their commitments to complete them very seriously.  Understanding the continued strength of the science case for these projects is quite valuable, and the panel should provide its assessment of these projects in this context.

- A successful plan should maintain a balance of large, medium, and small projects that can deliver scientific results throughout the decadal timeframe.  We do not expect the panel to consider the large number of possible small-scale projects individually, but advice on research areas where focused investments in smallscale projects can have a significant impact is welcome.

- There are elements of DOE HEP-operated infrastructure that are a stewardship responsibility for HEP.  Investments to maintain that infrastructure in a safe and reliable condition are an HEP responsibility and are outside the scope of the panel.  Major infrastructure upgrades that create new science capabilities are within the scope of the charge and should be considered by the panel.

- Successfully exploiting a newly built project requires funding for the commissioning and operation of the project and to support the researchers who will use these new capabilities to do world-leading science.  Funding is also needed for research and development (R&D) that develops new technologies for future projects.  Scientists and technical personnel working in experimental particle physics often contribute to all these project phases, while theoretical physics provides both the framework to evolve our fundamental understanding of the known universe as well as the innovative concepts that will expand our knowledge into new frontiers.  The panel should deliver a research portfolio that will balance all these factors and consider related issues such as training and workforce development.

- Both NSF and DOE are deeply committed to diversity, equity, inclusion, and accessibility principles in all the scientific communities they support.  Creating a more diverse and inclusive workforce in particle physics will be necessary to implement the plan that this panel recommends, and the panel may further recommend strategic actions that could be taken to address or mitigate barriers to achieving these goals.

- Broad national initiatives relevant to the science and technology of particle physics have been developed by the administration and are being implemented by the funding agencies. These include, but are not limited to, investments in advanced electronics and instrumentation, artificial intelligence and machine learning, and quantum information science. Potential synergies between these initiatives and elements of the recommended portfolio should be considered.

We request that the panel include these considerations in their deliberations and discuss how they affect their recommendations in the report narrative.

The panel's report should identify priorities and make recommendations for an optimized particle physics program over 10 years, FY 2024–FY 2033, under the following budget scenarios:

1) Increases of 2.0 percent per year during fiscal years 2024 to 2033 with the FY 2024 level calculated from the FY 2023 President's Budget Request for HEP.
2) Budget levels for HEP for fiscal years 2023 to 2027 specified in the Creating Helpful Incentives to Produce Semiconductors and Science Act of 2022, followed by increases of 3.0 percent per year from fiscal years 2028 to 2033.

The recommended projects and initiatives should be implementable under reasonable assumptions and be based on generally accepted estimates of science reach and capability. Estimated costs for future projects and facility operations should be given particular scrutiny and may be adjusted if the panel finds it prudent to do so. Given the long timescales for realizing these initiatives, we expect the funding required to enable the priorities the panel identifies may extend well past the 10-year budget profile, but any recommendation should be technically and fiscally plausible to execute in a 20-year timeframe.

In addition to articulating the scientific opportunities that can and cannot be pursued in the various scenarios, the panel may provide their opinions on the approximate overall level of support that is needed for core particle physics research and advanced technology R&D programs to be successful in the context of the science goals of the recommended plan.

We expect the "Snowmass" community planning reports and HEPAP's 2022 study on international benchmarking of scientific resources and capabilities will be useful inputs and that the panel will make efforts to maximize community input and participation in the overall process. Coordination and congruence with the National Academies of Sciences, Engineering, and Medicine's recent and ongoing decadal studies in astronomy, astrophysics, and particle physics are also important considerations.

Finally, effective communication about the excitement, impact, and vitality of particle physics that can be shared with a general audience and other disciplines continues to be critical when advocating the strategic plan. It would be particularly valuable if the panel could re-state the key scientific questions that drive the field so that they are accessible to non-specialists and crisply articulate the value of basic research and the broader benefits of particle physics on other sciences and society.

We would appreciate the panel's preliminary comments by August 2023 and a final report by October 2023. We recognize that this is a challenging task; nevertheless, your assessments will be an essential input to planning at both the DOE and NSF.

Sincerely,

*Asmeret Asefaw Berhe*

Asmeret Asefaw Berhe
Director, Office of Science
U.S. Department of Energy

*Sean L. Jones*

Sean L. Jones
Assistant Director
Directorate for Mathematical and
   Physical Sciences
National Science Foundation

# HEPAP Transmittal Letters

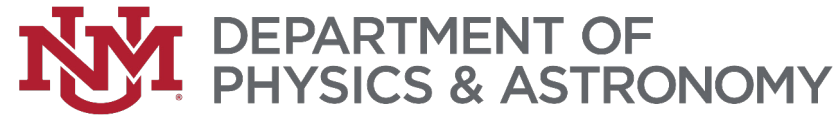

8 December 2023

Dr. Asmeret Asefaw Berhe
Director, Office of Science
U.S. Department of Energy

Dr. C. Denise Caldwell
Acting Assistant Director
Directorate for Mathematical and Physical Sciences
National Science Foundation

Dear Dr. Berhe and Dr. Caldwell:

The report of the Particle Physics Project Prioritization Panel (P5), "Exploring the Quantum Universe," was presented to HEPAP at its meeting on Dec 7 and 8, 2023. This report addresses the charge "to develop a strategic plan for U.S. high-energy physics that can be executed over a 10-year timeframe in the context of a 20-year, globally aware strategy for the field." At the meeting, P5 Chair Hitoshi Murayama and P5 Deputy Chair Karsten Heeger reviewed the report and its recommendations and responded to questions.

Following discussion and deliberation, HEPAP approved the P5 report. The HEPAP members commend P5 for the success with which, starting with input from the particle physics community, it developed a strategic plan for the field. HEPAP also notes the quality of the report in addressing the charge and expresses its appreciation to the members of P5 for the effort that the subpanel devoted to the process. HEPAP strongly endorses the strategic plan presented in the P5 report and supports its immediate implementation.

With this letter, on behalf of HEPAP, I submit for your consideration the final report of P5.

Respectfully

*Sally Seidel*

Sally Seidel
Interim Chair, High Energy Physics Advisory Panel

On behalf of the members of HEPAP:

Halina Abramowicz        Thomas Giblin        Marcelle Soares-Santos
Luis Anchordoqui         Sudhir Malik         Philip Tanedo
Ayana Arce               Reina Maruyama       Jesse Thaler
Kenneth Bloom            Yasuhiro Okada       Natalia Toro
R. Sekhar Chivukula      Mayly Sanchez
Sarah Cousineau          Heidi Schellman
Brenna Flaugher          Monika Schleier-Smith

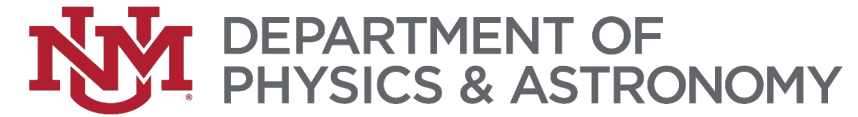

8 May 2024

Dr. Harriet Kung
Acting Director, Office of Science
U.S. Department of Energy

Dr. C. Denise Caldwell
Acting Assistant Director
Directorate for Mathematical and Physical Sciences
National Science Foundation

Dear Dr. Kung and Dr. Caldwell:

The report of the Particle Physics Project Prioritization Panel (P5), "Exploring the Quantum Universe," was presented to HEPAP at its meeting on December 7 and 8, 2023. This report addresses the charge "to develop a strategic plan for U.S. high-energy physics that can be executed over a 10-year timeframe in the context of a 20-year, globally aware strategy for the field." At the meeting, P5 Chair Hitoshi Murayama and P5 Deputy Chair Karsten Heeger reviewed the report and its recommendations and responded to questions.

Following discussion and deliberation, HEPAP approved the P5 report pending some small corrections intended to clarify without revising the recommendations. The HEPAP members commend P5 for the success with which, starting with input from the particle physics community, it developed a strategic plan for the field. HEPAP also notes the quality of the report in addressing the charge and expresses its appreciation to the members of P5 for the effort that the subpanel devoted to the process. HEPAP strongly endorses the strategic plan presented in the P5 report and supports its immediate implementation. With this letter, on behalf of HEPAP, I submit for your consideration the final report of P5.

Respectfully,

*Sally Seidel*

Sally Seidel
Chair, High Energy Physics Advisory Panel

On behalf of the members of HEPAP:

Halina Abramowicz        Brenna Flaugher      Heidi Schellman
Luis Anchordoqui         Thomas Giblin        Monika Schleier-Smith
Ayana Arce               Sudhir Malik         Marcelle Soares-Santos
Kenneth Bloom            Reina Maruyama       Philip Tanedo
R. Sekhar Chivukula      Yasuhiro Okada       Jesse Thaler
Sarah Cousineau          Mayly Sanchez        Natalia Toro

# Panel Members

Shoji Asai
*University of Tokyo*

Amalia Ballarino
*CERN*

Tulika Bose
*University of Wisconsin–Madison*

Kyle Cranmer
*University of Wisconsin–Madison*

Francis-Yan Cyr-Racine
*University of New Mexico*

Sarah Demers
*Yale University*

Cameron Geddes
*Lawrence Berkeley National Laboratory*

Yuri Gershtein
*Rutgers University*

Karsten Heeger, Deputy Chair
*Yale University*

Beate Heinemann
*DESY*

JoAnne Hewett, HEPAP Chair,
ex officio until May 2023
*SLAC National Accelerator Laboratory*

Patrick Huber
*Virginia Tech*

Kendall Mahn
*Michigan State University*

Rachel Mandelbaum
*Carnegie Mellon University*

Jelena Maricic
*University of Hawaii at Manoa*

Petra Merkel
*Fermi National Accelerator Laboratory*

Christopher Monahan
*William & Mary*

Hitoshi Murayama, Chair
*University of California, Berkeley*

Peter Onyisi
*University of Texas at Austin*

Mark Palmer
*Brookhaven National Laboratory*

Tor Raubenheimer
*SLAC National Accelerator Laboratory/
Stanford University*

Mayly Sanchez
*Florida State University*

Richard Schnee
*South Dakota School of Mines
& Technology*

Sally Seidel, Interim HEPAP Chair,
ex officio since June 2023
*University of New Mexico*

Seon-Hee Seo
*IBS Center for Underground Physics,
Fermi National Accelerator Laboratory*

Jesse Thaler
*Massachusetts Institute of Technology*

Christos Touramanis
*University of Liverpool*

Abigail Vieregg
*University of Chicago*

Amanda Weinstein
*Iowa State University*

Lindley Winslow
*Massachusetts Institute of Technology*

Tien-Tien Yu
*University of Oregon*

Robert Zwaska
*Fermi National Accelerator Laboratory*

# Process and Meetings

### Information gathering phase

February 6, 2023. Kickoff meeting with the Department of Energy (DOE) and the National Science Foundation (NSF)

Open Town Halls, all with short remarks. Live captioning and American Sign Language interpretation was provided at the town halls.

a. Berkeley National Lab: February 22, 23. 513 registrants

b. Fermilab/Argonne National Lab: March 21, 22, 23. 797 registrants, partially overlapped with Elementary Particle Physics 2024 (EPP2024) town hall

c. Brookhaven: April 12, 13. 666 registrants

d. SLAC: May 3, 4. 512 registrants.

Virtual Town Halls

a. University of Texas at Austin: June 5. 159 registrants, exclusive session for early career physicists.

b. Virginia Tech, June 27. 119 registrants

### Keeping the community informed

1. American Physical Society's (APS's) Division of Particles and Fields session on P5 (April 15)

2. Early Career Network Workshop (June 8,9)

3. CE Science Workshop (June 14,15)

4. CEPC Workshop (July 6)

5. ICFA (July 15)

6. HEPAP (August 7)

7. HEP-PI Meeting (August 15)

8. CEPC Workshop (October 22)

9. ICFA Seminar (November 30)

We used the APS DPF and Division of Physics and Beams mailing lists, Snowmass mailing list, and the P5 website to send out information about the P5 activities.

### Deliberation Phase
(closed meetings)

1. May 31 to June 2, Austin, TX

2. June 21 to 23, Gaithersburg, MD

3. July 11 to 14, Santa Monica, CA

4. August 1 to 4, Denver, CO

Meetings by working groups, with additional input from the following:

*Agencies:* Asmeret Berhe, Harriet Kung, Sean Jones, Saúl González, DOE/HEP, NSF/PHY, NSF/AST (Debra Fisher, Nigel Sharp), NSF/OPP (Jim Ulvestad)

*Government:* Cole Donovan (Department of State, Office of Science and Technology Policy)

*Community:* International Benchmarking Panel, computing frontier, DPF leadership, previous P5 (Steve Ritz, Andy Lankford), Committee of Visitors' reports (Ritchie Patterson, Dmitry Denisov)

*National Laboratories:* Oak Ridge National Laboratory, Thomas Jefferson National Accelerator Facility (Jefferson Lab)

## Writing Phase
(Zoom meetings)

5. Five full-day meetings
(August 11, 18, 25, September 1, 8)

6. Short meetings
(weekly since September 14)

## Peer-Review Process

The draft report was sent to the following people for a peer review on October 25 with a very tight deadline of October 31. We received many invaluable comments. The input and comments we received did not change the contents of the recommendations, but helped us improve the clarity and presentation of our report significantly. Hereby we acknowledge their help.

Elena Aprile
*Columbia University*

William Barletta
*Massachusetts Institute of Technology*

John Carlstrom
*University of Chicago*

Mu-Chun Chen
*University of California, Irvine*

Sekhar Chivukula
*University of California, San Diego*

André de Gouvêa
*Northwestern University*

Josh Frieman
*University of Chicago*

Elizabeth Hayes
*NASA Goddard Space Flight Center*

Katrin Heitmann
*Argonne National Laboratory*

David Hertzog
*University of Washington, Seattle*

Mark Messier
*Indiana University*

Laura Reina
*Florida State University*

Andrei Seryi
*Jefferson Lab*

Hirohisa Tanaka
*SLAC National Accelerator Laboratory*

John Womersley
*University of Edinburgh*

# Subcommittee on Cost, Risks, and Schedule

## Charge
(March 1, 2023)

The cost/schedule/risk subcommittee to P5 is asked to obtain and clarify the cost, schedule, and risk information from the proponents of high cost (>$250M in FY23$) HEP projects funded or being considered for funding by DOE and/or NSF. The subcommittee will not prepare its own estimates. The subcommittee should assess this information at a high level, noting key assumptions, risks, and cost and schedule uncertainties including the risk from non-DOE/NSF funding sources, international partners making in-kind contributions and collaborations, and missing costly items, if any. The subcommittee is also asked to comment on the operation costs for projects during commissioning and when the resulting facilities are in steady-state operation. This subcommittee will provide P5 with the expert opinions on the uncertainty ranges for the projects that P5 needs in order to develop a strategy for the field within assumed budgetary constraints. The subcommittee will submit their preliminary report to P5 in early summer.

## Members

Jay Marx, Chair
*California Institute of Technology*

Gil Gilchriese, Matthaeus Leitner
*Lawrence Berkeley National Laboratory*

Giorgio Apollinari, Doug Glenzinski
*Fermi National Accelerator Laboratory*

John Seeman, Mark Reichanadter, Nadine Kurita
*SLAC Accelerator National Laboratory*

Jon Kotcher, Srini Rajagopalan
*Brookhaven National Laboratory*

Allison Lung
*Jefferson Lab*

Harry Weerts
*Argonne National Laboratory*

## Process

1. Single contact person for each project determined

2. Set of 20 questions sent to project with answers requested in 1 week

3. Divided committee into eight subgroups of three members each with a designated lead for each project

   - Each member involved in two of the eight projects

   - Possible conflicts of interest taken account in assignment to subgroups

   - Assignment matrix

Based on the inputs from the proponents and experience of the subcommittee members, they provided three possible funding and schedule scenarios:

- The low estimate is meant to be possible but optimistic—most things have to go right.

- The mid-range estimate is meant to be moderately probable.

- The high estimate is meant to be pessimistic but not unlikely.

**The subcommittee report was received on June 30, 2023.**

# Important References

# Conflict of Interest Resolution

A vote will be held on December 8, 2023, by the members of the HEPAP panel, regarding acceptance of the P5 Report. On the basis of recommendations received from the U.S. DOE General Counsel, which are based on review of the financial disclosures of the HEPAP members, the following members of the panel have been determined to have conflicts of interest due to their employment at US national laboratories; they will not participate in the vote for or against acceptance of the P5 Report.

BRENNA FLAUGHER, FNAL employee
HEIDI SCHELLMAN, FNAL employee
MONIKA SCHLEIER-SMITH, SLAC employee
NATALIA TORO, SLAC employee

Each member of HEPAP will abstain from discussion of topics as indicated below. These abstentions to not affect that member's permission to participate in the vote on the report.

LUIS ANCHORDOQUI will have participation restrictions regarding his employer, Lehman College, City University of New York. He may participate in HEPAP matters that affect his employer, so long as they are policy matters that do not affect the employer uniquely and specifically.

AYANN ARCE will have participation restrictions regarding her employer, Duke University. She may participate in HEPAP matters that affect her employer, so long as they are policy matters that do not affect the employer uniquely and specifically.

KENNETH BLOOM will have participation restrictions regarding his employer, the University of Nebraska-Lincoln. He may participate in HEPAP matters that affect his employer, so long as they are policy matters that do not affect the employer uniquely and specifically.

SARAH COUSINEAU will have participation restrictions regarding her employers, Oak Ridge National Laboratory/UT-Battelle, LLC and the University of Tennessee. She may participate in HEPAP matters that affect her employers, so long as they are policy matters that do not affect the employers uniquely and specifically.

BRENNA FLAUGHER will have participation restrictions regarding her employer, Fermi National Accelerator Laboratory/Fermi Research Alliance, LLC. She may participate in HEPAP matters that affect her employer, so long as they are policy matters that do not affect the employer uniquely and specifically.

THOMAS GIBLIN will have participation restrictions regarding his employer, Kenyon College. He may participate in HEPAP matters that affect his employer, so long as they are policy matters that do not affect the employer uniquely and specifically.

SUDHIR MALIK will have participation restrictions regarding his employer, the University of Puerto Rico. He may participate in HEPAP matters that affect his employer, so long as they are policy matters that do not affect the employer uniquely and specifically.

REINA MARUYAMA will have participation restrictions regarding her employer, Yale University. She may participate in HEPAP matters that affect her employer, so long as they are policy matters that do not affect the employer uniquely and specifically.

MAYLY SANCHEZ will have participation restrictions regarding her employer, Florida State University. She may participate in HEPAP matters that affect her employer, so long as they are policy matters that do not affect the employer uniquely and specifically.

HEIDI SCHELLMAN will have participation restrictions regarding her employers, Fermi National Accelerator Laboratory/Fermi Research Alliance, LLC and Oregon State University. She may participate in HEPAP matters that affect her employers, so long as they are policy matters that do not affect the employers uniquely and specifically.

MONICA SCHLEIER-SMITH will have participation restrictions regarding her employer, SLAC National Accelerator Laboratory/Stanford University. She may participate in HEPAP matters that affect her employer, so long as they are policy matters that do not affect the employer uniquely and specifically.

SALLY SEIDEL will have participation restrictions regarding her employer, the University of New Mexico. She may participate in HEPAP matters that affect her employer, so long as they are policy matters that do not affect the employer uniquely and specifically.

MARCELLE SOARES-SANTOS will have participation restrictions regarding her employer, the University of Michigan. She may participate in HEPAP matters that affect her employer, so long as they are policy matters that do not affect the employer uniquely and specifically.

PHILIP TANEDO will have participation restrictions regarding his employer, the University of California, Riverside. He may participate in HEPAP matters that affect his employer, so long as they are policy matters that do not affect the employer uniquely and specifically.

JESSE THALER will have participation restrictions regarding his employer, the Massachusetts Institute of Technology. He may participate in HEPAP matters that affect his employer, so long as they are policy matters that do not affect the employer uniquely and specifically.

NATALIA TORO will have participation restrictions regarding her employer, Stanford University/SLAC National Accelerator Laboratory. She may participate in HEPAP matters that affect her employer, so long as they are policy matters that do not affect the employer uniquely and specifically. She will also have participation restrictions regarding TRIUMF and TRIUMF's Particle Physics Experiment Advisory Committee.

Submitted on 5 December 2023,

*Sally Seidel*

Sally Seidel
HEPAP Interim Chair

# Acronyms & Glossary

**Accelerator:** A particle accelerator, which accelerates a beam of particles such as electrons, positrons, muons, or protons.

**ACE:** Accelerator Complex Evolution. A proposed major upgrade to the accelerator complex at Fermilab.

**ACE-BR:** Accelerator Complex Evolution Booster Replacement. A proposed major upgrade to the accelerator complex at Fermilab, including a replacement of the booster synchrotron, as part of ACE.

**ACE-MIRT:** Main Injector Ramp and Target is a proposed upgrade to Fermilab accelerator complex to upgrade the beam to power of 2.1 MW; part of the re-envisioned DUNE Phase II. Included as a component of ACE.

**ACORN:** Accelerator Controls Operations Research Network. A project to modernize accelerator controls at Fermilab.

**ADMX:** Axion Dark Matter eXperiment. An experiment aiming to detect a hypothetical particle called axion as a candidate of dark matter built at University of Washington, Seattle. Its second generation is called ADMX-G2.

**AI/ML:** Artificial intelligence/machine learning.

**ALP:** Axion-like particle. A class of hypothetical pseudoscalar particles, beyond the Standard Model. A generalization of the axion that relaxes the relationship between particle mass and coupling to the Standard Model.

**AMF:** Advanced Muon Facility. A proposed muon facility at Fermilab to produce bright beams of muons for charged lepton-flavor violation physics and other future experiments.

**ANL:** Argonne National Laboratory.

**ASIC:** Application-specific integrated circuits.

**ASTAE:** Advancing Science and Technology through Agile Experiments. A portfolio of small-scale DOE experiments proposed in this report.

**AST:** Division of Astronomical Sciences. A division in NSF Mathematical and Physical Sciences focused on astronomical sciences.

**AS&T:** Accelerator science and technology.

**ATLAS:** An operating experiment at LHC, CERN. One of the two experiments responsible for the discovery of the Higgs boson in 2012.

**Atom interferometer:** A quantum sensor that uses a wave-like behavior of atoms. It is used for wave-like dark matter searches.

**Atomic clock:** A high-precision clock that measures time using the resonant frequency of atoms, the basis of the definition of a second. It is used for dark matter searches through variations in fundamental constants.

**Antiparticle:** A particle that has the same mass and the opposite electric charge of a given particle.

**ATF:** Accelerator Test Facility. A user facility at Brookhaven National Laboratory that supports experiments for advanced accelerator and laser research including research on technologies.

**AUP:** High-Luminosity LHC Accelerator Upgrade Project. The accelerator portion of the US contribution to HL-LHC providing critical components such as high-field "triplet" magnets using the Nb3Sn technology.

**Automatic differentiation:** A set of techniques for evaluating derivatives automatically; widely used for optimization.

**AWA:** Argonne Wakefield Accelerator Facility. A facility to demonstrate electron-beam-driven wakefield acceleration technologies.

**Axion:** A hypothetical particle postulated to account for the rarity of processes that break charge-parity symmetry. It is very light, electrically neutral, and pseudoscalar.

**BELLA:** Berkeley Lab Laser Accelerator. A facility at Lawrence Berkeley National Laboratory that uses intense lasers to drive wakefields in plasma for research on technologies.

**Belle, Belle II:** Belle was an experiment that studied the properties of the bottom quark bound states and discovered CP violation in those systems. Belle II is a successor experiment that explores quantum imprints of new phenomena in these sys-tems. Both Belle and Belle II are hosted by KEK, a national accelerator laboratory in Japan.

**BES:** Beijing Spectrometer. A series of detectors mounted at Beijing Electron Positron Collider (BEPC). Both BEPC and its upgrade, BEPC II, are hosted by Institute of High Energy Physics (IHEP) in China. The current detector is BES III, which explores quantum imprints.

**BES:** Basic Energy Sciences. An office within the DOE Office of Science performing basic science investigations into materials, chemistry, and geoscience.

**BICEP:** Background Imaging of Cosmic Extragalactic Polarization. An experiment that measures cosmic microwave background (CMB) polarization, the oldest light in the universe, to understand questions about the beginning of the universe.

**Big Bang:** The unofficial name for the beginning of the universe as a small patch of space with a huge amount of energy that expanded in an explosive fashion.

**BNL:** Brookhaven National Laboratory.

**Booster:** A proton synchrotron accelerator, part of the accelerator complex at Fermilab.

**BSM:** Beyond the Standard Model. A realm of particle physics not encompassed by the Standard Model of Particle Physics.

**CAD:** Computer-aided design software. A widely used mechanical engineering design tool.

**CCC:** Cool Copper Collider, a concept for a future particle accelerator using normal conducting copper accelerators cooled to cryogenic temperatures.

**CERN:** Conseil Européen pour la Recherche Nucléaire. An organization for nuclear research situated on the border of Switzerland and France.

**CEvNS:** Coherent elastic neutrino-nucleus scattering. A process of neutrino-nucleus scattering in which the nucleus acts like a single particle. First experimentally observed in 2017 by the COHERENT experiment.

**CHIPS:** Creating Helpful Incentives to Produce Semiconductors (CHIPS) and Science Act of 2022. New large-scale funding to boost domestic research and manufacturing of semiconductors in the US.

**CLIC:** Compact Linear Collider. A proposed accelerator designed as an addition to the CERN accelerator complex.

**CMB:** Cosmic microwave background, the oldest light in the universe, emitted 380,000 years after the Big Bang. It has become microwave radiation by the expansion of the universe, and it permeates the entire observable universe.

**CMB-S4:** Cosmic Microwave Background Stage IV. An experiment set to make precision measurements of the polarization of the CMB. It will constrain the physics of inflation in the early universe, light relics, and the properties of neutrinos among other science goals. It has two sites, the South Pole site and the Atacama desert in Chile.

**CMS:** Compact Muon Solenoid. An operating experiment performed at the LHC, CERN; one of the two experiments that discovered the Higgs boson in 2012.

**Collider:** A particle accelerator that accelerates beams of particles in opposite directions and then collides them to achieve very high-energy collisions.

**CPAD:** The Coordinating Panel for Advanced Detectors. A panel that seeks to promote, coordinate, and assist in the research and development of instrumentation and detectors for high energy physics experiments under the auspices of the Division of Particles and Fields of the American Physical Society.

**CP violation:** Subtle differences between the behaviors of matter and antimatter. Under most circumstances, antimatter behaves almost exactly like matter in the mirror. If that behavior is not the case, it is said that the charge-parity symmetry (CP) is violated.

**CTA:** Cherenkov Telescope Array. An array of telescopes that study high-energy gamma rays from the universe. It is under construction on the Canary Islands, Spain, and at Paranal, Chile. The US is a member of the 25-nation consortium.

**Dark energy:** An unknown form of energy that makes up about 69% of energy in the universe today. Its existence has been inferred from the observation that the expansion of the universe is accelerating today. One possibility is that it is the energy of the vacuum, called the cosmological constant, designated as Λ. Another possibility is that it is the slowly evolving energy density of a field.

**Dark matter:** An unknown type of matter that makes up 84% of matter in the universe today. Its existence has been inferred from the rotation speed of stars and gas in galaxies, gravitational lensing due to clusters of galaxies, large-scale structure, and CMB measurements. Its gravitational pull was necessary to form stars and galaxies from the primordial gas.

**Dark Matter (DM) G2, G3:** Generation 2 and Generation 3 dark matter experiments. Generation 2 WIMP-search experiments comprise no more than a few tens of tons of target mass and are currently operating or under construction. Generation 3 WIMP-search experiments will have tens to hundreds of tons of target mass, increasing the WIMP sensitivity to neutrino fog level.

**Dark photons:** Hypothetical particles from hidden sectors proposed as force carriers similar to photons, but potentially associated with dark matter.

**DarkSide-20k:** An experiment aiming at direct detection of dark matter, focusing on the WIMP dark matter particle search. It is located in the Gran Sasso Underground Laboratory in Italy.

**DARPA:** Defense Advanced Research Projects Agency.

**DES:** Dark Energy Survey. An experiment that mapped out hundreds of millions of galaxies and detected supernovae, looking for patterns in cosmic structure in order to reveal the nature of the dark energy that is accelerating the expansion of the universe.

**DESC:** Dark Energy Science Collaboration. A group of scientists that study fundamental physics using data from The Vera C. Rubin Observatory's Legacy Survey of Space and Time (LSST).

**DESI, DESI-II:** Dark Energy Spectroscopic Instrument. A spectroscopic galaxy survey at the Mayall 4m telescope at Kitt Peak National Observatory in Arizona.

**DMNI:** Dark Matter New Initiatives. A suite of innovative agile dark matter experiments set to search for various forms of dark matter.

**DOE:** Department of Energy. Its Office of Science is the largest funding agency for physical sciences and includes the Office of High Energy Physics, which supports particle physics research.

**DUNE:** Deep Underground Neutrino Experiment. An experiment that studies the oscillation of neutrinos over 800 miles, a macroscopic quantum phenomenon. The beam of neutrinos is created artificially at Fermilab in Illinois and directed toward the Sanford Underground Research Facility in South Dakota, where the neutrinos are detected and studied in the DUNE Far Detectors.

**eBOSS:** Extended Baryon Oscillation Spectroscopic Survey. A project carried out as part of the Sloan Digital Sky Survey (SDSS-IV) on a 2.5m telescope at Apache Point Observatory in New Mexico.

**EDM:** Electric dipole moment, a possible interaction of an electric field with spin of an elementary particle. If it exists for an elementary particle, it would break the time reversal symmetry and is a form of CP violation.

**EIC:** <u>Electron-Ion Collider</u>. A new facility colliding electrons and protons, under construction at BNL in support of the DOE Nuclear Physics program.

**FACET-II:** <u>Facility for Advanced Accelerator Experimental Tests</u>. An accelerator facility at SLAC National Accelerator Laboratory that provides high-energy electron beams for researching particle accelerator technologies.

**FAST:** <u>Fermilab Accelerator Science and Technology</u>. A facility including a fully-equipped R&D accelerator chain to support research and development of accelerator technology for the next generation of particle accelerators.

**FASER2:** Forward Search Experiment at the HL-LHC. A proposed experiment to look for new long-lived particles. It builds on the success of the FASER experiment and is proposed to be housed at the FPF.

**FCC:** <u>Future Circular Collider</u>. A future particle accelerator complex planned at CERN to support the FCC-ee and FCC-hh in a new underground tunnel with 91 km circumference.

**FCC-ee:** A proposed electron-positron Higgs factory in the FCC tunnel, possibly an intermediate step toward the FCC-hh collider.

**FCC-hh:** A proposed proton-proton collider in the FCC tunnel that will push the energy about seven times higher than that of the current LHC.

**FD3, FD4:** The third and fourth far detector modules of the DUNE experiment, located at the Sanford Underground Research Facility (SURF) in South Dakota.

**Fermilab:** <u>Fermi National Accelerator Laboratory</u>. A US National Laboratory focused on particle physics research located in Batavia, Illinois.

**FES:** <u>Fusion Energy Sciences</u>. An office in the DOE Office of Science performing investigations into fusion energy sources, and the corresponding science of matter at very high temperatures and densities.

**Fifth force:** A new hypothetical force, beyond the four fundamental forces of nature (gravitational, electromagnetic, weak, and strong force).

**FORMOSA:** An experiment to search for millicharged particles in the LHC collisions. It builds on the experience of the MilliQan experiment and is proposed to be housed at the FPF.

**FPF:** <u>Forward Physics Facility</u>. A proposed underground facility to operate several hundred meters away from ATLAS during the LHC-HL running, with several new experiments aligned with ATLAS collision axis.

**FPF trio:** The bundled request for US funding for three of the proposed experiments at FPF, including FASER2, FORMOSA, and the Forward Liquid Argon Experiment (FLArE)

**FY23:** Fiscal Year 2023, spanning from October 1, 2022 to September 30, 2023.

**GARD:** General Accelerator Research and Development program with the DOE Office of High Energy Physics. GARD develops advanced technologies for the acceleration of particles for HEP and other applications.

**Geometric deep learning:** A broad class of machine learning (ML) approaches that take into account symmetries and invariances.

**GeV:** Giga-electron-volt. A unit of energy that is approximately equal to the proton's mass times the speed of light squared ($E=mc^2$).

**Graph neural networks:** A method in AI inspired by the way the human brain functions, applied to graph represented data.

**Gravitational lensing:** Focusing of light due to the gravity of mass concentrations bending light such that it is magnified over large distances. Gravitational lensing is used in the investigation of dark matter, dark energy, and other astrophysical phenomena.

**Gravitational waves:** Ripples of spacetime caused by extremely energetic events in the universe such as pairs of black holes or neutron stars that coalesce on inspiraling orbit. Gravitational waves have been recently discovered with terrestrial detectors and are an evolving astrophysical probe of particle physics.

**HELEN:** Higgs-Energy Lepton. An electron-positron linear collider based on advances in superconducting radio frequency technology.

**HEP:** <u>High Energy Physics</u>. An office in the DOE Office of Science.

**HEPAP:** <u>High Energy Physics Advisory Panel</u>. A committee that reports to the associate director of DOE HEP and assistant director of NSF's Directorate of Mathematical and Physical Sciences (DMP).

**HL-LHC:** <u>High-Luminosity LHC</u>. High-luminosity upgrade to the Large Hadron Collider.

**Hidden sectors:** Particles and quantum fields described by their mediator particles, that have been hypothesized, but have not been observed yet, possibly due to their small couplings.

**Higgs boson:** A fundamental particle that is believed to be condensed throughout the universe as the Higgs field, the heart of the Standard Model of particle physics. It is responsible for giving mass to fundamental particles such as quarks and leptons.

**Higgs factory:** A particle accelerator that collides beams of electrons and positrons, to produce about a million Higgs bosons in order to study the particle's precise properties.

**IceCube:** <u>An astrophysical neutrino observatory</u>, located at the South Pole, utilizing $1 km^3$ of Antarctic ice as a target for neutrino detection. IceCube has produced a picture of the Milky Way galaxy in neutrinos.

**IceCube-Gen2:** A <u>proposed expansion of IceCube</u> that will employ optical and radio detection methods to cover a broad range of neutrino energies.

**ICFA:** <u>International Committee for Future Accelerators</u>. A body of the International Union of Pure and Applied Physics created to facilitate international collaboration in the construction and use of accelerators for particle physics.

**ILC:** <u>International Linear Collider</u>. A proposed future electron positron collider using superconducting radio frequency technology.

**ITN:** ILC Technology Network. A network of international institutions set to advance ILC-related technology in selected areas toward engineering design and to explore opportunities for other accelerator applications.

**IMCC:** <u>International Muon Collider Collaboration</u>. An international collaboration formed to develop the R&D and planning as a pathway toward a collider of muons and anti-muons.

**Inflation:** A period of exponential expansion of the universe in a fraction of a second after the Big Bang. It is hypothesized to have expanded the size of the universe more than a billion trillion times and seeded quantum fluctuations that became stars and galaxies.

**INSPIRE:** A one-stop <u>information platform</u> for the HEP community, comprising eight interlinked databases on literature, conferences, institutions, journals, researchers, experiments, jobs, and data. Run in collaboration by CERN, DESY, Fermilab, IHEP, IN2P3, and SLAC, it has been serving the scientific community for almost 50 years.

**IOTA:** Integrable Optics Test Accelerator ring. The primary focus of the FAST accelerator R&D facility, IOTA is a circular accelerator testing nonlinear and other advanced approaches to particle beam technology.

**J-PARC:** The Japan Proton Accelerator Research Complex, jointly operated by KEK and the Japan Atomic Energy Association. J-PARC provides proton beams for T2K, COMET, and other particle physics experiments.

**kBELLA:** A <u>proposed</u> technology test facility of new plasma wakefield acceleration techniques that use high-power lasers to accelerate particles within very short distances.

**KEK:** <u>High Energy Accelerator Research Organization</u> in Japan, which hosts many particle physics experiments and co-operates J-PARC with the Japan Atomic Energy Agency.

**LAr:** Liquid argon (cooled to cryogenic temperatures) used as a detector medium for neutrino, collider, and dark matter experiments.

**LARP:** <u>US LHC Accelerator Research Program</u>. A program in the US that studied possible improvements on the LHC, which led to AUP.

**LArTPC:** Liquid argon time-projection chamber, a detector technology for neutron experiments and other particle physics.

**LQCD:** Lattice quantum chromodynamics. A tool to carry out non-perturbative theoretical calculations to describe properties of hadrons, which are composed of quarks and gluons.

**LBNF:** <u>Long Baseline Neutrino Facility</u>. A world-class facility hosting the DUNE experiment and providing the long-baseline neutrino beam.

**LBNL:** <u>Lawrence Berkeley National Laboratory</u>.

**ΛCDM:** The current paradigm to describe the evolution of the universe from the Big Bang to today, based on the cosmological constant (Λ, lambda) and cold dark matter (CDM), with fluctuation of densities seeded by inflation.

**LCLS:** <u>Linac Coherent Light Source</u>. The world's first hard X-ray free-electron laser located at SLAC National Accelerator Laboratory supported by DOE Basic Energy Sciences.

**LCLS-II, LCLS-II-HE:** <u>Linac Coherent Light Source-II and its upgrade</u>. The world's first MHz rate hard X-ray free-electron laser located at SLAC National Accelerator Laboratory.

**Lepton:** A collective name for elementary matter particles that do not participate in the strong interaction—namely electron, muon, tau, three neutrinos, and their anti-matter counterparts.

**LHC:** <u>Large Hadron Collider</u>. The world's largest and highest-energy accelerator, located at CERN, Switzerland.

**LHCb:** <u>Large Hadron Collider beauty</u>. An experiment that primarily studies systems of bottom quark produced by the collisions in the LHC.

**LIM:** <u>Line intensity mapping</u>. A proposed method to create 3D maps of the universe using emission lines from atoms or molecules redshifted to longer wavelengths.

**LSND:** Liquid Scintillator Neutrino Detector. A completed neutrino-oscillation experiment that reported an anomaly, suggesting existence of more than three neutrino species. Currently tested by SBN experiments at Fermi National Laboratory.

**LSST:** <u>Legacy Survey of Space and Time</u>. A 10-year survey that the Vera C. Rubin Observatory will carry out.

**LZ:** <u>LUX-ZEPLIN</u>. An experiment for direct detection of dark matter at SURF, searching for WIMPs. It is a TPC filled with 7,000 kg of liquified xenon.

**MagLab:** The national <u>High Magnetic Field Laboratory</u> at Florida State University.

**Magnetometer:** A device that precisely measures magnetic field and magnetic dipole moment.

**MATHUSLA:** <u>Massive Timing Hodoscope for Ultra Stable Neutral Particles</u>. A proposed surface detector at HL-LHC to search for hypothetical neutral long-lived particles (LLPs).

**MCND:** More Capable Near Detector. An upgraded DUNE Near Detector that is part of DUNE reinvisioned Phase II.

**MPS:** <u>Directorate of Mathematical and Physical Sciences</u>, within the National Science Foundation (NSF).

**MRI:** <u>Major Research Instrumentation program</u>. An NSF funding program for multi-user research instruments costing up to $4M.

**MSRI:** <u>Mid-Scale Research Infrastructure</u>. An NSF funding program to build experiments and facilities in the cost range of $4M–20M (MSRI-1) or $20M–100M (MSRI-2) (information for FY2023).

**Muon collider:** A circular particle accelerator that steers and collides beams of muons and anti-muons.

**Mu2e, Mu2e-II:** <u>Muon to electron conversion experiment</u> and its proposed upgrade, stationed at Fermilab, set to search for physics BSM.

**Muon g-2:** An <u>experiment</u> at Fermilab that performed the most precise measurement of the muon anomalous magnetic moment.

**MW:** Megawatt, a unit of power equal to one million watts, used to measure proton beam power as the production of protons per second, and the total energy per proton.

**Nanofabrication:** Manufacturing materials at the nanometer scale.

**NASA:** <u>National Aeronautics and Space Administration</u>.

**Nb₃Sn:** Niobium Three Tin. A superconducting material characterized by the ability to sustain high currents and magnetic fields; used in particle accelerators such as HL-LHC and nuclear magnetic resonance.

**New physics:** Phenomena that cannot be described by the currently known laws of physics, such as the Standard Model of particle physics or ΛCDM model of cosmology.

**Neutrino:** The most abundant known massive fundamental particles, a billion times more abundant than particles that make up the matter that surrounds us. Neutrinos are neutral, weakly interacting, and extremely light. They rarely interact with matter, so intense sources or huge detectors are required in order to observe their interactions and study their properties.

**Neutrino fog:** An irreducible background for WIMP dark matter search due to neutrino interactions in the dark matter detectors that mimic the WIMP signals. Due to the extremely small probability of neutrino interaction, only the G3 dark matter detectors will be large enough to start seeing neutrino interactions from the atmospheric neutrino flux passing through detectors.

**Neutrino Platform:** <u>CERN's main contribution</u> to a globally coordinated program of neutrino research, providing access for European researchers. It includes an R&D facility at CERN to develop and prototype the next generation of neutrino detectors, including Icarus, Baby MIND and ProtoDUNE. It also includes direct contributions to experiments, such as the cryostats for the DUNE detectors.

**NNSA:** <u>National Nuclear Security Agency</u>, an agency within the US Department of Energy.

**NOvA:** <u>NuMI Off-axis νₑ Appearance</u>. A long-baseline neutrino oscillation experiment, operating in Ash River, Minnesota, detecting the neutrino beam sent from Fermilab, 500 miles away.

**NSAC:** <u>Nuclear Science Advisory Committee</u>, which provides official advice to DOE and NSF on the national program for basic nuclear science research.

**NSF:** <u>National Science Foundation</u>. A US federal funding agency for basic and applied science.

**NSF/PHY:** <u>Division of Physics</u> in the <u>NSF Directorate of Mathematical and Physical Sciences</u> (MPS).

**nuSTORM:** <u>Neutrinos from Stored Muons</u>. A proposed facility that can provide electron- and muon-neutrino beams from the decay of low-energy muons confined in a storage ring.

**Offshore:** Located outside the US.

**Onshore:** Located in the territory of the US.

**OPP:** <u>Office of Polar Programs</u>, an NSF program responsible for the logistics and much of the operations of the South Pole Station.

**ORISE:** <u>Oak Ridge Institute for Science and Education</u>. Office that provided logistical support for P5 panel meetings and travel needs.

**P5:** Particle Physics Project Prioritization Panel, jointly commissioned by DOE and NSF, under the auspices of HEPAP, to develop a detailed strategic plan for US particle physics over the next 10 years, with a 20-year vision.

**PDG:** <u>Particle Data Group</u>. An international collaboration led by LBNL that publishes a biannual compilation of important data in particle physics.

**Parton:** Point-like, seemingly fundamental, constituent of a hadron (a composite particle consisting of two or more quarks held together by a strong force). Sometimes also used to describe other fundamental particles.

**Positron:** Anti-electron. A positively charged elementary particle with mass identical to an electron. It is an antiparticle to the electron.

**pCM:** Parton center-of-momentum. A measure of how much energy is available for the creation of heavy particles in particle collisions. For elementary particle collisions, such as between an electron and positron, the parton center-of-momentum and beam center-of-momentum are essentially the same. For composite particle collisions, like those between protons, the parton center-of-momentum is substantially lower than the beam center-of-momentum because the actual collisions are of the partonic constituents (quarks and gluons in the case of protons).

**PIP, PIP-II:** <u>Proton Improvement Plan</u> and its successor. An enhancement to the Fermilab accelerator complex, powering the world's most intense high-energy neutrino beam.

**Plasma wakefield:** A technology developed for compact, high-gradient accelerators whereby a witness bunch is accelerated by the oscillating electric field in a plasma, which is created by other particle or laser beams.

**ProtoDUNE:** Kiloton-scale prototype LArTPC detectors that serve as technology demonstrators for DUNE Far Detectors and are roughly one-tenth the size of far detectors generally. ProtoDUNE detectors are part of the CERN Neutrino Platform program.

**QCD:** Quantum chromodynamics. A theory of strong interaction between quarks, mediated by particles called gluons.

**QIS:** Quantum information science, technologies for computation, information processing, and detection that elude classical limitations through the use of quantum effects.

**Quantum calorimeter:** A quantum sensor for thermal measurement of quanta of energy for dark photon, dark matter, and other hidden sector searches.

**Quark:** A collective name for elementary matter particles that do participate in the strong interaction—namely, up, down, strange, charm, bottom, and top quark, and their antimatter counterparts.

**R&D:** Research and development. Efforts to develop technology and techniques for new experiments and facilities that extend capability, reduce costs, or reduce risk.

**RDCs:** Research and development collaborations, which develop technologies and techniques for future experiments.

**RF:** Radio frequency, a portion of the electromagnetic spectrum ranging from kHz to GHz. In particle physics, RF is used to directly accelerate particles and as a detection technique.

**RHIC:** Relativistic Heavy Ion Collider, particle collider, currently operating at Brookhaven National Laboratory.

**Rubin:** Vera C. Rubin Observatory. An 8m telescope built on Cerro Pachón ridge in Chile that will perform the LSST.

**SLAC:** SLAC National Accelerator Laboratory.

**Snowmass:** A scientific study conducted by the particle physics community to plan a scientific vision for the future.

**SBN:** Short Baseline Neutrino Program is an experimental program at Fermilab.

**SNB:** Supernova neutrino burst, a phenomenon in which many neutrinos are produced in a short period of time before a supernova explosion occurs.

**SNOLAB:** A deep underground research laboratory located near Sudbury, Ontario, Canada.

**Spec-S5:** A proposed Stage-V ground-based spectroscopic experiment to measure inflation parameters and dark energy from medium galaxy redshifts. It follows the Stage-IV DESI experiment.

**srEDM:** Electric Dipole Moment using Storage Rings. A proposed experiment to measure EDM of protons using an electrostatic storage ring.

**SRF:** Superconducting radio-frequency, also SCRF, where the RF resonators are fashioned of superconducting materials, such that the energy dissipation is lowered.

**SURF:** Sanford Underground Research Facility. A facility hosting a suite of experiments for rare event searches including DUNE, located in Lead, South Dakota.

**SuperCDMS:** Super Cryogenic Dark Matter Search experiment, currently under construction at SNOLAB, set to search for low-mass WIMPs.

**SuperKEKB:** An electron-positron collider located at KEK, Tsukuba, Japan.

**SWGO:** The Southern Wide-field Gamma-ray Observatory. A ground-based gamma-ray detector for mapping large-scale emission and providing access to the full sky for transient and variable multi-wavelength and multi-messenger phenomena.

**Synchrotron:** A type of circular particle accelerator in which the resonant radio frequency is synchronized to the changing velocity of the particles.

**T2K:** Tokai to Kamioka. A long-baseline neutrino experiment operating in Japan. A neutrino beam is sent from the J-PARC proton accelerator facility to the Super-Kamiokande detector in the Kamioka mine over a distance of 295 km.

**TeV:** Tera-electron-volt. A unit of energy that is 1000 times larger than a GeV, or giga electron volt.

**TPC:** Time projection chamber. A type of tracking detector of particles. It drifts ionization electrons to the end of the chamber by a uniform electric field and measures the distance from the end plates by the drift time.

**WIMP:** Weakly interacting massive particle. One of the most promising dark matter candidates.

**XENON-nT:** A direct detection dark matter experiment, focusing on WIMP dark matter particle search. It is operating in the Gran Sasso Underground Laboratory in Italy.

**ZEUS:** Zettawatt-Equivalent Ultrashort pulse laser System, a facility at the University of Michigan that uses intense lasers to drive high intensity science including wakefields in plasma and research on technologies.

# Full List of Recommendations

For convenience, we gather here the full list of recommendations from the report, with the caveat that some meaning may be lost when taken out of context.

**Recommendation 1:** As the highest priority independent of the budget scenarios, complete construction projects and support operations of ongoing experiments and research to enable maximum science.

We reaffirm the previous P5 recommendations on major initiatives:

a.   HL-LHC (including ATLAS and CMS detectors, as well as Accelerator Upgrade Project) to start addressing why the Higgs boson condensed in the universe (*reveal the secrets of the Higgs boson*, section 3.2), to *search for direct evidence for new particles* (section 5.1), to *pursue quantum imprints of new phenomena* (section 5.2), and to *determine the nature of dark matter* (section 4.1).

b.   The first phase of DUNE and PIP-II to open an era of precision neutrino measurements that include the determination of the mass ordering among neutrinos. Knowledge of this fundamental property is a crucial input to cosmology and nuclear science (*elucidate the mysteries of neutrinos*, section 3.1).

c.   The Vera C. Rubin Observatory to carry out the Legacy Survey of Space and Time (LSST), and the LSST Dark Energy Science Collaboration, to *understand what drives cosmic evolution* (section 4.2).

In addition, we recommend continued support for the following ongoing experiments at the medium scale (project costs > $50M for DOE and > $4M for NSF), including completion of construction, operations and research:

d.   NOvA, SBN, T2K, and IceCube (*elucidate the mysteries of neutrinos*, section 3.1).

e.   DarkSide-20k, LZ, SuperCDMS, and XENONnT (*determine the nature of dark matter*, section 4.1).

f.   DESI (*understand what drives cosmic evolution*, section 4.2).

g.   Belle II, LHCb, and Mu2e (*pursue quantum imprints of new phenomena*, section 5.2).

**Recommendation 2:** Construct a portfolio of major projects that collectively study nearly all fundamental constituents of our universe and their interactions, as well as how those interactions determine both the cosmic past and future.

These projects have the potential to transcend and transform our current paradigms. They inspire collaboration and international cooperation in advancing the frontiers of human knowledge. Plan and start the following major initiatives in order of priority from highest to lowest:

a.   CMB-S4, which looks back at the earliest moments of the universe to probe physics at the highest energy scales. It is critical to install telescopes at and observe from both the South Pole and Chile sites to achieve the science goals (section 4.2).

b. A re-envisioned second phase of DUNE with an early implementation of an enhanced 2.1 MW beam—ACE-MIRT—a third far detector, and an upgraded near-detector complex as the definitive long-baseline neutrino oscillation experiment of its kind (section 3.1).

c. An offshore Higgs factory, realized in collaboration with international partners, in order to reveal the secrets of the Higgs boson. The current designs of FCC-ee and ILC meet our scientific requirements. The US should actively engage in feasibility and design studies. Once a specific project is deemed feasible and well-defined (see also Recommendation 6), the US should aim for a contribution at funding levels commensurate to that of the US involvement in the LHC and HL-LHC, while maintaining a healthy US onshore program in particle physics (section 3.2).

d. An ultimate Generation 3 (G3) dark matter direct detection experiment reaching the neutrino fog, in coordination with international partners and preferably sited in the US (section 4.1).

e. IceCube Gen-2, for study of neutrino properties complementary to DUNE and for indirect detection of dark matter covering higher mass ranges, using non-beam neutrinos as a tool. (section 4.1).

**Recommendation 3:** Create an improved balance between small-, medium-, and large-scale projects to open new scientific opportunities and maximize their results, enhance workforce development, promote creativity, and compete on the world stage.

To achieve this balance across all project sizes we recommend the following:

a. Implement a new small-project portfolio at DOE, Advancing Science and Technology through Agile Experiments (ASTAE), across science themes in particle physics with a competitive program and recurring funding opportunity announcements. This program should start with the construction of experiments from the Dark Matter New Initiatives (DMNI) by DOE-HEP (section 6.2).

b. Continue Mid-Scale Research Infrastructure (MSRI) and Major Research Instrumentation (MRI) programs as a critical component of the NSF research and project portfolio.

c. Support DESI-II for cosmic evolution, LHCb upgrade II and Belle II upgrade for quantum imprints, and US contributions to the global CTA Observatory for dark matter (sections 4.2, 5.2, and 4.1).

The Belle II recommendation includes contributions towards the Super-KEKB accelerator.

**Recommendation 4:** Support a comprehensive effort to develop the resources—theoretical, computational, and technological—essential to our 20-year vision for the field. This includes an aggressive R&D program that, while technologically challenging, could yield revolutionary accelerator designs that chart a realistic path to a 10 TeV pCM collider.

Investing in the future of the field to fulfill this vision requires the following:

a. Support vigorous R&D toward a cost-effective 10 TeV pCM collider based on proton, muon, or possible wakefield technologies, including an evaluation of options for US siting of such a machine, with a goal of being ready to build major test facilities and demonstrator facilities within the next 10 years (sections 3.2, 5.1, 6.5, and Recommendation 6).

b. Enhance research in theory to propel innovation, maximize scientific impact of investments in experiments, and expand our understanding of the universe (section 6.1).

c. Expand the General Accelerator R&D (GARD) program within HEP, including stewardship (section 6.4).

d. Invest in R&D in instrumentation to develop innovative scientific tools (section 6.3).

e. Conduct R&D efforts to define and enable new projects in the next decade, including detectors for an $e^+e^-$ Higgs factory and 10 TeV pCM collider, Spec-S5, DUNE FD4, Mu2e-II, Advanced Muon Facility, and line intensity mapping (sections 3.1, 3.2, 4.2, 5.1, 5.2, and 6.3).

f. Support key cyberinfrastructure components such as shared software tools and a sustained R&D effort in computing, to fully exploit emerging technologies for projects. Prioritize computing and novel data analysis techniques for maximizing science across the entire field (section 6.7).

g. Develop plans for improving the Fermilab accelerator complex that are consistent with the long-term vision of this report including neutrinos, flavor, and a 10 TeV pCM collider (section 6.6).

**Recommendation 5:** Invest in initiatives aimed at developing the workforce, broadening engagement, and supporting ethical conduct in the field. This commitment nurtures an advanced technological workforce not only for particle physics, but for the nation as a whole.

The following workforce initiatives are detailed in section 7:

a. All projects, workshops, conferences, and collaborations must incorporate ethics agreements that detail expectations for professional conduct and establish mechanisms for transparent reporting, response, and training. These mechanisms should be supported by laboratory and funding agency infrastructure. The efficacy and coverage of this infrastructure should be reviewed by a HEPAP subpanel.

b. Funding agencies should continue to support programs that broaden engagement in particle physics including strategic academic partnership programs, traineeship programs, and programs in support of dependent care and accessibility. A systematic review of these programs should be used to identify and remove barriers.

c. Comprehensive work-climate studies should be conducted with the support of funding agencies. Large collaborations and national laboratories should consistently undertake such studies so that issues can be identified, addressed, and monitored. Professional associations should spearhead field-wide work-climate investigations to ensure that the unique experiences of individuals engaged in smaller collaborations and university settings are effectively captured.

d. Funding agencies should strategically increase support for research scientists, research hardware and software engineers, technicians, and other professionals at universities.

e. A plan for dissemination of scientific results to the public should be included in the proposed operations and research budgets of experiments. The funding agencies should include funding for the dissemination of results to the public in operation and research budgets.

**Recommendation 6:** Convene a targeted panel with broad membership across particle physics later this decade that makes decisions on the US accelerator-based program at the time when major decisions concerning an offshore Higgs factory are expected, and/or significant adjustments within the accelerator-based R&D portfolio are likely to be needed. A plan for the Fermilab accelerator complex consistent with the long-term vision in this report should also be reviewed.

The panel would consider the following:

a. The level and nature of US contribution in a specific Higgs factory including an evaluation of the associated schedule, budget, and risks once crucial information becomes available.

b. Mid- and large-scale test and demonstrator facilities in the accelerator and collider R&D portfolios.

c. A plan for the evolution of the Fermilab accelerator complex consistent with the long-term vision in this report, which may commence construction in the event of a more favorable budget situation.

---

**Area Recommendation 1:** Increase DOE HEP-funded university-based theory research by $15 million per year in 2023 dollars (or about 30% of the theory program), to propel innovation and ensure international competitiveness. Such an increase would bring theory support back to 2010 levels. Maintain DOE lab-based theory groups as an essential component of the theory community.

**Area Recommendation 2:** For the ASTAE program to be agile, we recommend a broad, predictable, recurring, and preferably annual call for proposals. This ensures the flexibility to target emerging opportunities and fields. A program on the scale of $35 million per year in 2023 dollars is needed to ensure a healthy pipeline of projects.

**Area Recommendation 3:** To preserve the agility of the ASTAE program, project management requirements should be outlined for the portfolio and should be adjusted to be commensurate with the scale of the experiment.

**Area Recommendation 4:** A successful ASTAE experiment involves 3 phases: design, construction, and operations. A design phase proposal should precede a construction proposal, and construction proposals are considered from projects within the group that have successfully completed their design phase.

**Area Recommendation 5:** The DMNI projects that have successfully completed their design phase and are ready to be reviewed for construction should form the first set of construction proposals for ASTAE. The corresponding design phase call would be open to proposals from all areas of particle physics.

**Area Recommendation 6:** Increase the budget for generic Detector R&D by at least $20 million per year in 2023 dollars. This should be supplemented by additional funds for the collider R&D program.

**Area Recommendation 7:** The detector R&D program should continue to leverage national initiatives such as QIS, microelectronics, and AI/ML.

**Area Recommendation 8:** Increase annual funding to the General Accelerator R&D program by $10M per year in 2023 dollars to ensure US leadership in key areas.

**Area Recommendation 9:** Support generic accelerator R&D with the construction of small-scale test facilities. Initiate construction of larger test facilities based on project review and informed by the collider R&D program.

**Area Recommendation 10:** To enable targeted R&D before specific collider projects are established in the US, an investment in collider detector R&D funding at the level of $20M per year and collider accelerator R&D at the level of $35M per year in 2023 dollars is warranted.

**Area Recommendation 11:** To successfully deliver major initiatives and leading global projects, we recommend that:

a. National laboratories and facilities should work with funding agencies to establish and maintain streamlined access policies enabling efficient remote and on-site collaboration by international and domestic partners.

b. National laboratories should prioritize the facilitation of procurement processes and ensure robust technical support for experimenters.

c. National laboratories and facilities should prioritize the creation and maintenance of a supportive, inclusive, and welcoming culture.

**Area Recommendation 12:** Form a dedicated task force, to be led by Fermilab with broad community membership. This task force is to be charged with defining a roadmap for upgrade efforts and delivering a strategic 20-year plan for the Fermilab accelerator complex within the next five years for consideration (Recommendation 6). Direct task force funding of up to $10M should be provided.

**Area Recommendation 13:** Assess the booster synchrotron and related systems for reliability risks through the first decade of DUNE operation, and take measures to preemptively address these risks.

**Area Recommendation 14:** To provide infrastructure for neutrino and/or dark matter experiments, we recommend DOE fund the cavern outfitting of the SURF expansion.

**Area Recommendation 15:** Maintaining the capabilities of NSF's infrastructure at the South Pole, focused on enabling future world-leading scientific discoveries, is essential. We recommend continued and critically important direct coordination and planning between NSF-OPP and the CMB-S4 and IceCube-Gen2 projects.

**Area Recommendation 16:** Resources for national initiatives in AI/ML, quantum computing, and microelectronics should be leveraged and incorporated into research and R&D efforts to maximize the physics reach of the program.

**Area Recommendation 17:** Add support for a sustained R&D effort at the level of $9M per year in 2023 dollars to adapt software and computing systems to emerging hardware, incorporate other advances in computing technologies, and fund directed efforts to transition those developments into systems used for operations of experiments and facilities.

**Area Recommendation 18:** Through targeted investments at the level of $8M per year in 2023 dollars, ensure sustained support for key cyberinfrastructure components. This includes widely used software packages, simulation tools, information resources such as the Particle Data Group and INSPIRE, as well as the shared infrastructure for preservation, dissemination, and analysis of the unique data collected by various experiments and surveys in order to realize their full scientific impact.

**Area Recommendation 19:** Research software engineers and other professionals at universities and labs are key to realizing the vision of the field and are critical for maintaining a technologically advanced workforce. We recommend that the funding agencies embrace these roles as a critical component of the workforce when investing in software, computing, and cyberinfrastructure.

**Area Recommendation 20:** HEPAP, potentially in collaboration with international partners, should conduct a dedicated study aiming at developing a sustainability strategy for particle physics.

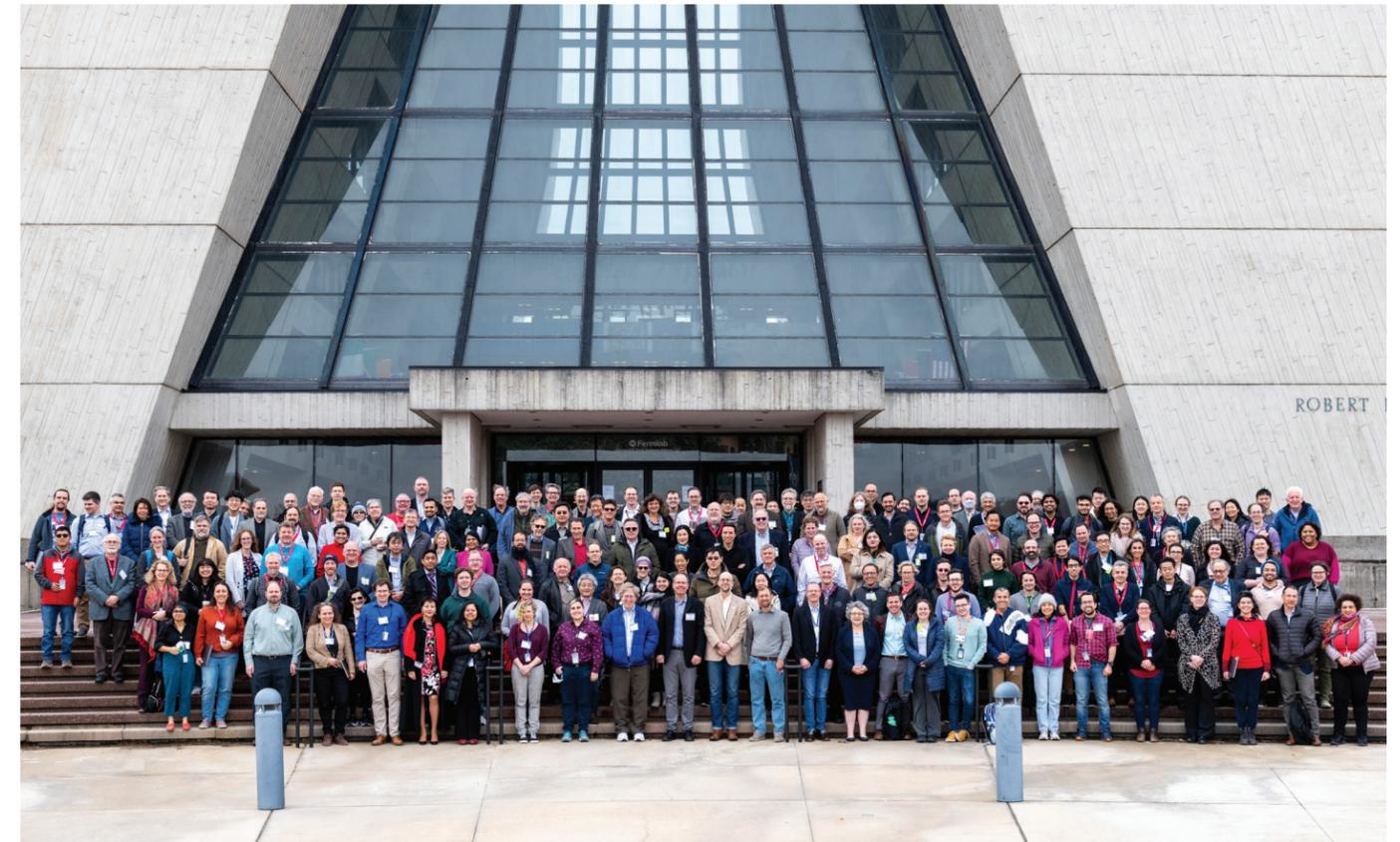

Members of the particle physics community at the Fermilab Town Hall, March 2023. *Photo: Ryan Postel, Fermilab*

We thank members of the cost subcommittee, in particular its chair, Jay Marx, for their timely and hard work. We also thank all the national laboratories that made their staff available for this important task. We thank people at funding agencies for providing us all necessary information and support throughout the process. We thank our peer reviewers for giving us constructive feedback under a tight deadline. We thank Lawrence Berkeley National Laboratory, Fermi National Accelerator Laboratory, Argonne National Laboratory, Brookhaven National Laboratory, SLAC National Laboratory, Virginia Tech University, and University of Texas Austin for hosting the town halls. We thank James Dawson and Marty Hanna for professional editing. We thank Michael Branigan, Brad Nagle, Olena Shmahalo, and Abigail Malate for providing beautiful graphics and layout. We thank the Yale Physics Department for supporting the development of the usparticlephysics.org website. We thank Kerri Fomby, Jody Crisp, and Taylor Pitchford at ORISE and Stephany Tone at LBNL for logistical support. We thank our families for supporting us during this year-long process. And most importantly, we thank the American Physical Society Division of Particles and Fields for organizing the Snowmass Community Study, and all members of our community for their bold and creative vision as well as their input to the process.

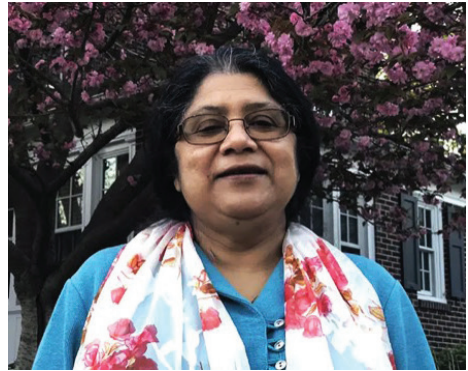

*Courtesy of Ulrich Heintz*

**Meenakshi Narain**, a highly respected member of the particle physics community, was a member of this panel. We were greatly saddened when she passed away on January 1, 2023. We express our deepest condolences to her family. Her passing was a devastating loss to our community.

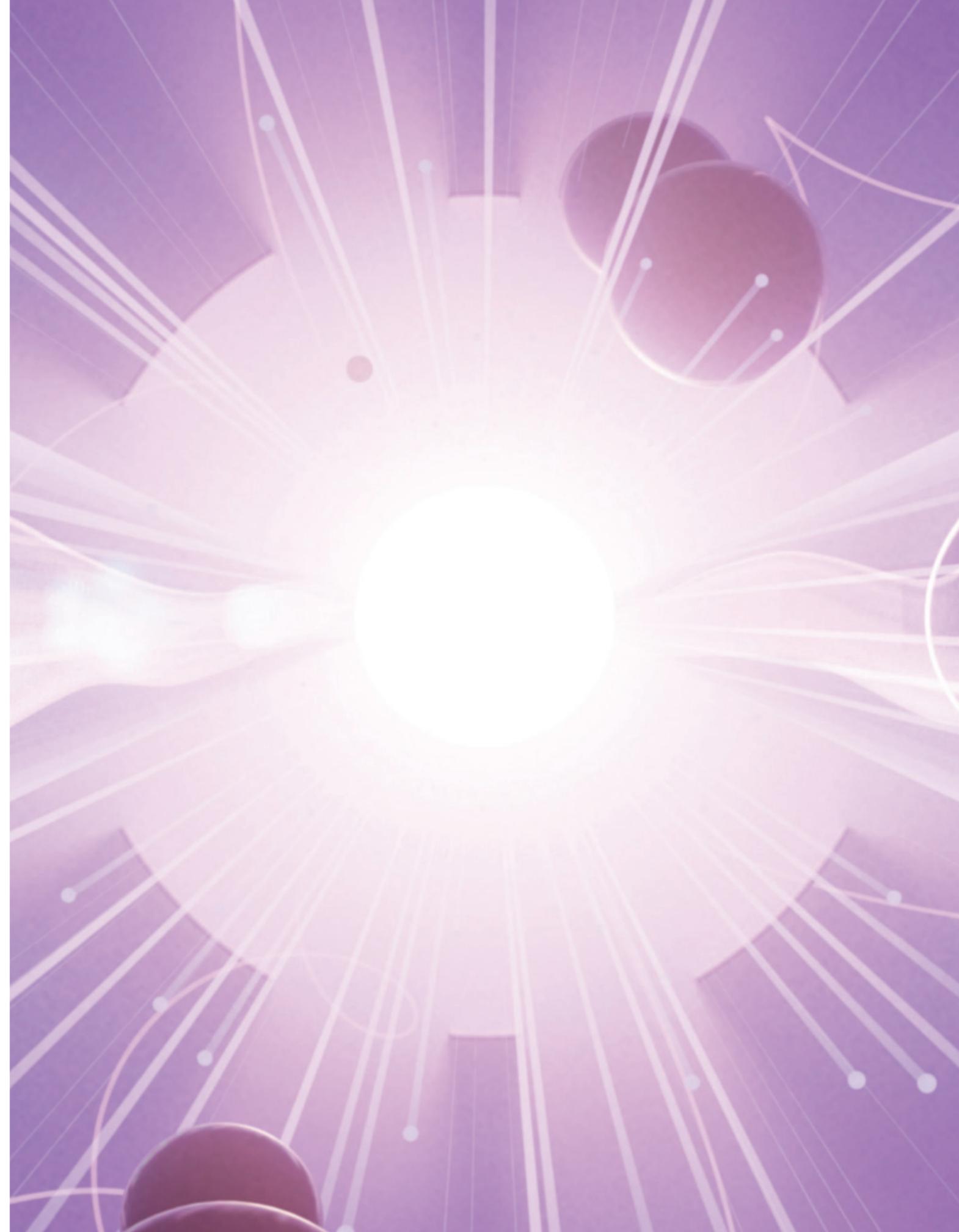

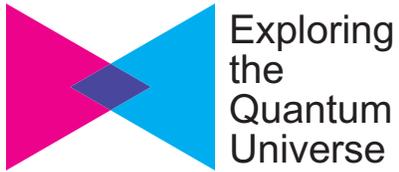

Pathways to Innovation
and Discovery
in Particle Physics

A strategic plan for the High Energy Physics Advisory Panel

2023p5report.org

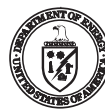
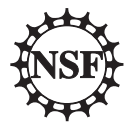